\documentclass[final,3p,times,fleqn]{elsarticle}

\usepackage{graphicx}
\usepackage{epsfig}

\usepackage{amssymb}
\usepackage{color}
\definecolor{lawngreen}  {rgb}{0.33,0.42,0.18}
\definecolor{olive}  {rgb}{0.64,0.80,0.35}
\definecolor{white}{rgb}{1.0,1.0,1.0}
\definecolor{black}{rgb}{0.0,0.0,0.0}
\definecolor{navajo}{rgb}{1,0.871,0.678}
\definecolor{tcol}  {rgb}{0,0.,0.}
\definecolor{emcol} {rgb}{1.,0.,0.}
\definecolor{black}  {rgb}{0.,0.,0.}
\definecolor{green}  {rgb}{0.0,1.,0.}
\definecolor{yellow}  {rgb}{1.,0.9,0.0}
\definecolor{LightGray} {rgb}{.824,.824,.824}

\journal{Journal of Computational Physics}

\begin{document}

\newcommand{\DontShow}[1] {}
\newcommand{\ToDo}[1] {\textbf{ToDo: #1}}
\def\tens#1{\bar{\bar{\ensuremath{\mathsf{#1}}}}}
\newcommand{\papa}[2] { \frac{\partial #1}{\partial #2} }
\newcommand{\papaco}[3] {{ \left. \frac{\partial #1}{\partial #2}\right|_{#3}}}
\newcommand{\tildepapaco}[3] {{ \left. \widetilde{\frac{\partial #1}{\partial #2}}\right|_{#3}}}

\newcommand{\COBOLD}{CO5BOLD}
\newcommand{\I}[1] {i{\rm #1}}
\newcommand{\m}[1] {m{\rm #1}}
\newcommand{\n}[1] {n{\rm #1}}
\newcommand{\mghost}[1] {m{\rm #1}_{\rm ghost}}
\newcommand{\nghost}[1] {n{\rm #1}_{\rm ghost}}
\newcommand{\dx}[1] {\Delta x{\rm_#1}}
\newcommand{\xc}[1] {x_{\mathrm{c} #1}}
\newcommand{\xb}[1] {x_{\mathrm{b} #1}}
\newcommand{\g}[1]{g{\rm_#1}}
\newcommand{\phic}[1]{\Phi_{\mathrm{c} ,\,  #1}}
\newcommand{\phib}[1]{\Phi_{\mathrm{b} ,\,  #1}}
\newcommand{\V}[1]{v_{#1}}
\newcommand{\B}[1]{B_{#1}}
\newcommand{\tildeV}[1]{\tilde{v}{\rm #1}}
\newcommand{\Press} {P}
\newcommand{\ei} {e_\mathrm{int}}
\newcommand{\eg} {e{\rm g}}
\newcommand{\eikb} {e{\rm ikb}}
\newcommand{\eikbg} {e{\rm ikbg}}
\newcommand{\etot} {e_\mathrm{tot}}
\newcommand{\tildeei} {\tilde{e}{\rm i}}
\newcommand{\tildeeik} {\tilde{e}{\rm ik}}
\newcommand{\tildeek} {\tilde{e}{\rm k}}
\newcommand{\tildeeip} {\tilde{e}{\rm ip}}
\newcommand{\eikp} {e{\rm ikp}}
\newcommand{\tildeeikp} {\tilde{e}{\rm ikp}}
\newcommand{\dPdei} {\papaco{P}{\ei}{\rho}}
\newcommand{\dPdrho} {\papaco{P}{\rho}{\ei}}
\newcommand{\drhoeidP} {\papaco{\rhoei}{P}{s}}
\newcommand{\drhoeidrho} {\papaco{\rhoei}{\rho}{P}}
\newcommand{\tildedrhoeidP} {\tildepapaco{\rhoei}{P}{s}}
\newcommand{\rhov}[1]{\rho v{\rm_#1}}
\newcommand{\rhoei} {\rho\ei}
\newcommand{\rhoeik} {\rho e{\rm ik}}
\newcommand{\rhoeip} {\rho e{\rm ip}}
\newcommand{\rhoek} {\rho e{\rm k}}
\newcommand{\rhoeg} {\rho e{\rm g}}
\newcommand{\rhoeikg} {\rho e{\rm ikg}}
\newcommand{\CCour} {C_{\rm Courant}}
\newcommand{\CCourmax} {C_{\rm Courant,max}}
\newcommand{\iup} {i_{\rm up}}
\newcommand{\cs} {c_\mathrm{s}}
\newcommand{\vc} {v_\mathrm{c}}
\newcommand{\wL} {w{\rm L}}
\newcommand{\wR} {w{\rm R}}
\newcommand{\Pred}{{P_{\rm st}}}
\newcommand{\Deltat}{\mbox{\small $\Delta$} t}
\newcommand{\dvdx}[2] { \frac{\Delta \V{#1}}{\dx{#2}} }

\newcommand{\Source}{S}
\newcommand{\dSource}{\Delta \Source}

\newcommand{\Int}{I}
\newcommand{\dInt}{\Delta \Int}
\newcommand{\ImS}{\hat{I}}
\newcommand{\dImS}{\Delta \hat{I}}
\newcommand{\Ired}{\tilde{I}}
\newcommand{\dIred}{\Delta \tilde{I}}

\newcommand{\Dx}{\Delta x}

\newcommand{\dtau}{\Delta \tau}
\newcommand{\dedtau}[1] { \frac{{\rm d} #1}{{\rm d} \tau} }
\newcommand{\dedtaunu}[1] { \frac{{\rm d} #1}{{\rm d} \tau_{\nu}} }
\newcommand{\dededtau}[1] { \frac{{\rm d}^2 #1}{{\rm d} \tau^2} }
\newcommand{\dededtaunu}[1] { \frac{{\rm d}^2 #1}{{\rm d} \tau_{\nu}^2} }

\newcommand{\dedx}[1] { \frac{{\rm d} #1}{{\rm d} x} }

\newcommand{\elt} {\widehat{e}}

\newcommand{\wII}{w(2)}
\newcommand{\wIII}{w(3)}

\newcommand{\Snu}{\ensuremath{S_\nu}}
\newcommand{\Jnu}{\ensuremath{J_\nu}}
\newcommand{\taunu}{\ensuremath{\tau_\nu}}

\newcommand{\aj}{AJ}                     
\newcommand{\apj}{ApJ}                   
\newcommand{\apjl}{ApJL}                 
\newcommand{\apjs}{ApJS}                 
\newcommand{\arfm}{Ann.\ Rev.\ Fluid Mech.}
\newcommand{\jfm}{J.\ Fluid Mech.}       
\newcommand{\aap}{A\mbox{\rm \&}A}       
\newcommand{\aaps}{A\mbox{\rm \&}AS}     
\newcommand{\apss}{APSS}
\newcommand{\araa}{Ann.\ Rev.\ Astron. \& Astrophys.}
\newcommand{\jcp}{J.\ Comput.\ Phys.}     
\newcommand{\jgr}{J.\ Geopys.\ Res.}      
\newcommand{\jqsrt}{J.\ Quant.\ Spectrosc. \& Rad.\ Transfer} 
\newcommand{\lnip}{Lecture Notes in Physics}
\newcommand{\memsai}{Mem.\ Soc.\ Astron.\ Italiana} 
\newcommand{\memsais}{Mem.\ Soc.\ Astron.\ Italiana Suppl.}
\newcommand{\mnras}{MNRAS}
\newcommand{\nat}{Nature}
\newcommand{\na}{New Astronomy}
\newcommand{\pasj}{Publ.\ Astron.\ Soc.\ Japan}
\newcommand{\solphys}{Solar Physics}
\newcommand{\sss}{Space Science Series}
\newcommand{\zap}{Zeitschrift f{\"u}r Astrophysik}
\newcommand{\an}{Astron.\ Nachr.}           
\newcommand{\ans}{Astron.\ Nachr.\ Suppl.}   

\begin{frontmatter}



\title{Simulations of stellar convection \textcolor{tcol}{with CO5BOLD}}


\author[cral,oac]{B.~Freytag}

\author[aip]{M.~Steffen}

\author[lsw]{H.-G.~Ludwig}

\author[ita,cma]{S.~Wedemeyer-B{\"o}hm}

\author[sop,kis]{W.~Schaffenberger}

\author[kis]{O.~Steiner}

\address[cral]{Centre de Recherche Astrophysique de Lyon,
UMR 5574: CNRS, Universit\'e de Lyon, \'Ecole Normale Sup\'erieure de Lyon,
46 all\'ee d'Italie,
F-69364~Lyon Cedex 07,
France}

\address[oac]{Istituto Nazionale di Astrofisica / Osservatorio Astronomico di Capodimonte,
Via~Moiariello~16,
I-80131~Naples,
Italy}

\address[aip]{Leibniz-Institut f\"ur Astrophysik Potsdam (AIP),
An der Sternwarte 16,
D-14482 Potsdam,
Germany}

\address[lsw]{ZAH, Landessternwarte K{\"o}nigstuhl,
D-69117~Heidelberg,
Germany}

\address[ita]{Institute of Theoretical Astrophysics
University of Oslo,
Postboks 1029 Blindern,
N-0315~Oslo, Norway}

\address[cma]{Center of Mathematics for Applications
University of Oslo,
Postboks 1053 Blindern,
N-0316~Oslo, Norway}

\address[sop]{School of Physics
University of Exeter,
Stocker Road,
Exeter, United Kingdom, EX4 4QL}

\address[kis]{Kiepenheuer-Institut f{\"u}r Sonnenphysik,
Sch{\"o}neckstrasse 6,
D-79104~Freiburg,
Germany}

\begin{abstract}
High-resolution images of the solar surface show
a granulation pattern of hot rising and cooler downward-sinking material --
the top of the deep-reaching solar convection zone.
Convection plays a role
for the thermal structure of the solar interior and the dynamo acting there,
for the stratification of the photosphere, where most of the visible light is emitted,
as well as for the energy budget of the spectacular processes in the chromosphere 
and corona.
Convective stellar atmospheres can be modeled by numerically solving the coupled 
equations of (magneto)hydro\-dynamics and non-local radiation transport in the presence of a 
gravity field. The \COBOLD\ code described in this article
is designed for so-called ``realistic'' simulations 
that take into account the detailed microphysics under the conditions in solar or
stellar surface layers (equation-of-state and optical properties of the matter).
These simulations indeed deserve the label ``realistic'' because they reproduce the 
various observables very well -- with only minor differences between different 
implementations.
The agreement with observations has improved over time and the simulations are now
well-established and have been performed for a number of stars.
Still, severe challenges are encountered when it comes to extending these simulations
to include ideally the entire star or substellar object:
the strong stratification leads to completely different conditions in the interior,
the photosphere, and the corona.
Simulations have to cover spatial scales from \textcolor{tcol}{the sub-granular level}
to the stellar diameter
and time scales from photospheric wave travel times to stellar rotation or
dynamo cycle \textcolor{tcol}{periods}.
Various non-equilibrium processes have to be taken into account.
Last but not least, realistic simulations are based on detailed microphysics and depend
on the quality of the input data, which can be the actual accuracy limiter.
This article provides an overview of the physical problem and the numerical solution
and the capabilities of \COBOLD, illustrated with a number of applications.  
\end{abstract}


\end{frontmatter}


\section{Introduction}
\label{sec:introduction}

In the core of the Sun, fusion of hydrogen to helium releases energy which is
transported outward, first by radiation only, but \textcolor{tcol}{further out} primarily 
by convection in the outer 30\,\% of the \textcolor{tcol}{radial} distance to the solar surface.
Most of this energy is emitted in \textcolor{tcol}{the} form of radiation in the photosphere
which is the \textcolor{tcol}{bottom layer of the solar atmosphere}.
\textcolor{tcol}{Furthermore}, a small part of the energy is carried by waves and 
\textcolor{tcol}{by}
magnetic fields, powering the dramatic phenomena visible in the solar
chromosphere and corona. In more massive and further evolved stars, the 
internal structure is more complex, with several shells \textcolor{tcol}{where} nuclear 
burning \textcolor{tcol}{takes place} and multiple convection zones.
 
The relatively thin solar photosphere (about 0.1\,\% of the solar radius)
therefore plays an important role for the inner as well as for the outer 
layers of the Sun.
The analysis of solar and stellar spectra can reveal surface properties
and the chemical composition, \textcolor{tcol}{and allows us} to draw conclusions about the
internal structure and evolutionary status.
For this purpose, physical models of stellar atmospheres
\textcolor{tcol}{with} a realistic treatment of both radiation
and convection are essential.

The classical analysis relies on \textcolor{tcol}{one-dimensional} (1D) stationary model atmospheres
(in most cases only the photosphere plus the very top layers
of the surface convection zone), where the average 
convective energy flux $F_{\rm conv}$ is computed from the 
so-called local mixing-length theory \cite{Vitense1953ZA.....32..135V},
\cite{BohmVitense1958ZA.....46..108B}, 
\cite{Mihalas1978stat.book.....M},
a heuristic recipe which assumes that $F_{\rm conv}$ can be determined 
from local properties \textcolor{tcol}{of the stratification}.
In the framework of this ``theory'', the mean 
thermal structure of a convective stellar atmosphere is found by the
requirement that the sum of radiative and convective flux equals the
total stellar flux, $F_{\rm rad}+F_{\rm conv}=\sigma\,T_\mathrm{eff}^4$, at all
depths. 
State-of-the-art radiative-convective equilibrium models of solar
and stellar atmospheres have been constructed with the classical model
atmosphere codes
ATLAS \cite{Kurucz1970SAOSR.309.....K, Kurucz2005MSAIS...8...14K},
MARCS \cite{Gustafsson1975A&A....42..407G, Gustafsson2008A&A...486..951G},
and PHOENIX \cite{Allard1995ApJ...445..433A, Allard2001ApJ...556..357A}, to 
name the most prominent examples.

\begin{figure}[tp]
\centering
\includegraphics[bb=50 10 420 400, width=5cm]{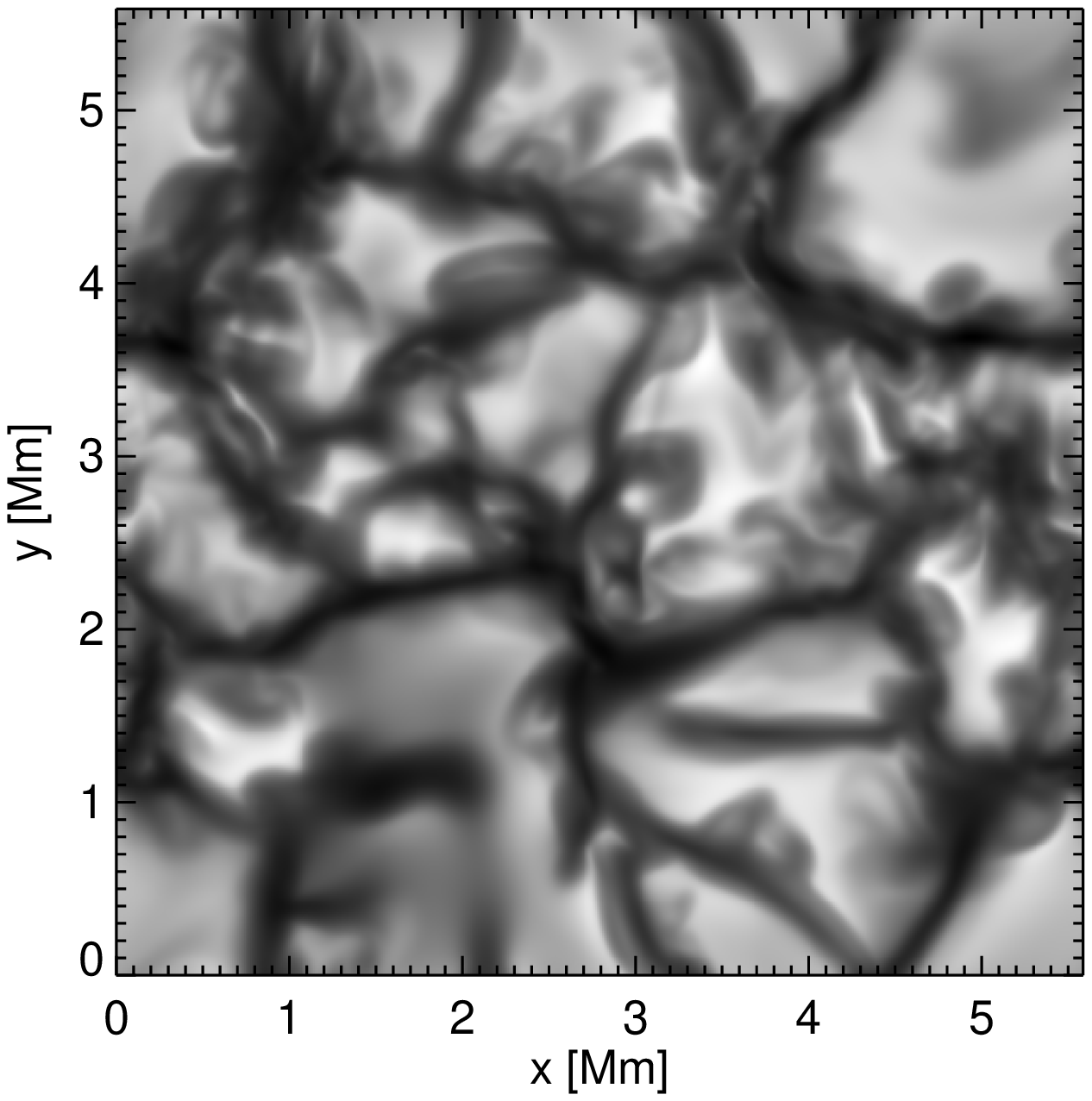}
\includegraphics[bb=20 5 290 280, width=7cm]{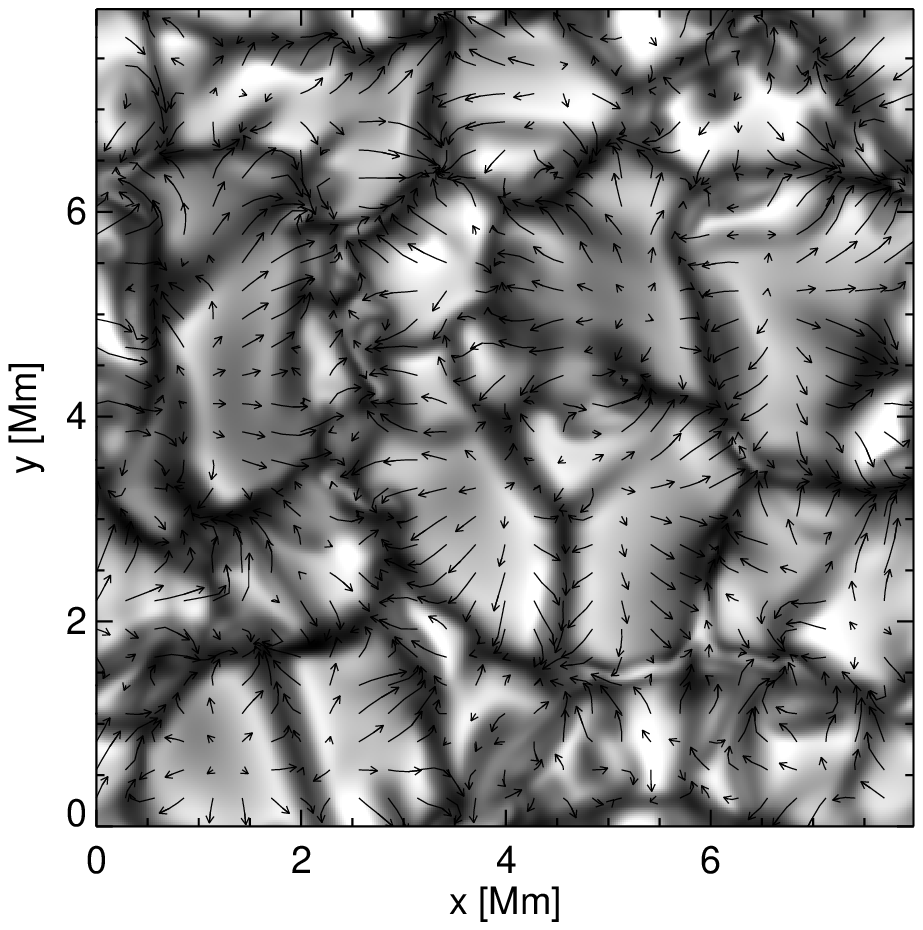}
\caption{Emergent continuum intensity at a wavelength of $\lambda = 500$\,nm,
synthesized for \textcolor{tcol}{a snapshot from a high-resolution ($400\times400\times300$) 
\COBOLD\ RHD simulation (left, see Sect.\,\ref{sec:codecomp}) and for a 
($286\times286\times266$) \COBOLD\ RMHD model (right, cf.\ 
Fig.\,\ref{fig:chromospherexz}), each representing
a small patch of solar surface granulation. The imposed average magnetic flux 
of the RMHD model is $\langle B_z\rangle=50$~G.} The arrows represent 
streamlines 
that follow the horizontal velocity on the surface of optical depth 
unity, i.e., at the bottom of the solar photosphere.}
\label{f:granulation}
\end{figure}

A severe drawback of these models is that the efficiency of the convective 
energy transport is controlled by a free parameter, the mixing-length 
parameter $\alpha_{\rm MLT}$, which is of the order unity but \emph{a priori} 
unknown. Therefore, $\alpha_{\rm MLT}$ must be calibrated against
observations. Unfortunately, different observables require different values 
of $\alpha_{\rm MLT}$ \cite{Steffen2000eaa..bookE5198S}.
\textcolor{tcol}{The best fit of the Balmer line profiles of solar-type stars is achieved with 
$\alpha_{\rm MLT}\approx 0.5$ \cite{Fuhrmann1993A&A...271..451F}, while 
continuum colors are better reproduced with $\alpha_{\rm MLT}$ in the range 
1--2, depending on the considered wavelength range
\cite{Steffen1999ASPC..173..217S}. The standard stellar-evolution 
calibration based on matching the current solar parameters calls for
$\alpha_{\rm MLT}\approx 2$ 
\cite{Christensen-Dalsgaard1996Sci...272.1286C}, 
\cite{Bahcall2001ApJ...555..990B}.
This disparity indicates that the underlying theoretical description 
is inadequate.}
In fact, the solar photosphere is neither homogenous nor static, since
it is influenced by the very top of the convection zone and shows
a granular pattern of bright upflow regions surrounded by darker
intergranular lanes of downflowing material, with a spatial scale of 
about 1\,Mm \textcolor{tcol}{($10^6$~m)} and evolving on a time scale of minutes 
(see Fig.\,\ref{f:granulation} for snapshots from \textcolor{tcol}{two \COBOLD\ simulations}
of the solar granulation). This motivated various efforts to overcome
the limitations of the 1D classical atmospheres and to develop instead 
self-consistent, parameter-free hydro\-dynamical models of stellar surface 
convection, accounting for the fact that convection is a non-local, 
time-dependent, and intrinsically three-dimensional phenomenon.

Early idealized numerical simulations of convection under stellar-like conditions
had to resort to severe simplifications (stationary 2D solutions on coarse grids) and could only
deliver qualitative results:
\citet{Latour1976ApJ...207..233L,Latour1981ApJ...248.1081L} and 
\citet{Toomre1976ApJ...207..545T}
used anelastic modal equations to study surface convection in A-type stars.
\citet{Musman1976ApJ...207..981M} and
\citet{Nelson1978SoPh...60....5N}
investigated convection in the Sun and some other stars with a similar method.
\textcolor{tcol}{
\citet{chan+wolff1982}
developed a code based on the alternating direction implicit (ADI) method
for the calculation of compressible convection.
\citet{Hurlburt1984ApJ...282..557H} carried out simulations
of compressible solar convection extending over multiple scale heights.  
\citet{Steffen1989A&A...213..371S} took (non-local) radiation transfer
into account in their 2D simulations of compressible solar convection.
}

The first realistic simulations of solar granulation were performed by
\citet{Nordlund1982A&A...107....1N} and included \textcolor{tcol}{three-dimensional} (3D) time-dependent
hydro\-dynamics (but anelastic and with moderate \textcolor{tcol}{spatial} resolution) and non-local
radiative energy transfer, \textcolor{tcol}{already then with} a simple treatment of the
frequency-dependence of the opacities. Hand-in-hand came the a posteriori detailed
spectrum synthesis by \citet{Dravins1981A&A....96..345D}.
\textcolor{tcol}{
Other 3D convection simulations relinquished the treatment
of radiation transfer
\cite{Toomre1990CoPhC..59..105T,
Malagoli1990ApJ...361L..33M,
Cattaneo1991ApJ...370..282C,
Hossain1991ApJ...380..631H,
Singh1993A&A...279..107S}.}
Current \textcolor{tcol}{radiation hydrodynamic} codes of
various groups use similar basic techniques -- in a significantly refined way
(compressible hydro\-dynamics, more grid points, more \textcolor{tcol}{opacity} bins, larger
computational domains, magnetic fields, \textcolor{tcol}{a chemical reaction network, dust, etc.).
For example, Stein and Nordlund have carried out \textcolor{tcol}{radiation hydrodynamics (RHD)}
simulations with
$2016 \times 2016 \times 500$ grid points with a spatial resolution of
24\,km in the horizontal direction and 12-80\,km in the vertical
direction.
Asplund et al.~\cite{Asplund2005ARA&A..43..481A, Asplund2009ARA&A..47..481A}
have computed chemical abundances using high spatial resolution
and accurate radiative transfer.}

Just like classical stellar atmospheres, the non-magnetic hydro\-dynamical
models are characterized by the average total energy flux per unit area and
time (effective temperature, $T_\mathrm{eff}$), surface gravity $g$, and chemical 
composition. But,
in contrast to the mixing-length models, there is no longer any free parameter
to adjust the efficiency of the convective energy transport. Similarly, the
fudge parameters micro- and macroturbulence, that have to be introduced in 1D
model atmospheres to match synthetic and observed shapes of spectral lines, are replaced by
the self-consistent hydro\-dynamical velocity field of the 3D simulations.
However, one has to keep in mind that the simulations are characterized by a 
large number of numerical parameters, e.g., the spatial resolution of the 
numerical grid, the size of the computational domain, the formulation of 
boundary conditions, and the parameters related to the numerical schemes
for solving the hydro\-dynamical and radiation transport equations. 
Of course, the hope is that the simulation results become essentially
independent of the choice of these numerical parameters, once a sufficiently 
high spatial, angular, and frequency resolution is achieved.

Hydro\-dynamical model atmospheres are not only computed for the Sun but also for
other stars, and are complementing and increasingly replacing classical
1D atmosphere models.
Important applications of convection simulations \textcolor{tcol}{with \COBOLD} and its predecessor include
the accurate spectroscopic determination of
solar and stellar chemical abundances and isotopic ratios
\citep[e.g., ][]{Caffau2008A&A...488.1031C,
Caffau2010SoPh..tmp...66C,
Cayrel2007A&A...473L..37C},
the theoretical calibration of the mixing-length parameter
\citep{Ludwig1999A&A...346..111L}, the study of
convective overshoot and mixing processes in stellar envelopes
\citep{Freytag1996A&A...313..497F},
and the excitation of waves by turbulent
convective flows \citep{Samadi2003A&A...404.1129S,
Samadi2007A&A...463..297S,
Straus2008ApJ...681L.125S}.


The presence of magnetic fields results in a wide range of additional complex 3D
phenomena. Small-scale concentrations of magnetic flux lead to enhanced
radiative losses, both in the photosphere and in the chromosphere. On the
other hand, large-scale magnetic structures can inhibit the convective energy
flux and produce the well-known dark sunspots. The interaction of convection
and magnetic fields can be modeled in the framework of (ideal)
magneto\-hydro\-dynamics (MHD).

In the purely hydro\-dynamical simulations described above, the resulting mean
flow is determined only by the prescribed physical quantities $T_\mathrm{eff}$,
$g$, and the assumed chemical composition, and is largely independent of the
formulation of the boundary conditions and details of the initial
configuration. This is no longer true for the more complex simulations of
solar magneto\-convection. In this case, the presence of a magnetic field
implies more freedom in setting up the problem: the initial 
configuration of the magnetic field and the magnetic boundary conditions have 
to be designed for the particular problem under consideration. In many
studies, the magnetic field is assumed to be vertical at the upper and lower
boundaries, such that the horizontally averaged  magnetic flux is fixed 
at a prescribed value. \textcolor{tcol}{For example,} $\langle B_z\rangle = 50$~G 
for the \COBOLD\ MHD simulation shown in Fig.\,\ref{f:granulation}, 
\textcolor{tcol}{which is}
representative of \textcolor{tcol}{the least magnetic solar-surface areas, 
the so-called}  quiet-Sun internetwork regions.
The velocity arrows in this figure show that the flow
converges towards the dark intergranular lanes, where cool gas returns into
the solar convection zone. This flow also leads to a concentration of 
magnetic flux in the downflow lanes, where is is visible as bright knots
or elongated features (e.g., near $x=5.5$~Mm, $y=1.5$~Mm in 
Fig.\,\ref{f:granulation}, right panel).

\textcolor{tcol}{
Early 2D MHD simulations of solar convection, which
include radiative transfer were presented by
\citet{Grossmann-Doerth1989ssg..conf..481G}
based on a adaptive moving finite element code, by
\citet{steiner+al1998}
with a finite-volume code based on automatic adaptive mesh refinement for MHD, described in
\citet{steiner+al1994},
and by
\citet{Atroshchenko1996KPCB...12...21A}
who used a method of approximate Eddington factors for the radiative transfer.
}

To our knowledge, the first realistic three-dimensional radiation
magneto\-hydro\-dynamic (RMHD) simulation of stellar magneto\-convection
was presented by \citet{Nordlund1986ssmf.conf...83N}.
\citet{nordlund+al_lrsp2009}
give a review on solar surface convection including results on
magneto\-convection.
Early two-dimensional \textcolor{tcol}{MHD simulations of stellar magneto-convection, which
dispense with detailed radiative transfer include
Galloway and Weiss \cite{Galloway1981ApJ...243..945G},
Deinzer et al.\ \cite{deinzer+al1984a},
Hurlburt and Toomre (1988) \cite{hurlburt+toomre1988},
Weiss et. al. (1990) \cite{Weiss1990MNRAS.245..434W}, and
Fox et al. (1991) \cite{Fox1991ApJ...383..860F}.}

The pioneering work of Nordlund and collaborators was only recently
followed up by others, also working in three spatial dimensions.
\textcolor{tcol}{Examples include} \citet{Hansteen2005ESASP.592..483H}
\textcolor{tcol}{and \citet{Gudiksen2011A&A...531A.154G}} with the Bifrost code,
\citet{2005ESASP.596E..65S} with \COBOLD,\footnote{See
  http://www.astro.uu.se/\~{}bf/co5bold\_main.html, http://www.co5bold.com} and
\citet{Voegler2005A&A...429..335V} with the MURaM code,\footnote{See
  http://www.mps.mpg.de/projects/solar-mhd/muram\_site/code.html} and
more recently, by \citet{heinemann+al2007} using the Pencil
code,\footnote{See http://www.nordita.org/software/pencil-code/}
\citet{jacoutot+al2008} with a code named SolarBox, developed by A. Wray, and
\citet{muthsam+al2010} with the Antares code. Recent impressive
large-scale 3D RMHD simulations include the supergranulation-size 
magneto\-convection simulations by  
\citet{Stein2009ASPC..416..421S}, using a variant of the STAGGER code
of \citet{Nordlund+Galsgaard1995},
the simulations of sunspots and solar active regions
described in \citet{cheung+al2010} and in \citet{rempel+al2009},
both works using the MURaM code, as well as the exploratory MHD 
models that span the entire solar atmosphere from the upper convection 
zone to the lower corona by \citet{Hansteen2007ASPC..368..107H}, 
\cite{Hansteen2010ApJ...718.1070H}, and
\citet{Martinez-Sykora2008ApJ...679..871M}, based
on Bifrost or an extended version of the STAGGER code.

Other three-dimensional simulations of stellar magneto\-convection use
approximations to the radiation transfer, like
\citet{2007ApJ...665.1469A,Abbett2010MmSAI..81..721A}, and \citet{isobe+al2008}.
Important results of solar magneto\-convection in
three spatial dimensions were also obtained by simply replacing the
radiation transfer with heat conduction, e.g., by
\citet{weiss+al1996}, \citet{tobias+al1998}, \citet{cattaneo1999},
\citet{ossendrijver+al2001}, or \citet{cataneo+al2003}.  For other
applications, radiative exchange or heat conduction is not as critical
as for convection, e.g., for the rise of buoyant magnetic flux tubes.
Such simulations were carried out, e.g., by \citet{archontis+al2005}
with the STAGGER code \cite{Nordlund+Galsgaard1995} or by
\citet{cheung+al2006} with the Flash code.\footnote{See
  http://flash.uchicago.edu/website/home/}

Simulations of global stellar convective dynamos have been started by
\citet{glatzmaier1984}.  More recent global MHD simulations of stellar
convection include \citet{browning+al2006} with the ASH-code
\cite{clune+al1999} and \citet{dobler+al2006} with the Pencil code.
\citet{ziegler2005} applied the \textcolor{tcol}{Nirvana}
code\footnote{See http://nirvana-code.aip.de/} to the problem of 
core collapse and fragmentation of a magnetized protostellar cloud.

Further MHD codes for potential application to realistic stellar 
convection simulations, which have been developed in an astrophysical 
context are the A-MAZE code,\footnote{See
  http://www.the-a-maze.net/people/folini/research/a\_maze/a\_maze.html}
the Enzo code,\footnote{See http://lca.ucsd.edu/portal/software/enzo}
the VAC code,\footnote{See http://grid.engin.umich.edu/\~{}gtoth/VAC/}
or the Zeus code,\footnote{See
  http://lca.ucsd.edu/portal/codes/zeusmp2} for a non exhaustive
list.

For reviews on solar magneto\-convection see
\citet{nordlund+al_lrsp2009, Nordlund2009AIPC.1171..242N,
Carlsson2009MmSAI..80..606C, steiner2010}.

\section{Basics}

\subsection{Basic considerations about convective scales}
\label{sec:BasisConsiderations}

Ideally, hydrodynamical models of stellar convection should comprise the entire
convection zone in a spherical shell with sufficient spatial resolution, and 
should cover all relevant time scales. In general, such a global approach is 
not feasible, however, for the reasons outlined in the following basic 
considerations.

\subsubsection{Spatial scales}

Presently, realistic models of stellar convection are restricted to a small
representative volume located near the surface, including both the top layers 
of the convection zone and the photosphere, where most of the stellar radiation
is emitted. In this context, it is important to realize that convection is 
driven by entropy fluctuations generated near the surface by radiative
cooling. The deeper layers approach an adiabatic mean state and have little 
direct influence on the small-scale granular flows at the surface. For this 
reason, it is possible to obtain physically consistent ab initio models of 
stellar surface convection from local-box simulations
that cover only a small fraction of the geometrical depth of the whole
convection zone. \textcolor{tcol}{Since the lower boundary is thus located right inside the 
convection zone where the total stellar luminosity is entirely carried 
by the convective flow, it is essential to employ an \emph{open} lower boundary
condition that impedes the flow as little as possible (details are given
in Sect.\,\ref{sec:LocalModels}).}

As a typical example, let us consider a local-box simulation 
of the solar granulation measuring $L_x \times L_y = 10$~Mm $\times 10$~Mm in 
the horizontal directions with periodic lateral boundary conditions in $x$ and 
$y$. In the vertical direction, open boundaries are imposed, and the extension 
of the box is assumed to be $L_z=4$~Mm, with \textcolor{tcol}{$L_z^- \approx 3$~Mm 
($\Delta \ln P \approx 7$) below and $L_z^+ \approx 1$~Mm 
($\Delta \ln P \approx 8$)} above the optical surface, where 
\textcolor{tcol}{$\Delta \ln P$ is the number of gas pressure e-foldings.}
A box of this size covers
only $1.5$\% of the total depth of the solar convection zone, but is large
enough to accommodate several surface convection cells called granules
(cf.\ Fig.\,\ref{f:granulation}), ensuring that the periodic boundary
conditions do not have a critical influence on the resulting flow pattern.
The minimum spatial resolution of the numerical grid is set by the requirement
to cover one pressure scale height by at least $10$ \textcolor{tcol}{grid} cells.
In the following, we assume that a typical grid comprises 
$N_x\times N_y\times N_z = 250\times 250\times 200$ cells,
where the horizontal cell size is constant ($\Delta x = \Delta y \approx 
40$~km), while the vertical cell size increases with depth \textcolor{tcol}{(in proportion to
the local pressure scale height $H_p$, see below)} from about $10$~km at the 
surface to about $50$~km near the bottom of the computational domain 
(for some actual examples see Table~\ref{T1}).
\begin{figure}[tp]
\centering
\mbox{\includegraphics[bb=40 40 580 360,width=6cm]{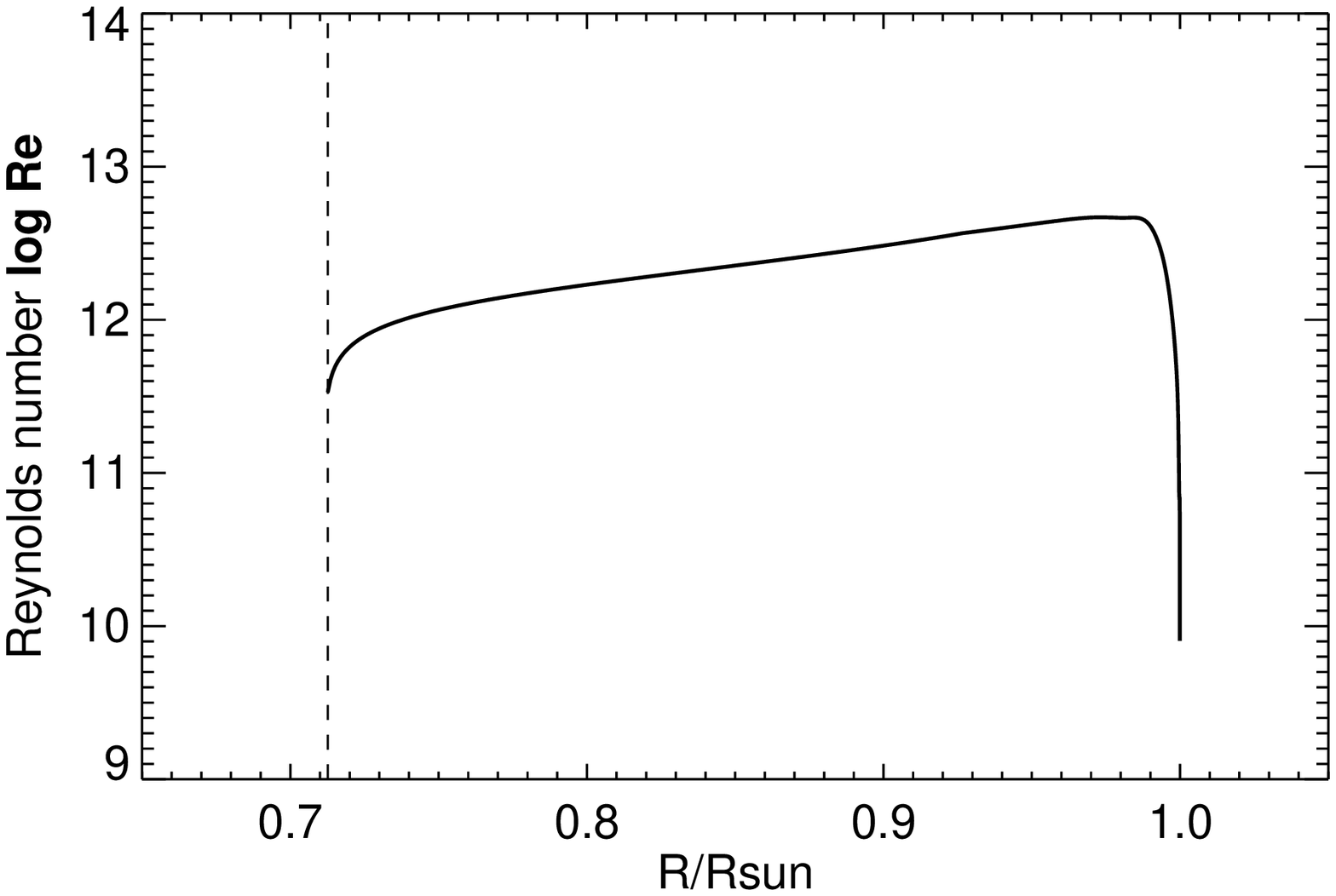}}
\mbox{\includegraphics[bb=40 40 580 360,width=6cm]{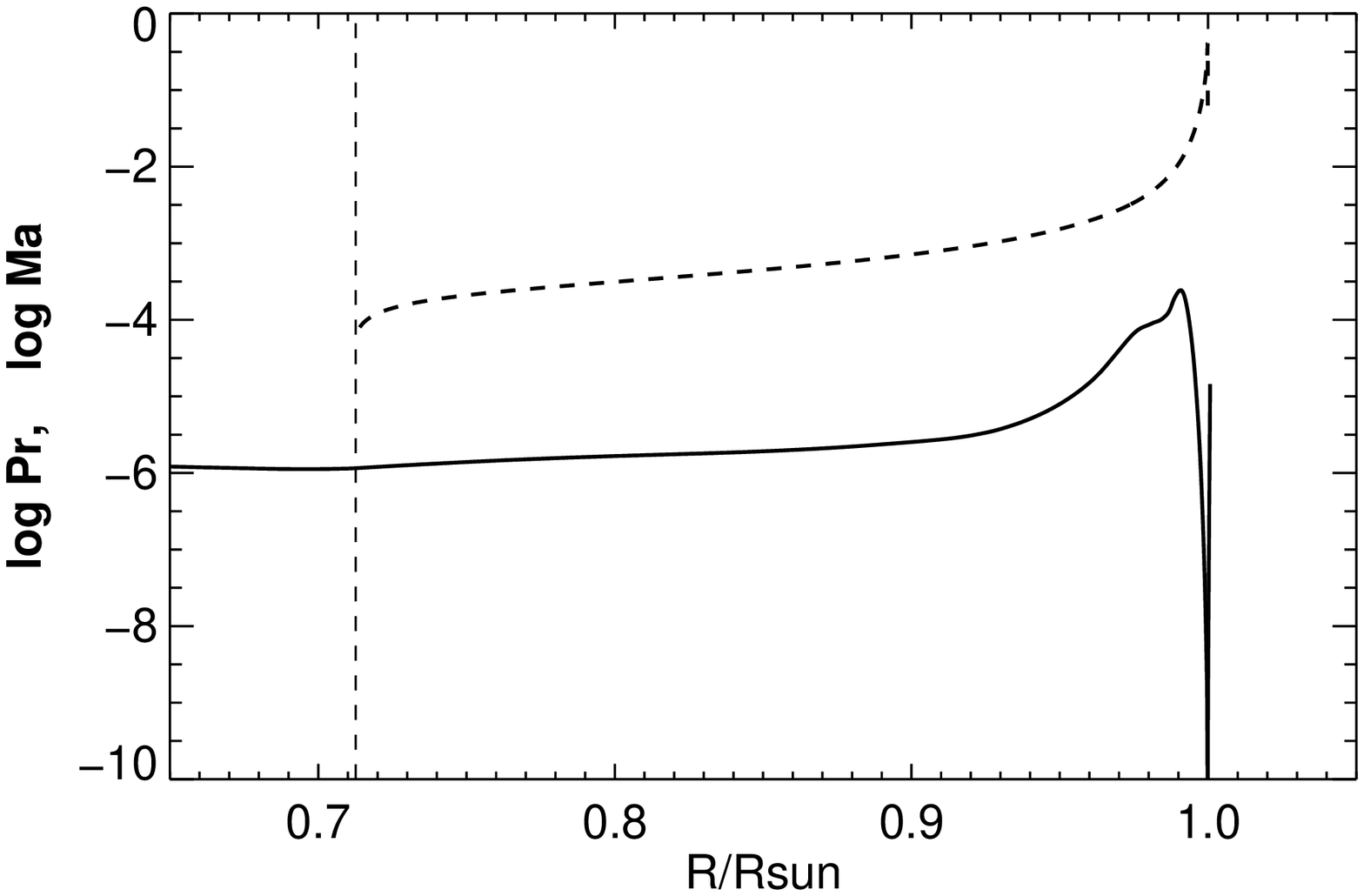}}
\caption{Reynolds number Re (left) \textcolor{tcol}{according to Eq.\,(\ref{eq:Re}),} 
and Prandtl number Pr (right) \textcolor{tcol}{computed from  Eqs.\,(\ref{eq:Pr2}) and
(\ref{eq:taurad}) with $k=k_0=10\,\pi/H_p$,} as function of radius in the 
envelope and atmosphere of the Sun, using the solar model by 
\citet{Christensen-Dalsgaard1996Sci...272.1286C}. In addition, the 
upper \textcolor{tcol}{(dashed)} curve in the right panel refers to the Mach number,
$\mathrm{Ma}=\vc/\cs$. The vertical dashed line marks the bottom of 
the solar convection zone.
}
\label{f:RePr}
\end{figure}

It is well known that the convective envelope of the Sun is characterized by
very large flow Reynolds numbers, Re. Based on the standard solar model of 
\citet{Christensen-Dalsgaard1996Sci...272.1286C}, we have evaluated this 
dimensionless number locally as
\begin{equation}\label{eq:Re}
\mathrm{Re} = \frac{\vc\,H_p}{\nu}
  \enspace ,
\end{equation}
\textcolor{tcol}{where  $H_p=-(\mathrm{d} \ln P/\mathrm{d} z)^{-1}$
is the local pressure scale height
(e-folding length of the gas pressure $P$),}
$\vc$ is the 
characteristic convective velocity according to classical mixing-length 
theory \cite{Vitense1953ZA.....32..135V, BohmVitense1958ZA.....46..108B},
and $\nu$ is the microscopic (atomic plus radiative) kinematic viscosity, 
$\nu=(\eta_a+\eta_r)/\rho$, with $\eta_a$ and $\eta_r$ calculated according to 
\citet{Spitzer1962pfig.book.....S} and \citet{Thomas1930QJMat...1..239T},
respectively. The depth dependence of Re in the solar envelope is displayed 
in the left panel of Fig.\,\ref{f:RePr}, showing that Re~$> 10^{10}$ in the
entire convection zone. This implies that the flow is highly turbulent
wherever convection occurs (see however \cite{Nordlund1997A&A...328..229N}).
The turbulent kinetic energy is dissipated into heat at the Komolgorov 
microscale, $\ell \approx H_p\,\mathrm{Re}^{-3/4}$, which varies between
$0.05$ and $10$~cm from the top to the base of the solar convection zone.
Clearly, the spatial resolution of the numerical simulations sketched above
is insufficient by more than 6 orders of magnitude to properly resolve
the complete turbulent cascade. All realistic stellar convection 
simulations therefore follow the so-called large-eddy approach, where only 
the largest flow structures, including the driving scales, are resolved, and 
the small-scale kinetic energy is dissipated at the grid scale, either by the
numerical scheme or by a subgrid-scale model. Consequently, the effective 
numerical viscosity in such models is at least 8 orders of magnitude larger 
than in reality.

\textcolor{tcol}{In addition to the Reynolds number, the properties of the flow 
are further characterized  by the (dimensionless) Prandtl number,}
\begin{equation}\label{eq:Pr}
\mathrm{Pr} = \frac{\nu}{\chi}
  \enspace ,
\end{equation}
\textcolor{tcol}{the ratio of the coefficients describing the diffusion of
momentum, $\nu$, and heat, $\chi$. In the stellar interior and atmosphere, 
heat transfer is dominated by radiation, which in the optically 
thick layers can be described as a diffusion process. The radiative 
diffusivity is given by}
\begin{equation}\label{eq:chi}
\chi = \frac{16\,\sigma\,T^3}{3\,\kappa\,\rho^2\,c_v}
  \enspace 
\end{equation}
($\sigma$: Stefan-Boltzmann constant, $T$: temperature, 
$\kappa$: radiative opacity per unit mass,
$\rho$: mass density, $c_v$: specific heat at constant volume).
\textcolor{tcol}{Pr depends only on the thermodynamic state of the stellar gas.
In the optically thin layers (photosphere), radiative heat exchange cannot 
be described as a diffusion process, and hence 
the definition of Pr via Eqs.\,(\ref{eq:Pr}) and (\ref{eq:chi})
is no longer meaningful. Instead, Pr can be defined more generally as 
\begin{equation}\label{eq:Pr2}
\mathrm{Pr} = \frac{t_{\rm rad}}{t_{\rm vis}}
  \enspace ,
\end{equation}
the ratio of radiative time scale ($t_{\rm rad}$, see Eq.\,(\ref{eq:taurad}) 
below) to viscous time scale  ($t_{\rm vis}^{-1} = \nu\,k^2$). However, the 
Prandtl number then becomes a function of wavenumber $k$ for optically thin 
conditions.} In the solar convection zone and atmosphere, Pr ranges between 
\textcolor{tcol}{$10^{-4}$ and $10^{-10}$} (see Fig.\,\ref{f:RePr}, right panel), indicating that 
\textcolor{tcol}{the} radiative energy diffusion is much more efficient
than the viscous diffusion of momentum, in other words,
the dynamical lifetime of a turbulent vortex is 
much longer than its thermal relaxation time.

\textcolor{tcol}{
In large-eddy simulations, diffusion is provided by an explicit
artificial viscosity and/or by the numerical advection scheme, which leads
to a diffusive cutoff at the scale of the grid resolution. In general, the
effective viscosity depends on the grid resolution $\Delta x$, and on
the wavenumber (and amplitude) of the local velocity perturbation. 
For small-scale structures close to the grid resolution, the coefficients 
characterizing the \emph{numerical} diffusion of momentum, $\tilde{\nu}$, 
and heat, $\tilde{\chi}$, are of similar size, and hence the Prandtl number
is of the order unity, $\mathrm{Pr}=\tilde{\nu}/\tilde{\chi}\approx 1$,
as long as the radiative diffusivity is much smaller 
than the numerical one, $\chi \ll \tilde{\chi} \approx \vc\,\Delta x$. 
This condition always holds in the bulk of the solar convection zone
(assuming $\Delta x \approx H_p/10$).
On the other hand, the effective artificial/numerical diffusion can be 
significantly smaller for well resolved smooth structures, such that 
$\chi > \tilde{\chi}$, and $\mathrm{Pr}\approx \tilde{\nu}/\chi < 1$.
This is especially true for the near-surface layers where the radiative
diffusivity is high. Large-eddy simulations of solar-type surface
convection can therefore achieve moderately low Prandtl numbers,
in the sense that the physical radiative energy transport dominates over
numerical diffusion of heat. In the bulk of the convection zone, however,
the radiative diffusivity is too low, and hence numerically 
$\mathrm{Pr}\approx 1$ on all resolved scales. 
}

\begin{figure}[tp]
\centering
\mbox{\includegraphics[bb=40 40 580 360,width=6cm]{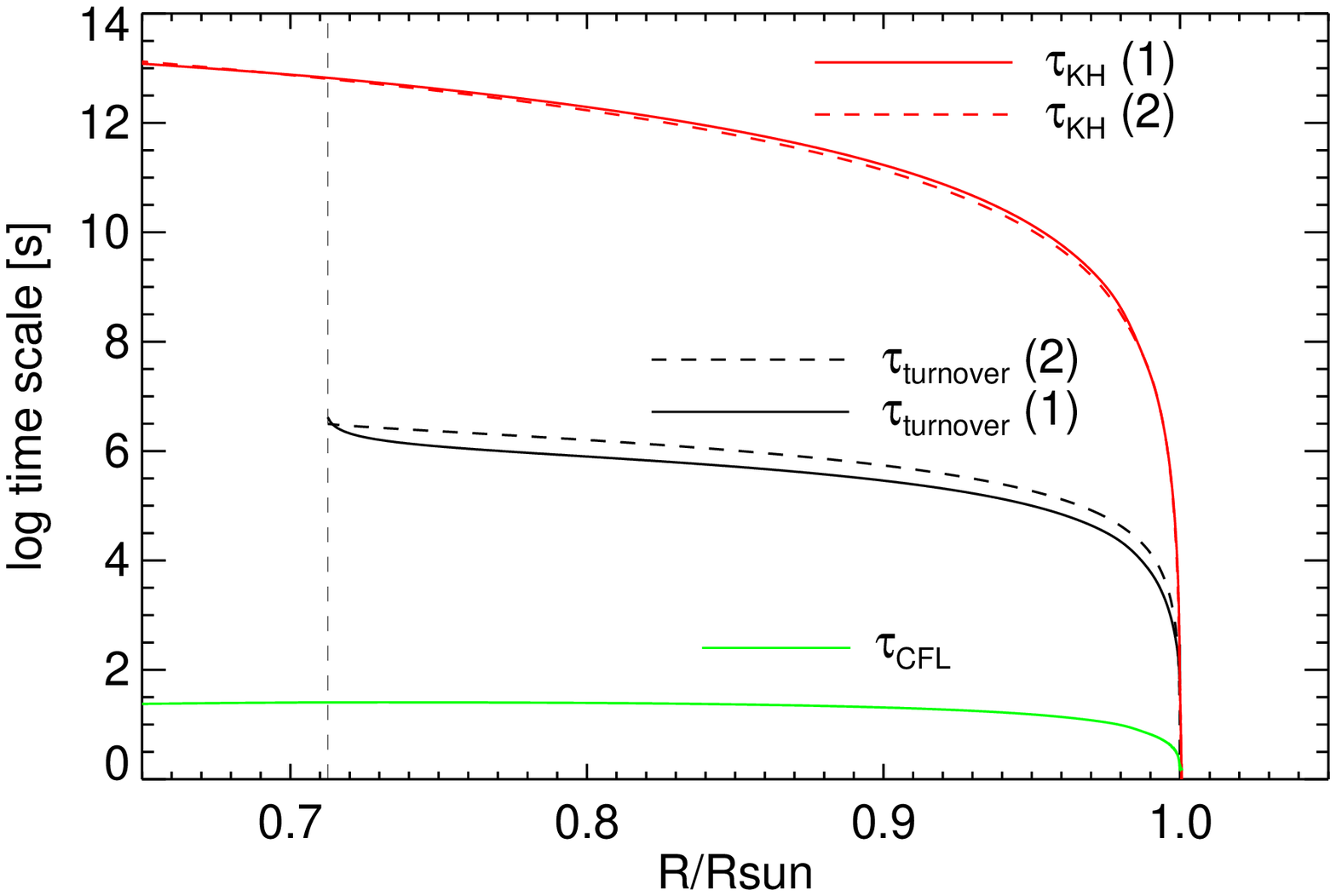}}
\mbox{\includegraphics[bb=40 40 580 360,width=6cm]{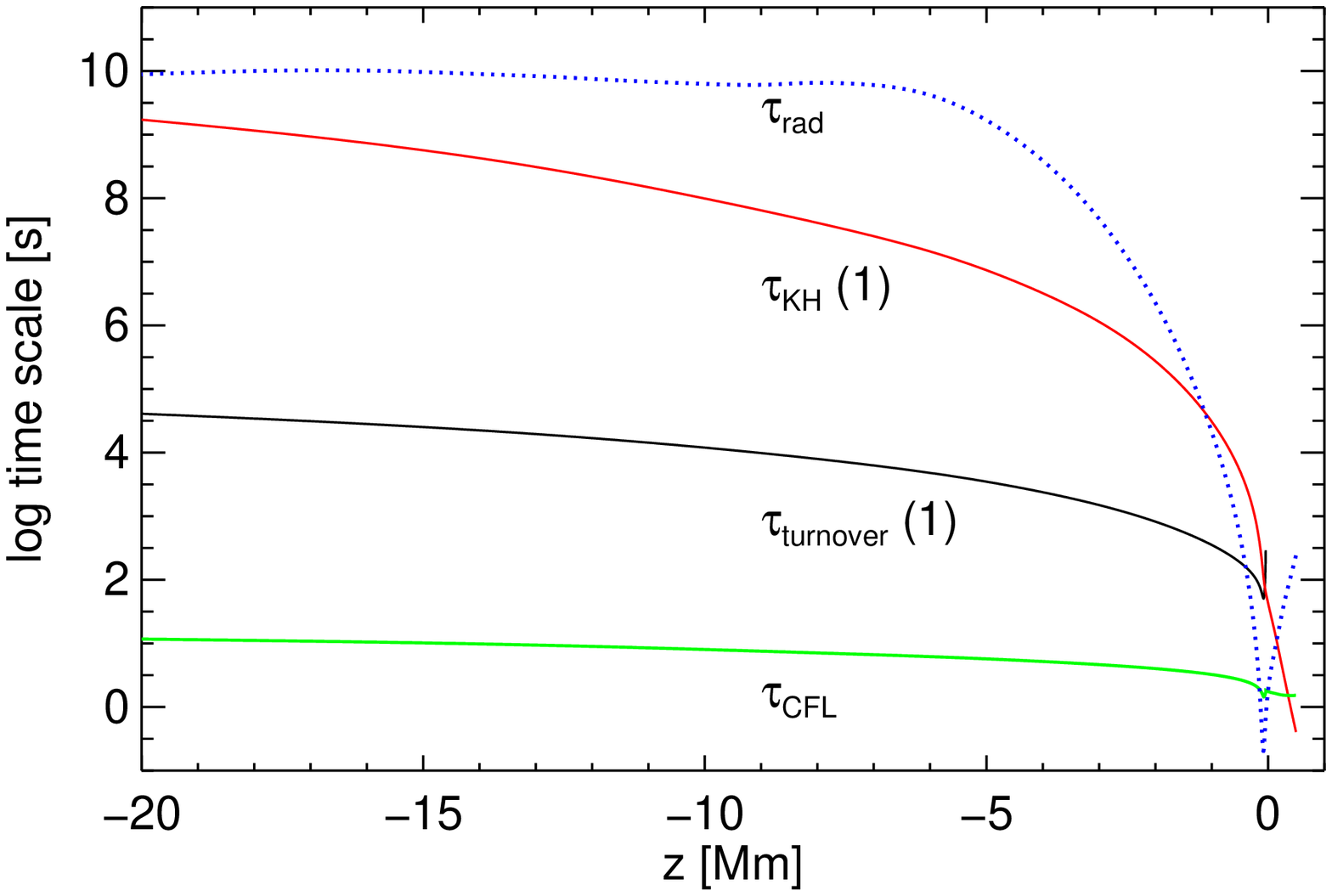}}
\caption{Left: Kelvin-Helmholtz time scale $\tau_{\rm KH}$ (upper set of curves, 
computed from Eqs.\,(\ref{eq:tauKH1}), (\ref{eq:tauKH2})), convective turnover 
time scale $\tau_{\rm turnover}$ (middle set of  curves,
computed according to Eqs.\,(\ref{eq:tto1}), (\ref{eq:tto2})),
and the CFL time scale $\tau_{\rm CFL}$ (lower curve, Eq.\,(\ref{eq:tauCFL})), 
as a function of radius in the solar envelope, based on the solar model by 
\citet{Christensen-Dalsgaard1996Sci...272.1286C}. The vertical dashed line 
in the left panel indicates the bottom of the solar convection zone. The 
right panel zooms into the upper $5$\% of the convection zone, and shows in
addition the radiative time scale $\tau_{\rm rad}$ calculated from 
Eq.~(\ref{eq:taurad}) with $k=k_0=10\,\pi/H_p$ (dotted).}
\label{f:timescales}
\end{figure}

\subsubsection{Time scales}

The time span covered by a numerical convection simulation must be 
sufficiently long to ensure that the whole structure contained in the
computational box can reach a thermally relaxed state. Thermal relaxation
by radiative diffusion proceeds on the Kelvin-Helmholtz time scale, defined as
the thermal energy content (per unit area) divided by the total energy flux:
$\tau_{\rm KH} = E_{\rm th}/F_{\rm tot}$.
In Fig.\,\ref{f:timescales}, we show the depth-dependence of $\tau_{\rm KH}$
in the solar convection zone (upper curves), computed as 
\begin{equation}\label{eq:tauKH1}
\tau_{\rm KH}^{(1)}(r) = \frac{4\pi\,r^2\,\rho\,c_p\,T\,H_p}{L_\odot}
  \enspace ,
\end{equation}
and
\begin{equation}\label{eq:tauKH2}
\tau_{\rm KH}^{(2)}(r) = \frac{4\pi}{L_\odot}\,\int_r^{R_\odot}
\,\rho\,c_p\,T\,r^2\,\mathrm{d}r
  \enspace ,
\end{equation}
respectively ($R_\odot$ and $L_\odot$ denote the solar radius and
luminosity). Both expressions give essentially identical results,
indicating that at a depth of $z=-3$~Mm below the solar surface, 
$\tau_{\rm KH}\approx 10^6$~s or $\approx 280$~h. Fortunately, it turns out
that relaxation is significantly faster than expected from this estimate.
Since the energy flux is carried by convection, a few convective turnover
times are sufficient to establish a self-consistent equilibrium state.
The convective turnover time scales, calculated as
\begin{equation}\label{eq:tto1}
\tau_{\rm turnover}^{(1)}(r) = \frac{H_p}{\vc}
  \enspace ,
\end{equation}
and
\begin{equation}\label{eq:tto2}
\tau_{\rm turnover}^{(2)}(r) = \int_r^{R_\odot} \frac{1}{\vc}\,\mathrm{d}r
  \enspace ,
\end{equation}
respectively, are also shown in Fig.\,\ref{f:timescales} (left, middle curves).
The plot shows that $\tau_{\rm turnover} \approx 2000$~s at $z=-3$~Mm,
a factor $500$ smaller than $\tau_{\rm KH}$. Roughly, the simulation
needs to be advanced for about 10 turnover times, $t_{\rm sim} \approx
10\, \tau_{\rm turnover} \approx 20\,000$~s to obtain a relaxed model.
This number has to be related to the numerical time step $\Delta t$
applicable to the hydrodynamics scheme.
The well-known Courant-Friedrichs-Lewy (CFL) condition
for the stability of \textcolor{tcol}{an} explicit numerical method states that   
$\Delta t < \tau_{\rm CFL}$, where $\tau_{\rm CFL}$ is given by the travel time of
the fastest wave across a grid cell. For the present non-magnetic example we 
can use the approximation
\begin{equation}\label{eq:tauCFL}
\tau_{\rm CFL}(r) = \frac{\Delta x}{(\cs+\vc)}\approx \frac{H_p}{10\,(\cs+\vc)}
  \enspace ,
\end{equation}
where $\cs$ is the adiabatic sound speed. Evaluation of Eq.~(\ref{eq:tauCFL}) 
shows that $\tau_{\rm CFL} \approx 1$~s in the upper layers (see 
Fig.\,\ref{f:timescales}).

The numerical time step is not only limited by the CFL condition. In 
addition, $\Delta t$ must be smaller than the characteristic radiative 
time scale $\tau_{\rm rad}$ that rules the decay of local temperature 
perturbations at the smallest possible spatial scale \textcolor{tcol}{(wavenumber 
$k_0=10\,\pi/H_p$)}. To a good approximation, $\tau_{\rm rad}$ can be 
calculated as
\begin{eqnarray}\label{eq:taurad}
\tau_{\rm rad}(r) = \frac{c_v}{16\,\sigma\,\kappa\,T^3}\,
\left(1+3\,\frac{\rho^2\,\kappa^2}{k^2}\right) =
\frac{1}{\chi}\,\left(\frac{1}{3\,\rho^2\,\kappa^2} + k^{-2}\right)
\enspace ,
\end{eqnarray}
which is valid in both optically thick and thin regions
\textcolor{tcol}{\cite{Spiegel1957ApJ...126..202S,Unno1966PASJ...18...85U}.} 
As illustrated in
Fig.\,\ref{f:timescales} (right), $\tau_{\rm rad}(k_0)$ reaches a sharp local
minimum of \textcolor{tcol}{$\approx 0.2$~s} close to the optical surface. The time step of the 
numerical simulation is thus set by the radiative time scale, 
\textcolor{tcol}{$\Delta t < 0.2$~s}, and the total number of required time steps is
$N_t=t_{\rm sim}/\Delta t \approx \textcolor{tcol}{10^5}$.
Assuming for reference a
processor that can update $N_c=10^6$ grid cells per CPU second,
the total CPU time required for this standard simulation 
would be $(N_x\times N_y\times N_z\times N_t)/N_c \approx 10^{12}/10^6$~s or 
about $12$ days, which is well feasible \textcolor{tcol}{even without a high degree of
parallelization}. However, it is also clear that much larger models 
\textcolor{tcol}{(e.g., $10$ times better spatial resolution in each direction) are out of
reach without massive parallelization.}

As an example, consider the solar supergranulation which has a typical 
horizontal scale of 20--30~Mm. Numerical simulations of this phenomenon 
thus require a horizontal cross section of at least $100\times 100$~Mm$^2$. 
Since the spatial resolution cannot be reduced much if the granular scale
still is to be resolved, such a horizontally extended 
simulation would take a factor $100$ more CPU time than the standard case 
outlined above. In addition, the simulation box would need to be extended to 
deeper layers for this kind of modeling. Assume that the lower boundary is 
moved from a depth of $z=-3$~Mm to $z=-20$~Mm, which means extending the model 
by about 6 more pressure scale heights. Adding $100$ grid cells in the 
vertical direction could be sufficient to cover the extra $17$~Mm. In terms of
computing time, these extra cells are relatively cheap, because radiative 
transfer can be treated by the diffusion approximation in these deep layers. 
Note, however, that keeping the horizontal resolution at $\Delta x = 
\Delta y \approx 40$~km to resolve
the granulation at the surface, the aspect ratio of the cells near the bottom
of the box becomes rather extreme, $\Delta x/\Delta z \approx 1/10$.
But the real problem is that the turnover time increases by a factor $20$.
Since $\Delta t$ is set by the surface layers, the number of time steps 
increases by the same factor. In summary, a supergranulation simulation
will take roughly a factor $2000$ more time than a standard granulation 
model, about \textcolor{tcol}{$65$}~years of CPU time. With massively parallel computers,
such models are becoming marginally feasible 
(cf.\ \cite{Stein2009ASPC..416..421S}).

\subsubsection{Global convection simulations}

Simulations of the entire solar convection zone are much more expensive:
the turnover time increases by another factor $100$, while the surface area
is about $600$ times larger with respect to the above supergranulation model.
In terms of the numbers quoted above, such a global convection simulation, 
which ideally should be carried out in a rotating spherical coordinate 
system, would take of the order of $4$ million years 
of CPU time, but still would cover only one year of solar time. In order 
to study the solar magnetic dynamo action, it would certainly be desirable 
to run the simulation over several $22$-year cycles, say a period 100 solar 
years, which is equivalent to \textcolor{tcol}{$400$} million CPU years.

Since the surface layers set the numerical time step and spatial resolution, 
the computational cost can be much reduced by restricting the simulations to
the deeper layers of the convection zone: here the flow Mach number is small
(see Fig.\,\ref{f:RePr}), and the so-called anelastic approximation can 
be employed to avoid the time step limitation by the CFL condition;
moreover, the radiative time step is very large (see Fig.\,\ref{f:timescales})
and does not impose any additional limitation.
This approach has been adopted in the global simulations of the solar
convection zone with the ASH-code by
\textcolor{tcol}{\citet{Brun2004ApJ...614.1073B}.}
However, the direct link between model and observation is necessarily broken in 
such kind of modeling.

While realistic simulations of global solar convection remain
phantasmal, prospects can be better for other type of stars:
realistic global star-in-a-box simulations have already been 
performed successfully for red supergiants, where only a few \textcolor{tcol}{huge}
convection cells occupy the surface of the star (see 
Sect.\,\ref{sec:SupergiantsAGBstars}).

\subsubsection{From the upper convection zone to the lower corona}

The essential physics necessary for realistic simulations of solar
surface convection includes compressible hydrodynamics describing
transonic flows of a partially ionized gas in a gravitationally 
stratified atmosphere, coupled with non-local, frequency-dependent 
radiative energy exchange. In the subsurface layers, the flow
becomes strongly subsonic and can be described in the anelastic
approximation, while the radiative transfer becomes local and can
be treated by the gray diffusion approximation. In contrast, physics
becomes more complicated when considering the outer solar atmosphere.

Simulations comprising the chromosphere and lower corona must include
magnetic fields. Since the magnetic field tends to form localized flux 
concentrations in the intergranular lanes (cf.\ Fig.\,\ref{f:granulation},
\textcolor{tcol}{right panel}),
the spatial resolution of MHD simulations needs to be better than that of
non-magnetic granulation models. In addition, the time step is dictated
by the Alfv{\'e}n speed
\begin{equation}\label{eq:va}
v_\mathrm{A} = \frac{B}{\sqrt{\mu_0\,\rho}}
  \enspace ,
\end{equation}
which can become much larger than the sound speed in places where
the plasma-$\beta$ is low, i.e., \textcolor{tcol}{where} the magnetic field $B$ is large 
and the density $\rho$ is small.
Typically, $\Delta t_{\rm MHD} \approx \Delta t_{\rm HD}/100$.

The low density of the outer atmosphere has also consequences for 
the radiation transport. Since the collision frequency is reduced,
the simplifying assumption of local thermodynamic equilibrium (LTE)
tends to break down, and photon scattering becomes important. This 
implies that the source function is no longer a function of the local
temperature, but depends also on the angle-averaged radiation field. 
In contrast to the photospheric absorption line spectrum,
the chromospheric spectrum \textcolor{tcol}{contains strong emission lines,
which dominate the energetics in the chromosphere}.
Under these circumstances, the solution of the radiation transfer
problem becomes very time consuming.

\textcolor{tcol}{Heat transfer by thermal conduction becomes important above gas 
temperatures of a few $10^4$\,K, i.e., in the transition region and in the 
corona above \citep[see, e.g.,][]{2001SoPh..198..325S,2006SoPh..234...41K}. 
Thermal conduction is usually modeled by means of the Spitzer formula but can 
result in a significant increase of the computational costs. }

Further complications arise due to the fact that the ionization of
hydrogen (and other elements) is no longer in thermal equilibrium 
in the low density regions, and cannot be obtained from precomputed 
look-up tables. Rather, the degree of hydrogen ionization, and hence 
the electron density, has to be derived from the solution of the 
time-dependent rate equations of a multi-level atom, which poses
severe challenges.

For further  discussion of these problems see 
Sect.\,\ref{sec:swb_solarchromchall}, as well as
\citet{Hansteen2007ASPC..368..107H}, 
\cite{Hansteen2010ApJ...718.1070H},
\citet{Martinez-Sykora2008ApJ...679..871M}, and
\citet{Gudiksen2011A&A...531A.154G}.

\subsection{Equations}

The hydro\-dynamics equations are expressed as conservation relations
plus source terms for
\begin{equation}\label{eq:conservedq}
  \rho, \;
  \rhov{1}, \rhov{2}, \rhov{3}, \;
  \etot
\enspace ,
\end{equation}
the mass density,
the three momentum densities,
and the total energy density (per volume), respectively.
The coordinate axes are simply numbered, in this case and in the code itself.
In some sections, we use the more standard notation $x$, $y$, and $z$, though.

The three-dimensional \textbf{hydro\-dynamics equations},
including source terms due to gravity, 
are the \emph{mass conservation equation}
\begin{equation} \label{eq:3dhydrorho}
  \papa{\rho}{t} + \papa{\; \rho \; \V{1}}{x_1}
                 + \papa{\; \rho \; \V{2}}{x_2}
                 + \papa{\; \rho \; \V{3}}{x_3} = 0 \enspace ,
\end{equation}
the \emph{momentum equation}
\begin{equation} \label{eq:3dhydrov}
  \papa{}{t}
    \left( \!
      \begin{array}{c}
        \rhov{1} \\
        \rhov{2} \\
        \rhov{3}
      \end{array}
    \! \right)
  +
  \papa{}{x_1}
    \left( \!
      \begin{array}{l}
        \rhov{1}  \; \V{1} + P \! \\
        \rhov{2}  \; \V{1} \\
        \rhov{3}  \; \V{1}
      \end{array}
    \right)
+
  \papa{}{x_2}
    \left( \!
      \begin{array}{l}
        \rhov{1}  \; \V{2} \\
        \rhov{2}  \; \V{2} + P \!\! \\
        \rhov{3}  \; \V{2}
      \end{array}
    \right)
  +
  \papa{}{x_3}
    \left( \!
      \begin{array}{l}
        \rhov{1}  \; \V{3} \\
        \rhov{2}  \; \V{3} \\
        \rhov{3}  \; \V{3} + P \!
      \end{array}
    \right)
  \! = \!
    \left( \!
      \begin{array}{c}
        \rho \; \g{1} \\
        \rho \; \g{2} \\
        \rho \; \g{3}
      \end{array}
    \! \right)\enspace ,
\end{equation}
and the \emph{energy equation} 
\begin{equation}  \label{eq:3dhydroe}
   \papa{\rho\etot}{t}  + \papa{\; (\rho\etot \! + \! P) \; \V{1}}{x_1}
                        + \papa{\; (\rho\etot \! + \! P) \; \V{2}}{x_2}
                        + \papa{\; (\rho\etot \! + \! P) \; \V{3}}{x_3}
{}+\papa{F_{\rm 1 rad}}{x_1} + \papa{F_{\rm 2 rad}}{x_2} + \papa{F_{\rm 3 rad}}{x_3} 
                     = 0 \enspace .
\end{equation}
Here $F_{\rm 1 rad}$, $F_{\rm 2 rad}$, $F_{\rm 3 rad}$ are the components
of the radiative energy flux (see below).
The gas pressure $P$ is computed from the density $\rho$ and the internal 
energy, $\ei$, via an \emph{equation of state}, usually available to the 
program in tabulated form, 
\begin{equation}
\label{eq:EOS}
  P = P (\rho, \ei)  \enspace .
\end{equation}
$\etot$ is given by the equation for the total energy,
\begin{equation}
\label{eq:TotalEnergy}
\rho\etot=  \rho\ei 
          + \rho\frac{\V1^2+\V2^2+\V3^2}{2} 
          + \rho\Phi \enspace , 
\end{equation}
where $\V{1}$, $\V{2}$, $\V{3}$ are the components of the velocity vector,
and $\Phi$ is the gravitational potential.
\textcolor{tcol}{In \COBOLD, a prescribed,
time-independent gravitational potential is used, so far.
Self-gravity is not accounted for.}
The gravity field is given by
\begin{equation} \label{eq:Gravitation}
    \left(
      \begin{array}{c}
        \rule{0mm}{5mm}\g{1} \\
        \rule{0mm}{5mm}\g{2} \\
        \rule{0mm}{5mm}\g{3} \\
      \end{array}
    \right) =
   - \left(
      \begin{array}{c}
         \rule{0mm}{5mm}\papa{}{x_1} \\
         \rule{0mm}{5mm}\papa{}{x_2} \\
         \rule{0mm}{5mm}\papa{}{x_3}
      \end{array}
     \right)
       \Phi
  \enspace .
\end{equation}
With \COBOLD, Eqs.~(\ref{eq:3dhydrorho})-(\ref{eq:3dhydroe}) are 
solved with the hydro\-dynamics module described in Sect.\,\ref{sec:HD}.

The \textbf{equations of ideal magneto\-hydro\-dynamics} (MHD),
including gravity and radiative energy exchange,
are written in the more compact vector notation as
\begin{equation}
\label{eq:MHD-equations}\setlength{\arraycolsep}{2pt}
\begin{array}{cllll}
  \displaystyle\papa{\rho}{t}
  & + & \mathbf{\nabla}\cdot\left(\rho\mathbf v\right)                                 & = & 0              \enspace, \\[2ex]
  \displaystyle\papa{\rho\mathbf{v}}{t} 
  & + & \mathbf{\nabla}\cdot\left(\rho\mathbf{vv} +
        \left(P+\displaystyle\frac{\mathbf{B}\cdot\mathbf{B}}{2}\right) \mathbf{I} -
                               \mathbf{BB}\right)                                      & = & \rho\mathbf{g} \enspace , \\[2ex]
  \displaystyle\papa{\mathbf{B}}{t}
  & + & \mathbf{\nabla}\cdot\left(\mathbf{vB} - \mathbf{Bv}\right)                     & = & 0              \enspace , \\[2ex]
  \displaystyle\papa{\rho\etot}{t}
  & + & \mathbf{\nabla}\cdot\left( \left(\rho\etot+P+
        \displaystyle \frac{\mathbf{B}\cdot\mathbf{B}}{2}\right)\mathbf{v} -
        \left(\mathbf{v}\cdot\mathbf{B}\right)\mathbf{B} + \mathbf{F_{\rm rad}}\right) & = & 0             \enspace .
\end{array}
\end{equation}
Here,
$\mathbf{B}$ is the magnetic field vector,
where we have chosen the units such that the magnetic permeability $\mu$ is equal to one.
$\mathbf{I}$ is the identity matrix and
$\mathbf{a}\cdot\mathbf{b} = \sum_k a_k b_k$ the scalar product
of the two vectors $\mathbf{a}$ and $\mathbf{b}$.
The dyadic tensor product of two vectors 
$\mathbf{a}$ and 
$\mathbf{b}$ is the tensor 
$\mathbf{ab} = \mathbf{C}$ with elements
$c_{mn} = a_m b_n$
and \textcolor{tcol}{the $n$th component of} the divergence of the tensor $\mathbf{C}$ 
\textcolor{tcol}{is 
$\left(\mathbf{\nabla} \cdot \mathbf{C}\right)_n  = \sum_m \partial c_{mn}/\partial x_m$.}
In this case, the total energy is given by
\begin{equation}
\label{eg:TotalEnergy}
\rho\etot=  \rho\ei 
          + \rho\frac{\mathbf{v}\cdot\mathbf{v}}{2} 
          + \frac{\mathbf{B}\cdot\mathbf{B}}{2}
          + \rho \Phi \enspace , 
\end{equation}
where $\ei$ is again the internal energy per unit mass.
The additional solenoidality constraint,
\begin{equation}
\label{eq:NoMonopoles}
\mathbf{\nabla}\cdot \mathbf{B}=0 \enspace ,
\end{equation}
must also be fulfilled. The equation of state and the equation for the 
gravitational field are given by Eq.\,(\ref{eq:EOS}) and Eq.\,(\ref{eq:Gravitation}),
respectively.
With \COBOLD, the equation system, Eq.\,(\ref{eq:MHD-equations}), 
is solved with the MHD module described in Sect.\,\ref{sec:MHD}.

In addition, there are equations for the \textbf{non-local radiation transport}
solved with \COBOLD\ with the modules described in
Sect.\,\ref{sec:LongCharRad} and Sect.\,\ref{sec:ShortCharRad}.
These modules account for the frequency dependence of the opacities by the
multi-group technique described in Sect.\,\ref{sec:OpacityBinning}.
In the following equations, the subscript $\nu$ refers to the index of the 
frequency group.

The variation of the intensity $\Int_\nu$ along a ray with direction $\mathbf{n}$
can be described by the radiative transfer equation
\begin{equation} \label{eq:deqdIdtau}
  \frac{1}{\rho\kappa_\nu}\mathbf{n}\cdot\mathbf{\nabla}\Int_\nu + \Int_\nu = 
  \Source_\nu \enspace .
\end{equation}
The \textcolor{tcol}{group-averaged} opacities $\kappa_\nu$ are typically given as a function
of temperature $T$ and gas pressure $P$,
\begin{equation} \label{eq:opacitygrey}
  \kappa_\nu = \kappa_\nu ( T, P ) \enspace ,
\end{equation}
and the \textcolor{tcol}{group-integrated} source function, \textcolor{tcol}{$\Source_\nu(T)$, is normalized 
such that
\begin{equation} \label{eq:Source}
  \sum_\nu\,\Source_\nu = B(T) = \frac{\sigma}{\pi} T^4 \enspace ,
\end{equation}
where $B(T)$ is the frequency-integrated Planck function.}
Introducing the optical depth $\tau_\nu$ according to
\begin{equation}
\rm{d}\tau_\nu = \rho\kappa_\nu\,\mathbf{n}\cdot \rm{d}\mathbf{x} \enspace ,
\end{equation}
where $\mathbf{n}\cdot \rm{d}\mathbf{x}$ is the path increment along the ray, 
the radiative transfer equation can be written as
\begin{equation}
\dedtaunu{\Int_\nu} + \Int_\nu = \Source_\nu \enspace .
\end{equation}
The frequency-integrated radiative energy flux vector  in direction $\vec{n}$ is given by angular 
integration over the full sphere, and summation over frequency groups
\begin{equation} \label{eq:Frad}
  \mathbf{F_{\rm rad}} = \sum_\nu \, \int_{4\pi}
                   \Int_{\nu}(\Omega) \, \mathbf{n}  
                \;{\rm d}\Omega \enspace .
\end{equation}
The energy change due to radiative transfer can then be computed from the flux divergence as
\begin{equation} \label{eq:divFrad}
  			 Q_{\rm rad}  = - \mathbf{\nabla}\cdot\mathbf{F_{\rm rad}} \enspace .
\end{equation}

\textcolor{tcol}{
To include additional physics such as chemical reactions (Sect.\,\ref{sec:chemnet}),
dynamic hydrogen ionization (Sect.\,\ref{sec:hionmodule}) 
or dust (Sect.\,\ref{sec:Dust}) the above equations are augmented by
\begin{equation}\label{eq:ni}
  \displaystyle\papa{n_i}{t}   +  \mathbf{\nabla}\cdot\left(n_i\mathbf v\right)      =  S_i
     \enspace ,
\end{equation}
where the number densities $n_i$ represent the densities of chemical species,
ionization states, or dust particles. The source term $S_i$ accounts
for chemical reactions, ionization and recombination, or dust formation.}

\subsection{Basic numerics}

The numerical simulations described here are performed with
\COBOLD\  (COnservative COde for the COmputation of COmpressible COnvection
in a BOx of L Dimensions, L=2,3).
It uses operator splitting \cite{Strang1968SJNA...53..506S} to separate the
various (usually explicit) operators:
the hydro\-dynamics (Sect.\,\ref{sec:HD}) or magneto\-hydro\-dynamics (Sect.\,\ref{sec:MHD}),
the tensor viscosity (Sect.\,\ref{sec:TensorViscosity}),
the radiation transport
\textcolor{tcol}{(different for the two setups, see below;
local models: Sect.\,\ref{sec:LongCharRad} or
global models: Sect.\,\ref{sec:ShortCharRad}),}
and optional source steps
(e.g., due to time-dependent dust formation or hydrogen ionization, Sect.\,\ref{sec:OptionalModules}).
The tabulated equation of state accounts for the partial ionization of hydrogen and helium 
\textcolor{tcol}{and a representative metal} (Sect.\,\ref{sec:EOS}).
The opacities can be either gray or \textcolor{tcol}{can}
account for \textcolor{tcol}{the} frequency dependence via an opacity-binning scheme 
(Sect.\,\ref{sec:OpacityBinning}).
Parallelization is done with OpenMP.

\COBOLD\ is used for two different types of model geometries,
\textcolor{tcol}{which} are characterized by different
gravitational potentials,
boundary conditions,
and modules for the radiation transport:
in the \emph{local-box} \textcolor{tcol}{(or \emph{box-in-a-star})} setup (Sect.\,\ref{sec:LocalModels}),
used to model small patches of a stellar surface,
the gravitation is constant, the lateral boundaries are periodic,
and the radiation transport module relies on a Feautrier scheme applied to a system
of long rays \textcolor{tcol}{(Sect.\,\ref{sec:LongCharRad})}.
In contrast, supergiant simulations employ
the \emph{global} \textcolor{tcol}{or} \emph{star-in-a-box} setup 
(Sect.\,\ref{sec:GlobalModels}) \textcolor{tcol}{for which}
the computational domain is a cube, and the grid is equidistant in all directions.
All outer boundaries are open for matter and radiation.
The prescribed gravitational potential is spherical.
\textcolor{tcol}{For this setup, a different radiation-transport module is used, 
\textcolor{tcol}{which implements}
a short-characteristics method (Sect.\,\ref{sec:ShortCharRad}).}

Some more technical informations can be found in the
\COBOLD\ Online User Manual.\footnote{See
http://www.astro.uu.se/\~{}bf/co5bold\_main.html, http://www.co5bold.com}

\section{Detailed numerics}

In this section, we present some numerical details of the code that
are adapted to the conditions found in stellar atmospheres.

\subsection{Numerical grid \textcolor{tcol}{and independent variables}}

Instead of the conserved quantities, Eq.\,(\ref{eq:conservedq}), we choose
\textcolor{tcol}{the primitive variables}
\begin{equation}\label{eq:primitiveq}
  \rho, \; 
  \V{1}, \V{2}, \V{3}, \;
  \ei (, \;
  \B{1}, \B{2}, \B{3})
\end{equation}
as independent quantities,
using integer indices for the components of a vector.
\textcolor{tcol}{
Since the conserved variables are purely algebraic combinations of
the primitive variables, the primitive variables can be directly
updated using the conservation laws Eqs.~(\ref{eq:3dhydrorho})-(\ref{eq:3dhydroe}) or 
Eqs.\,(\ref{eq:MHD-equations}) without dismissing conservation-law principles.
This is explained in more details in Sects.~\ref{sec:HDUpdate} and \ref{sec:EintUpdate}.
}

The hydro\-dynamics variables
$
  \rho,
  \V{1}, \V{2}, \V{3}$, and
$
  \ei,
$
are cell centered with grid coordinates
($
  \xc{1}, \xc{2}, \xc{3}
$),
whereas
$
  \B{1}, \B{2}, \mathrm{and\;} \B{3}
$
are cell-boundary centered with coordinates
($
  \xb{1}, \xb{2}, \xb{3}
$).
The grid is Cartesian.
The grid spacing may be non-equidistant.
\textcolor{tcol}{Additional} subscripts are used to describe the grid indices.
The hydrodynamics variables $\rho, \V{1}, \V{2}, \V{3}, \mathrm{and\;} \ei$
must be thought of as cell-averaged quantities, while $\B{1}, \B{2}, 
\mathrm{and\;} \B{3}$ are mean magnetic flux densities through cell interfaces.
%

\subsection{Boundary conditions and setup}

Global models, that simulate an entire star-in-a-box
(typically a red supergiant, Sect.\,\ref{sec:SupergiantsAGBstars}),
differ \textcolor{tcol}{essentially in boundary conditions and the gravitational potential}
from local box-in-a-star models, that simulate only a small piece
of a star close to the main sequence.
The fundamental parameters \textcolor{tcol}{are the effective temperature, 
$T_\mathrm{eff}$, describing the radiative flux per area in local models,
or the luminosity in global models, the surface gravity, $g$,  and the 
chemical composition of the stellar material.}

\subsubsection{Local models}
\label{sec:LocalModels}

Local box-in-a-star models are designed to simulate a small patch at the surface of
a star, ignoring \textcolor{tcol}{effects of the spherical geometry} and variations in gravity.
The computational domain is a Cartesian box with constant, downwardly directed
gravitational acceleration given by
\begin{equation}
  \mathbf{g} =
    \left(0, 0, -g \right)       \enspace .
\end{equation}

The \emph{side boundaries} are usually periodic.
Closed walls are a rarely used option, as they tend to attract downdrafts.

The \emph{top boundary} is generally
either hit \textcolor{tcol}{under some finite} angle by an outgoing shock wave
or \textcolor{tcol}{it} lets material fall back \textcolor{tcol}{into the computational
domain} (often with supersonic velocities):
there is not much point in tuning the formulation for an optimum
transmission of small-amplitude waves
\cite{Wedemeyer2004A&A...414.1121W}.
Instead, a simple and stable prescription that lets the shocks pass
is sufficient.
It is implemented by \textcolor{tcol}{assigning typically two or more layers of ghost cells
(the number depending on the order of the reconstruction scheme), with
boundary values, for which}
the velocity components and the internal energy are kept constant.
The density is assumed to decrease exponentially \textcolor{tcol}{with height} in the
ghost layers, with a scale height set to a controllable fraction
of the local hydrostatic pressure scale height. \textcolor{tcol}{The layers of
ghost cells are located outside the computational domain proper.}
The control parameter allows \textcolor{tcol}{for the adjustment of} the mean mass flux through 
the open \textcolor{tcol}{top} boundary.

The \emph{bottom boundary} of a standard solar model is located well inside the
convection zone, where the material coming from below is assumed to have the
entropy of the adiabat of the deeper convective envelope
\citep{Ludwig1999A&A...346..111L}.
The corresponding boundary condition prescribes the entropy of the ascending
material, ensures a zero total mass flux, and reduces pressure fluctuations for
stability reasons.
Horizontal velocities are assumed to be constant with depth.
The values of $\rho$, $\ei$, and the vertical velocity $\V{3}$ in the lowermost grid layer
are actually modified during the application of this boundary condition.
Therefore, the conservation laws are only valid in the volume above the bottom layer.
For each cell
\textcolor{tcol}{in the bottom layer}
the following steps are performed:

\noindent The equation of state is solved,
\begin{equation}
  {\rm EOS}( \rho, \ei)
     \;\; \rightarrow \;\;
  s, \; P, \; T, \; \Gamma_1, \; \Gamma_3, \; \cs
  \enspace ,
\end{equation}
to get \textcolor{tcol}{the} entropy, pressure, temperature, first and third adiabatic coefficient, and \textcolor{tcol}{the} sound speed.
Horizontal averages
of \textcolor{tcol}{the} density and pressure
$
  \langle \rho \rangle^{(0)} , \enspace
  \langle P \rangle
$
over the entire bottom layer
are computed,
where the superscripts $ ^{(0)},\ldots,^{(3)}$ here and in the following equations
denote the sub step.
A characteristic time scale is estimated by
\begin{equation}
  \langle t_{\rm char} \rangle =
    \dx{3} /
    \langle \cs +  \left| \V{3} \right| \rangle
  \enspace .
\end{equation}
In cells with an upflow ($\V{3} > 0$), mass and energy are modified according to
\begin{equation}
  \rho^{(1)} = \rho + \mbox{C$_{s{\rm Change}}$} \frac{\Deltat}{t_{\rm char}}
                      \frac{-\rho^2 \; T \, \left( \Gamma_3 - 1 \right)}{P \; \Gamma_1} \,
                      \left( s_{\rm inflow} - s \right)
  \enspace ,
\end{equation}
\begin{equation} \label{eq:eischangebot}
  \ei^{(1)} = \ei + \mbox{C$_{s{\rm Change}}$} \frac{\Deltat}{t_{\rm char}}
                    T \left( \! 1 - \frac{\Gamma_3 - 1}{\Gamma_1} \right)
                    \left( s_{\rm inflow} - s \right)
\end{equation}
with the two external parameters $\mbox{C$_{s{\rm Change}}$}$ ($\sim$0.1)
and $s_{\rm inflow}$. 
The latter controls the effective temperature $T_\mathrm{eff}$.
To reduce deviations of the pressure from the horizontal mean, the following
corrections are applied to all cells in the bottom layer:
\begin{equation}
  \rho^{(2)} = \rho^{(1)} + \mbox{C$_{P{\rm Change}}$} \frac{\Deltat}{t_{\rm char}}
                            \frac{1}{\cs^2} \,
                            \left( \langle P \rangle - P \right)
  \enspace ,
\end{equation}
\begin{equation}
  \ei^{(2)} = \ei^{(1)} + \mbox{C$_{P{\rm Change}}$} \frac{\Deltat}{t_{\rm char}}
                          \frac{1}{\Gamma_1 \; \rho}
                          \left( \langle P \rangle - P \right)
  \enspace ,
\end{equation}
adding another parameter $\mbox{C$_{P{\rm Change}}$}$ ($\sim$0.3).
To keep the total mass in the model volume unaltered, the density in the
bottom layer is corrected with
\begin{equation}
  \rho^{(3)} = \rho^{(2)} + \langle \rho \rangle^{(0)} -
                            \langle \rho^{(2)} \rangle
  \enspace .
\end{equation}
Because of this step, this boundary condition acts
as a closed boundary for plane-parallel waves.
Finally, the vertical velocity is modified to ensure a zero-average vertical
mass flux,
\begin{equation}
  \V{3}^{(1)} = \V{3} - \frac{\langle \rho^{(3)} \, \V{3} \rangle}
                             {\langle \rho \rangle^{(0)}}
  \enspace .
\end{equation}
Now, the old values are replaced by the new ones,
\begin{equation}
  \rho^{\rm (new)} = \rho^{(3)}   \enspace , \enspace 
  \ei^{\rm (new)}  = \ei^{(2)}    \enspace , \enspace 
  \V{3}^{\rm (new)} = \V{3}^{(1)}
  \enspace .
\end{equation}
Later, during the hydro\-dynamics step, the ghost cells are simply filled with
constantly extrapolated values from the bottom layer while keeping the
gravitational potential constant in these layers.

\subsubsection{Global models}
\label{sec:GlobalModels}

For global models, the gravitational potential depends on the radius $r$ \textcolor{tcol}{only}.
\textcolor{tcol}{The $1/r$ potential is a good approximation for the outer layers of supergiant stars,
which have a small massive core surrounded by an extended low-density envelope.
To avoid the central singularity the potential is smoothed near the center.
The potential can also be flattened at large distances to artificially enlarge the pressure
(and density) scale height preventing extremely low pressures and densities
in the corners of the simulation box.} The potential is given by
\begin{equation} \label{eq:Phir0r1}
  \Phi \left( r \right) = -{G\,M_\ast}{\left( r_0^4+r^4/\sqrt{1+(r/r_1)^8}\, \right)^{-1/4}}  \enspace ,
\end{equation}
where $M_\ast$ is the mass of the star to be modeled and
$r_0$ and $r_1$ are smoothing parameters in the core and the outer envelope, respectively.
\textcolor{tcol}{Within the sphere $r < r_0$, a source term to the internal energy provides
the stellar luminosity.
Motions in the core are damped by a drag force to suppress dipolar oscillations.}

All six surfaces of the computational box employ the same open
boundary condition,
which is also used for the top boundary in the local models
(Sect.\,\ref{sec:LocalModels}).


For global models the temperature/pressure range of the photospheric opacity tables
is insufficient.
It is therefore merged at around 12\,000\,K from
high-temperature OPAL data \cite{Iglesias1992ApJ...397..717I}
and low-temperature PHOENIX data \cite{Hauschildt1997ApJ...483..390H}.

\subsection{Initial conditions}
\label{sec:InitialConditions}

Due to the chaotic nature of stellar convection
\cite{Steffen1995CSF.....5.1965S}
and the primary interest in averaged \textcolor{tcol}{or statistical} properties,
the details of the initial conditions hardly matter,
except for initial strong magnetic field configurations.
On the other hand, the total mass within the computational domain is of
main importance.
However, choosing a pressure and temperature distribution \textcolor{tcol}{too far} 
off from the (usually close to hydrostatic) mean conditions 
\textcolor{tcol}{requires} an unnecessarily
long time, until plane-parallel pulsations have settled down and the
stratification is thermally relaxed.
It is often advisable to start with a standard 1D atmosphere model
(e.g., produced with PHOENIX as in \cite{Freytag2010A&A...513A..19F}),
to expand it trivially into the second and third dimension and to add
small velocity fluctuations to it as seed for convective motions.
An even better alternative is to use an existing 3D snapshot
with similar parameters -- if available -- and scale it to the desired model properties.

Even with a careful construction of the start model, transient plane-parallel pulsations
are common. \textcolor{tcol}{These pulsations} are generated by tiny deviations from 
\textcolor{tcol}{the} exact numerical hydrostatic
equilibrium in the deeper layers, \textcolor{tcol}{causing larger amplitudes} in the
\textcolor{tcol}{tenuous top} layers. 
To damp them out, a vertical drag force acting only on the horizontal average of
the vertical mass flux can be applied in the initial phase of a simulation.

\subsection{Equation of state}
\label{sec:EOS}

Under the conditions of cool stellar surfaces,
a lot of energy can go into the ionization of hydrogen and helium.
In \COBOLD,
the equation of state (EOS) accounts for the ionization balance of
HI, HII,
H$_2$,
HeI, HeII, HeIII,
and a representative metal.
Pre-tabulated values as functions of density
and internal energy are used
($\log\rho, \log\ei \! \rightarrow \! \log{P}, \log{T}, s$).
In fact, the coefficients for a bicubic interpolation of ($\log{P}, \log{T}, s$)
are stored. Thermodynamic derivatives are computed from the corresponding
derivatives of the polynomials.

\subsection{Hydrodynamics}
\label{sec:HD}
%
\noindent In general, a hydro\-dynamics scheme should
\begin{enumerate}

\item be \emph{consistent} with the original hydro\-dynamics equations,

\item be \emph{stable},

\item solve the hydro\-dynamics equations in 3D with reasonable \emph{accuracy},
i.e., be of high order whenever possible and represent discontinuities with only a
few grid points,

\item be \emph{conservative} to handle shocks properly and give constant total
fluxes in stationary cases, which is particularly important for modeling convection,

\item include source terms due to gravity in a proper way
to allow \emph{static} solutions, so that especially the construction of
an exactly hydrostatic stratification in radiative equilibrium is possible,

\item handle a \emph{general equation of state} (from a table),

\item be \emph{fast}, e.g., easy to vectorize, to parallelize, and to make proper use of the various CPU caches,

\item handle \emph{various geometries} (in this case 1D, 2D, and 3D models),

\item be not too complex but stay fairly \emph{simple},

\item allow the \emph{coupling with additional physics} (especially radiation transport).

\end{enumerate}
Solvers differ in how close they get to the individual design goals.
For instance, total energy conservation might get sacrificed to improve the code stability
in cases of large Mach numbers.
And with detailed (read, time consuming) radiation transport modules, the performance
of the (usually comparably fast) hydro\-dynamics modules becomes unimportant.


The hydro\-dynamics scheme of \COBOLD\ uses a finite-volume approach.
By means of operator (directional) splitting \cite{Strang1968SJNA...53..506S},
the 2D or 3D problem is reduced to one dimension.
To compute the fluxes across each cell boundary
in every 1D column in $\xc{1}$ direction,
an approximate 1D Riemann solver of Roe type \cite{Roe1986ARFM...18..337R} is applied,
modified to account for
a realistic equation of state (Sect.\,\ref{sec:HDEOS}),
a non-equidistant grid (Sect.\,\ref{sec:HDGrid}),
and the presence of source terms due to an external gravity field (Sect.\,\ref{sec:HDGravity}).
The partial waves are reconstructed and advected with upwind-centered fluxes.
A slope limiter (MinMod, SuperBee, but usually van Leer)
\cite{Leveque1992book.....L}
or a reconstruction with monotonic parabolae (\citet{Colella1984JCoPh..54..174C})
is applied to decrease the order of the scheme
in the neighborhood of discontinuities \textcolor{tcol}{for keeping it stable}
while preserving higher-order accuracy
in the case of smooth \textcolor{tcol}{flows.}


The standard Roe solver has been extended in several ways to fit the particular
problem of stellar surface convection \textcolor{tcol}{as is explained in the
following subsections.}

\subsubsection{Non-equidistant grid}
\label{sec:HDGrid}

The hydro\-dynamics \textcolor{tcol}{scheme} handles Cartesian grids
\textcolor{tcol}{only. They} may be non-equidistant in any direction.
Without gravitation, the location of the cell centers $\xc{1}$ has not much relevance
as all quantities are either integral values within a cell 
(for instance \textcolor{tcol}{the} mass density)
or located at the cell boundaries $\xb{1}$ (for instance \textcolor{tcol}{the} mass flux).
In this simple case, a non-equidistant grid would only have an effect
on the reconstruction equations.

With the inclusion of gravity however, the potential energy within each cell
is located at $\xc{1}$. This means, that the pressure should also be
located there in order to allow for a correct balance of acceleration due to
the pressure gradient and the gradient of the gravitational potential.

The relative position of $\xc{1,\,i}$ within $\xb{1,\,i}$ and $\xb{1,\,i+1}$
is not set by the hydro\-dynamics scheme and can be chosen within
reasonable limits according to the requirements, e.g., of the
radiation-transport scheme.

\subsubsection{Update of the mass density and the velocity}
\label{sec:HDUpdate}

\textcolor{tcol}{
Given the update of the density in one coordinate direction for cell $i$
in conservation-law form,
\begin{equation}\label{eq:rho_update}
  \Delta\rho_i = - \frac{\Deltat}{\Dx}\left(f_{\rho,\,i+1} -
            f_{\rho,\,i}\right) \enspace , \qquad
            \rho_i^\mathrm{(new)} = \rho_i^\mathrm{(old)} + \Delta\rho_i \enspace ,
\end{equation}
where $f_{\rho,\,i}$ is the mass flux in this direction between cell $i$$-$$1$ and 
cell $i$, the update for the momentum in the 1-direction can be reformulated in 
terms of the update for the velocity as follows. From the conservation-law form
\begin{equation}\label{eq:rhov_update}
(\rho v_1)_i^\mathrm{(new)} = (\rho v_1)_i^\mathrm{(old)} - 
    \frac{\Deltat}{\Dx}\left(f_{\rho v_1,\,i+1} -
            f_{\rho v_1,\,i}\right) + \Deltat\, S_{\rho v_1,\,i}   \enspace ,
\end{equation}
where $f_{\rho v_1,\,i}$ is the 1-momentum flux in the considered direction
and $S_{\rho v_1,\,i}$ the source term for the 1-momentum,
we obtain
\begin{equation}\label{eq:v_update}
{v_1}_i^\mathrm{(new)} = {v_1}_i^\mathrm{(old)}
    - \left[\frac{\Deltat}{\Dx}\left(f_{\rho v_1,\,i+1} 
    - f_{\rho v_1,\,i}\right) - \Deltat\, S_{\rho v_1,\,i}
    + \Delta\rho_i\, {v_1}_i^\mathrm{(old)}\right]
    \frac{1}{\rho_i^\mathrm{(new)}} \enspace .
\end{equation}
$\Delta\rho_i$ and $\rho_i^\mathrm{(new)}$ on the right hand side of
Eq.~(\ref{eq:v_update}) are known from Eq.~(\ref{eq:rho_update}).
Then, the momentum $(\rho v_1)_i = \rho_i\, {v_1}_i$ is, up to the 
source term, a strictly conserved quantity. 
The fluxes $f_{\rho}$ and $f_{\rho v_1}$ are defined at the cell 
interfaces and determined by an approximate solution of the Riemann 
problem as explained in Sect.~{\ref{sec:HD}} and 
Sect.~{\ref{sec:HDEOS}}.
The advantage of his formulation becomes apparent when
treating the gravitational potential in the derivation of the 
discrete equation for the internal energy. This is explained in 
Sect.~\ref{sec:EintUpdate}.
}

\subsubsection{Gravity}
\label{sec:HDGravity}

The gravitational source term in \textcolor{tcol}{Eq.\,(\ref{eq:3dhydrov}) and}   
Eq.\,(\ref{eq:MHD-equations})
destroys the hyperbolic character of \textcolor{tcol}{the corresponding system of 
equations} and inhibits the
direct application of an (approximate) Riemann solver.
On the other hand, the separation via operator splitting is not a good idea in
this case,
because in stratified atmospheres the pressure gradient and the gravity
tend to cancel each other (nearly).  Their application in sequence --
and not together in a single step -- would cause spurious unwanted
accelerations back and forth.
On the other hand, the naive combination of the Roe solver \textcolor{tcol}{with} 
the source terms due to gravity
into a single operator \textcolor{tcol}{by simple addition}, leads to problems because the Roe
solver interpretes the strong pressure gradient in a stratified atmosphere as
indication of a shock wave, which is then treated as such,
causing spurious -- possibly large -- velocity fields.

There is some freedom in the choice of the exact reconstruction of
quantities inside the cells (\citet{Mellema1991A&A...252..718M}),
which is used to amalgamate the hydro\-dynamics with the gravity
operator by reducing the pressure jump across a cell boundary to the
deviation from hydrostatic stratification.
The latter is subtracted from the actual pressure inside the cells during the computation
of the amplitudes of the partial waves.
The idea is, that only pressure deviations from hydrostatic equilibrium --
and not just pressure gradients --
should give rise to fluxes of the partial waves.
In the exact hydrostatic case, the Roe solver should ``see'' no sound
waves.
This construction does not \textcolor{tcol}{supersede} the usual source terms due to gravitation.

\subsubsection{Update of the internal energy}
\label{sec:EintUpdate}

\textcolor{tcol}{The discrete form of Eq.\,(\ref{eq:3dhydroe}) for the total energy is 
given in conservation law form by
\begin{equation}
\left(\rho\ei 
      + \rho\frac{\V1^2+\V2^2+\V3^2}{2} + \rho\Phi_{\mathrm{c}}\right)_{i}^\mathrm{(new)} 
      \!\!\!=\left(\rho\ei 
      + \rho\frac{\V1^2+\V2^2+\V3^2}{2} + \rho\Phi_{\mathrm{c}}\right)_{i}^\mathrm{(old)} 
\!\!\!-\frac{\Deltat}{\Dx}\left((f_{\mathrm{e},\,i+1} + f_{\Phi,\,i+1}) 
      -  (f_{\mathrm{e},\,i} + f_{\Phi,\,i})\right) ,
\end{equation}
where $f_{\mathrm{e},\,i}$ is the 1D flux of the total energy without the
potential energy from cell $i$$-$$1$ into cell $i$ provided by the Roe solver, 
$f_{\Phi,\,i}$ is the flux of the potential energy and $\phic{i}$ is the gravity
potential in the center of the cell. Using Eq.~(\ref{eq:rho_update})
%
and defining the flux of the potential energy as
\begin{equation}
f_{\Phi,\,i} = \phib{i} \; f_{\rho,\,i} \enspace ,
\end{equation}
where $\phib{i}$ is the gravity potential at the interface between cell
$i - 1$ and cell $i$, all terms containing the gravity potential can
be combined to yield
\begin{eqnarray}\label{eq:updaterhoei}
\left(\rho\ei\right)_i^\mathrm{(new)} & = & \left(\rho\ei\right)_i^\mathrm{(old)} +          
            \left(\rho\frac{\V1^2+\V2^2+\V3^2}{2}\right)_{i}^\mathrm{(old)}
   - \left(\rho\frac{\V1^2+\V2^2+\V3^2}{2}\right)_{i}^\mathrm{(new)} -
            \frac{\Deltat}{\Dx}\left(f_{\mathrm{e},\,i+1} - f_{\mathrm{e},\,i}\right) \nonumber \\
   & &   - \frac{\Deltat}{\Dx}\left[\left(\phib{i+1} -
                \phic{i}\right)f_{\rho,\,i+1} - \left(\phib{i} -
                \phic{i}\right)f_{\rho,\,i}\right] \enspace .
\end{eqnarray}
The presence of the new kinetic energy at the right side of Eq.\,(\ref{eq:updaterhoei})
does not make the scheme implicit, since the velocity does not depend
on $\ei$ -- it is known from Eq.~(\ref{eq:v_update}) and corresponding equations for 
$\V2$ and $\V3$. 
We note that this update still conserves the total energy, Eq.~(\ref{eq:TotalEnergy}), 
to machine accuracy. The conservative inclusion of the radiative energy flux into
the energy equation is treated in Sect.~\ref{sec:LongCharRad}. 
}

\subsubsection{General equation of state including ionization}
\label{sec:HDEOS}

\textcolor{tcol}{
Several extensions of the Roe scheme for a general equation of state
have been proposed, see \cite{1997JCoPh.138..354M} and references therein. 
The differences compared to the case of
an ideal gas with constant $\gamma$
manifest themselves in
the need of additional Roe averages depending on which
variables are used in the equation of state.
For a general equation of state of the form  Eq.\,(\ref{eq:EOS}),
averages of $\rho$, $\ei$ and of the pressure derivatives $\dPdrho$ and
$\dPdei$ are required to build the Roe matrix. Choosing the usual Roe average 
for $\ei$ and the average $\sqrt{\rho_l\rho_r}$ for the density, where $\rho_l$ and
$\rho_r$ is the density on either side of the cell interface, the condition 
\begin{equation}\label{eq:pressurederivatives}
\Delta P = \dPdei\Delta\ei + \dPdrho\Delta\rho \enspace ,
\end{equation}
must be fulfilled by the averages of $\dPdrho$ and $\dPdei$ where
$\Delta P$, $\Delta\rho$ and $\Delta\ei$ are the jumps of the
pressure, density, and internal energy at the cell interface.
Eq.\,(\ref{eq:pressurederivatives}) is not sufficient to determine the
averages of the pressure derivatives.
\citet{Glaister1988JCP...74..382G} suggested formulas for $\dPdrho$ and
$\dPdei$ which meet Eq.\,(\ref{eq:pressurederivatives})  exactly. However, Glaister's formulae do not lead to the
same averaged sound speed as the original formulae of Roe in the simple case of constant $\gamma$.
Furthermore, they may also produce unphysical average states \cite{1997JCoPh.138..354M}. In \COBOLD, 
the averaging of the pressure derivatives is avoided. 
Instead, the dimensionless quantities $\Gamma_1$ and $\drhoeidP$ are averaged
with the usual Roe weights, which ensures the consistency with the simple gas case.
}

\textcolor{tcol}{
The pressure derivatives in the Roe matrix
lead to an additional term
\begin{equation}\label{eq:additionalenergyflux}
-\frac{1}{2}|\V1| \left(\ei -
  \rho \displaystyle{\dPdrho} \displaystyle{\dPdei}^{-1}
                  \right) \tilde\alpha^{(5)}
= -\frac{1}{2}|\V1|\tilde\alpha^{(6)} \enspace ,
\end{equation}
in the energy flux. This term, which vanishes in the perfect gas case, is treated
as a contribution from a sixth partial wave $\tilde\alpha^{(6)}$. 
The wave strength of the entropy wave $\tilde\alpha^{(5)}$ is given by
\begin{equation}
\tilde\alpha^{(5)} = \Delta\rho - \frac{\Delta P}{\cs^2} \enspace .
\end{equation}
The term $\dPdei$ in Eq.\,(\ref{eq:additionalenergyflux}) can become small,
possibly causing numerical errors. Using thermodynamic relations, 
$\tilde\alpha^{(6)}$ can be transformed into
\begin{equation}
  \tilde \alpha ^{(6)} = \Delta (\rho\ei) - \papaco{\rho\ei}{P}{s} \Delta P
  \enspace .
\end{equation}
which uses a better behaved derivative.
}

\subsubsection{Tensor viscosity}
\label{sec:TensorViscosity}

In addition to the stabilizing mechanism inherent in an upwind scheme with
monotonic reconstruction,
a 2D or 3D tensor viscosity can be activated.
It eliminates certain errors of Godunov-type methods
occurring in the case of strong velocity fields aligned with the grid \cite{Quirk1984IJNMF...18..555Q}.
Other types of problems can occur when e.g., a shock,
which has a strength that
could easily be handled by the hydro\-dynamics scheme alone,
gives rise to so large
opacity variations that the radiation-transport routines might get unstable (Sect.\,\ref{sec:Rad}).




To overcome such (possible) problems, an additional tensor-viscosity sub step
was included in the code, that can add dissipation in a way the Roe solver
by its own is not able to produce.
The kinematic viscosity is
\begin{eqnarray} \label{eq:nu}
  \nu & = &
      \frac{1}{3} \left( \dx{1} + \dx{2} + \dx{3} \right) C_{\rm linear} \, \cs
      + \min \left(  \dx{1}, \dx{2}, \dx{3} \right)
        \max \left(  \dx{1}, \dx{2}, \dx{3} \right) 
  \\ \nonumber
  & &
     \!\! \left\{ \rule{0mm}{7mm} 
             \; C_{\rm artificial}^2 \max(-\nabla v, 0) \right.
  \\ \nonumber
  & &
           + C_{\rm Smagorinsky}^2 \left[\rule{0mm}{6mm} 
             2 \left(\dvdx{1}{1}^2 + \dvdx{2}{2}^2 + \dvdx{3}{3}^2\right) \right.
 \;\; + \left. \left. 
               \left(\dvdx{1}{2} + \dvdx{2}{1}\right)^2 +
               \left(\dvdx{1}{3} + \dvdx{3}{1}\right)^2 +
               \left(\dvdx{2}{3} + \dvdx{3}{2}\right)^2  \right] ^{1/2} \right\}
\end{eqnarray}
with the parameters
$C_{\rm linear}$, $C_{\rm artificial}$, and $C_{\rm Smagorinsky}$
for the
linear,
artificial (von Neumann type) viscosity, and
turbulent \textcolor{tcol}{subgrid-scale} viscosity (\citet{Smagorinsky1963MWR....91..99S}),
respectively.
Typical values are (0, 0.5, 0.5), respectively.
Models of solar granulation and similar easy cases do not require
this extra viscosity. However, it is usually activated to avoid the necessity to
tune the numerical parameters individually for each stellar model.
For instance the models of the more dynamic atmospheres of red supergiants
(Sect.\,\ref{sec:SupergiantsAGBstars})
need some amount of extra dissipation provided by the tensor viscosity.
\textcolor{tcol}{Note that the tensor viscosity should not be mistaken for a
hyperviscosity. The task of hyperviscosities is in \COBOLD\ done by the
reconstruction schemes (MinMod, SuperBee, van Leer, PP, etc.).}

\subsection{Radiation transport}
\label{sec:Rad}

\subsubsection{Introduction}

In dynamical simulations which take time dependence and
coupling between radiative energy transfer and hydro\-dynamics equations
into account, the emitted intensity is only a by-product.
Important is instead the energy change per numerical grid cell due to
the difference of radiative gains and losses.
The requirement to solve the radiation-transport equations
for many grid points and many time steps calls for severe simplifications
as, e.g., the restriction to gray opacities or \textcolor{tcol}{to} a few frequency groups
(Sect.\,\ref{sec:OpacityBinning})
or the treatment of scattering as true absorption.
Actually, this can make code development easier.
But still, there are additional demands on the algorithm:
the scheme should conserve the total energy, i.e., internal sources and sinks of
energy minus losses through the surface should exactly sum up to zero.
The scheme has to be stable enough to
handle complex structures which may sometimes be
poorly resolved (e.g., chromospheric shocks, Sect.\,\ref{sec:swb_solarchrom}).

Some cases pose only low demands on the complexity of the algorithm:
if the entire model is optically thick, a diffusion approximation
using only differences between neighbor cells
is adequate to compute the radiative flux.
This results in a stable scheme, if the time step is properly limited.
If the whole numerical domain is optically thin and the radiation field
is simple enough, a local cooling function might be sufficient
to model radiative energy losses, calling for a scheme that
is stable if the time step is small enough.

However, stellar atmospheres are per definition at the transition between
optically thick and thin regions.
\textcolor{tcol}{The main form of energy transport switches from
convective plus radiative in the interior to mainly radiative in the
outer layers, where mechanical energy fluxes become very small due
to the low material density. Still, mechanical energy fluxes might
be sufficiently large to affect the temperature structure of the
chromosphere (Sect.\,\ref{sec:swb_solarchrom}), for example. While
radiative energy transport in the stellar interior can be properly
described by local physical quantities through the diffusion
approximation, in the outer layers, radiative energy exchange occurs
non-locally. This means that the local radiative flux depends on the
physical state of the material in the wider surroundings of a size
depending on the mean free path of the photons.}
Large opacity variations due to changes in the ionization states of major constituents
or due to shock waves can cause changes in the source function
on small spatial scales.
In numerical models, these two effects,
amplified by fluctuations in heat capacity,
can cause enormous jumps in the radiative relaxation time scale
from grid point to grid point.

Even if a standard scheme is able to overcome all these difficulties
and to compute accurate intensities for given opacities and source function,
there are still several possibilities how to derive the induced energy change per cell,
as detailed in the following:

%
%
The energy flux through the cell boundaries can be computed from the
intensity field, which then gives the divergence of the flux for each cell, 
and hence the energy change per cell according to Eq.\,(\ref{eq:divFrad}).
This would guarantee the conservativity of the scheme.
But unfortunately, the latter step requires an extreme numerical precision
in optically very thin regions where the relative flux changes
from cell to cell are tiny.
A high accuracy is also necessary in optically very thick regions
where only small deviations
of the intensity from the local source function contribute to the net flux.
Calculating a discrete derivative naturally amplifies noise and has
centering problems, when e.g., the intensity field is given at cell centers
and the divergence is required at the same position.

%
Another possibility consists in deriving the energy change per cell
from the difference of the angular mean of the intensity,
\begin{equation} \label{eq:J}
  J_{\nu} = \frac{1}{4\pi} \int_{4\pi} \Int_{\nu} \; \mathrm{d}\Omega
              = \textcolor{tcol}{\frac{1}{4\pi}} \int_{0}^{2\pi}
                  \int_{0}^{\pi}
                    \Int_{\nu} \sin \theta
                  \;{\rm d}\theta
                \;{\rm d}\varphi
  \enspace ,
\end{equation}
and the local source function $\Source_{\nu}$.
Using Eqs.~(\ref{eq:deqdIdtau}), 
(\ref{eq:Frad}), and (\ref{eq:divFrad}), 
one obtains
\begin{equation} \label{eq:J-S}
  Q_\mathrm{rad}
                 = \sum_{\nu} 4 \pi \kappa_{\nu} \rho \; \left( J_{\nu} - \Source_{\nu} \right)
  \enspace .
\end{equation}
This scheme does not have an explicit conservation form
and in fact it will most likely not be conservative.
This happens because the distribution of the source function within the cell 
\textcolor{tcol}{which} is used in the integration process for the intensity 
(where typically some high-order interpolation of the source function with 
optical depth is performed) \textcolor{tcol}{is not exactly the same distribution 
as is used} in computing the difference \textcolor{tcol}{$J_{\nu} - \Source_{\nu}$} 
(where the source function is assumed to be constant).
Another problem of this scheme is the accuracy in optically very thick regions,
where numerical cancellation may occur between $J_{\nu}$ and $\Source_{\nu}$.
%
A more indirect way is to derive from the intensity field some
geometrical information about the radiation field in the form of Eddington
factors. These are inserted into the equations describing the radiation
transport via the Eddington moments
(see, e.g., for two dimensions \citet{Stone1992ApJS...80..819S}
and for one dimension \citet{Hoefner1995A&A...297..815H}, \cite{Hoefner2003A&A...399..589H}).
This method requires a non-trivial solver to get the radiation field
in the first place
and later an algorithm to solve the huge system of Eddington equations.
This procedure might not suffer from the problems mentioned above.
However, it seems somewhat inefficient first to compute the intensity distribution
of the radiation field in some detail and then to throw away most of the
information and retain the Eddington factors only, which are used to solve
the radiative transfer equation again -- just in a different form.
The extra effort can be justified by gains due to, e.g., an elegant
handling of scattering processes or the achievement of large time steps
with an implicit operator \cite{Hoefner1995A&A...297..815H}, though.
%


In Sect.\,\ref{sec:LongCharRad} and Sect.\,\ref{sec:ShortCharRad},
we present two radiation-transport schemes
implemented in \COBOLD, that overcome the
aforementioned problems \textcolor{tcol}{in different ways.}
They compute the contribution to the energy change per cell on-the-fly
during the integration of the intensity for each direction.
\textcolor{tcol}{For standard \emph{local-box} models with periodic boundaries
(Sect.\,\ref{sec:LocalModels}),
we use a long-characteristics scheme, described in Sect.\,\ref{sec:LongCharRad},
while for \emph{star-in-a-box} models with all-open boundaries
(Sect.\,\ref{sec:GlobalModels}),
we use a short-characteristics method, outlined in Sect.\,\ref{sec:ShortCharRad}.
}

\subsubsection{Opacity binning}
\label{sec:OpacityBinning}


The rate of the radiative energy exchange is highly variable over the
relevant spectral range, since the absorption coefficient strongly varies with
frequency due to the presence of spectral lines, on top of the more gradual 
change of the continuous opacity. 
In cool stars like the Sun, spectral lines count in the millions so
that an exact treatment of the frequency dependence in a complex
multi-dimensional geometry is beyond present computer capacities, and one has
to resort to an approximate treatment. An important simplification stems from
the fact that one is not interested in the detailed frequency dependence of
the heat exchange between stellar plasma and radiation field but only in its
frequency-integrated total effect, $Q_{\rm rad}$.
Nowadays, all multi-dimensional hydro\-dynamical stellar atmosphere codes 
employ the so-called \textit{opacity-binning} technique. The method was first 
laid out by \citet{Nordlund1982A&A...107....1N}, and later refined in works by
\citet{ludwig1992}, \citet{Ludwig1994A&A...284..105L}, and
\citet{Voegler2004A&A...421..755V}. At present, opacity sampling -- a statistical technique
widely applied in standard 1D model atmospheres -- is discussed as possible
replacement of the opacity binning due to its better controlled accuracy and
greater flexibility.

The basic idea of opacity binning is the classification of frequency points
by the similarity of their associated $\Lambda$-operator -- the operator
relating source function~\Snu\ and mean intensity~\Jnu\ of the radiation field:
\begin{equation}
\Jnu = \Lambda_\nu\left[\Snu\right]\enspace .
\end{equation}
We added an index $\nu$ to the $\Lambda$-operator to emphasize that its form,
written in geometrical coordinates, is different for different frequencies due
to opacity variations. However, in cases where the operator happens to be
similar, its linearity allows to operate on the sum of the source functions to
obtain the integrated mean intensity, symbolically expressed as
\begin{equation}
J_{1+2}\equiv J_1 + J_2 = \Lambda_1\left[S_1\right] + \Lambda_2\left[S_2\right]
\approx \Lambda\left[S_1+S_2\right] \enspace ,
\end{equation}
where $\Lambda$ is some suitable mean of $\Lambda_1$ and $\Lambda_2$.
The problem now is to classify all frequencies into distinct sets $\Omega_i$
grouping together as similar as possible $\Lambda$-operators.  The
$\Lambda$-operator can be calculated from the monochromatic optical depth
scale~\taunu\ so that the classification can be equivalently done by grouping
frequencies with a similar relation between geometrical and optical depth
scales. This is in fact the way how one proceeds in practice.

When trying to classify the frequencies, one is confronted with the problem
that the optical depth scales depend on the atmospheric model under 
consideration, i.e., its geometry,
the ensuing thermal conditions, and velocities. One has to choose a reference
model for which the classification is performed. Naturally, this reference
model is chosen to be close to the stellar atmosphere to be
simulated, in the simplest case a 1D model of the atmosphere in
question. Other choices are possible, but in any case, the resulting
classification is optimized for a particular set of atmospheric parameters and
has to be repeated when numerical simulations in other parameter regimes are
conducted. Since even for fixed atmospheric parameters a large variety of
different thermodynamic conditions are met along various lines-of-sight in a
numerical model (with correspondingly different \taunu), limits to the
achievable accuracy by the opacity binning have to be expected. Thus only a
reasonable similarity among \taunu-scales within \textcolor{tcol}{an opacity bin} is aimed at 
in practice. Typically one is content 
if the \taunu-scales of a group of frequencies share the property to reach 
unity within a given range of depth -- usually defined via the 
frequency-independent Rosseland optical depth. This emphasizes the emergent 
radiation intensity as the primary quantity to be captured correctly,
obviously an important quantity linked to the overall flux properties of a 
stellar atmosphere. \textcolor{tcol}{Each opacity bin defines a corresponding frequency group.}

\begin{figure}[tb]
\centering
\mbox{\includegraphics[bb=14 50 580 355,width=0.6\textwidth,clip=TRUE]
{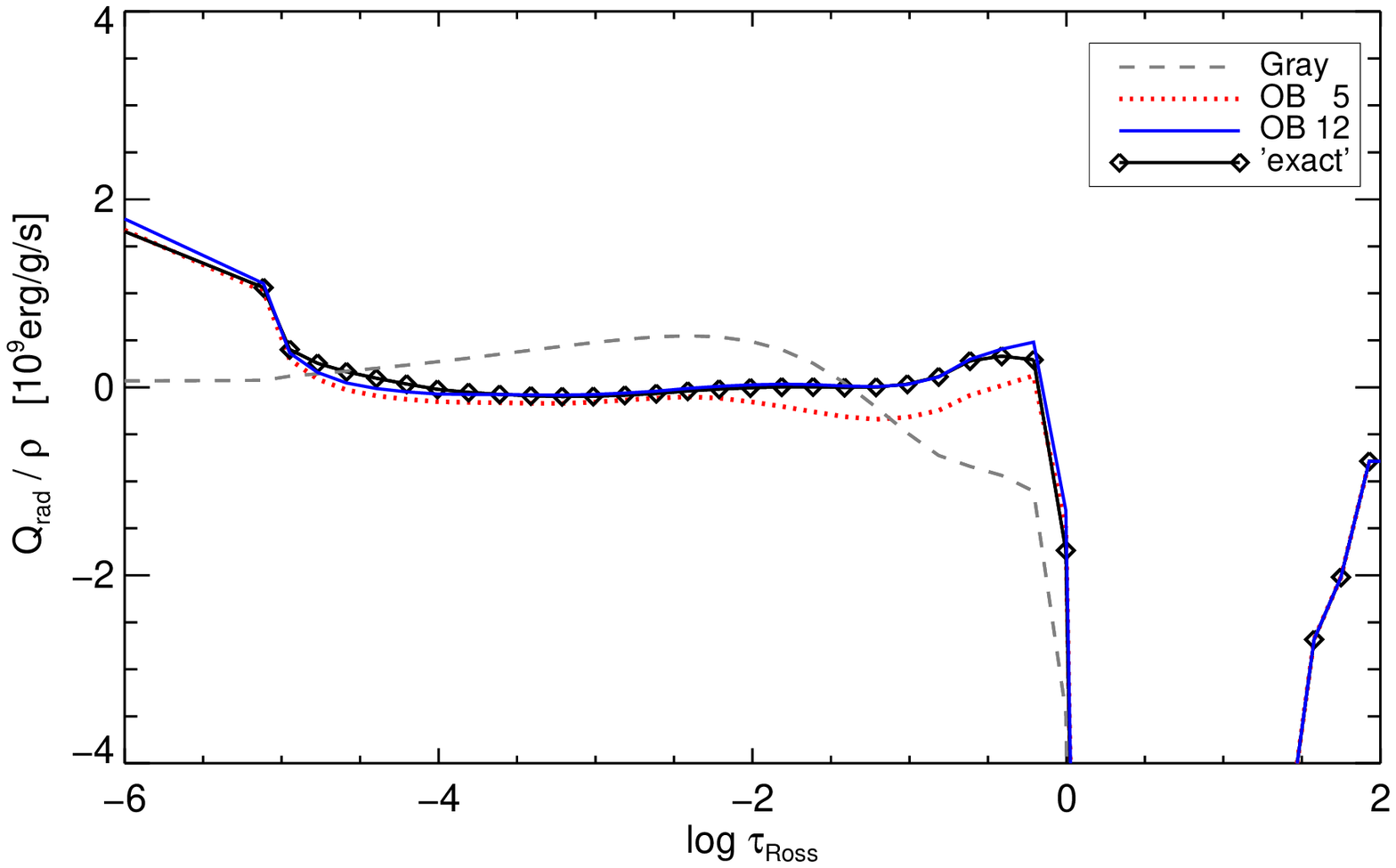}}
\mbox{\includegraphics[bb=14 28 580 355,width=0.6\textwidth]
{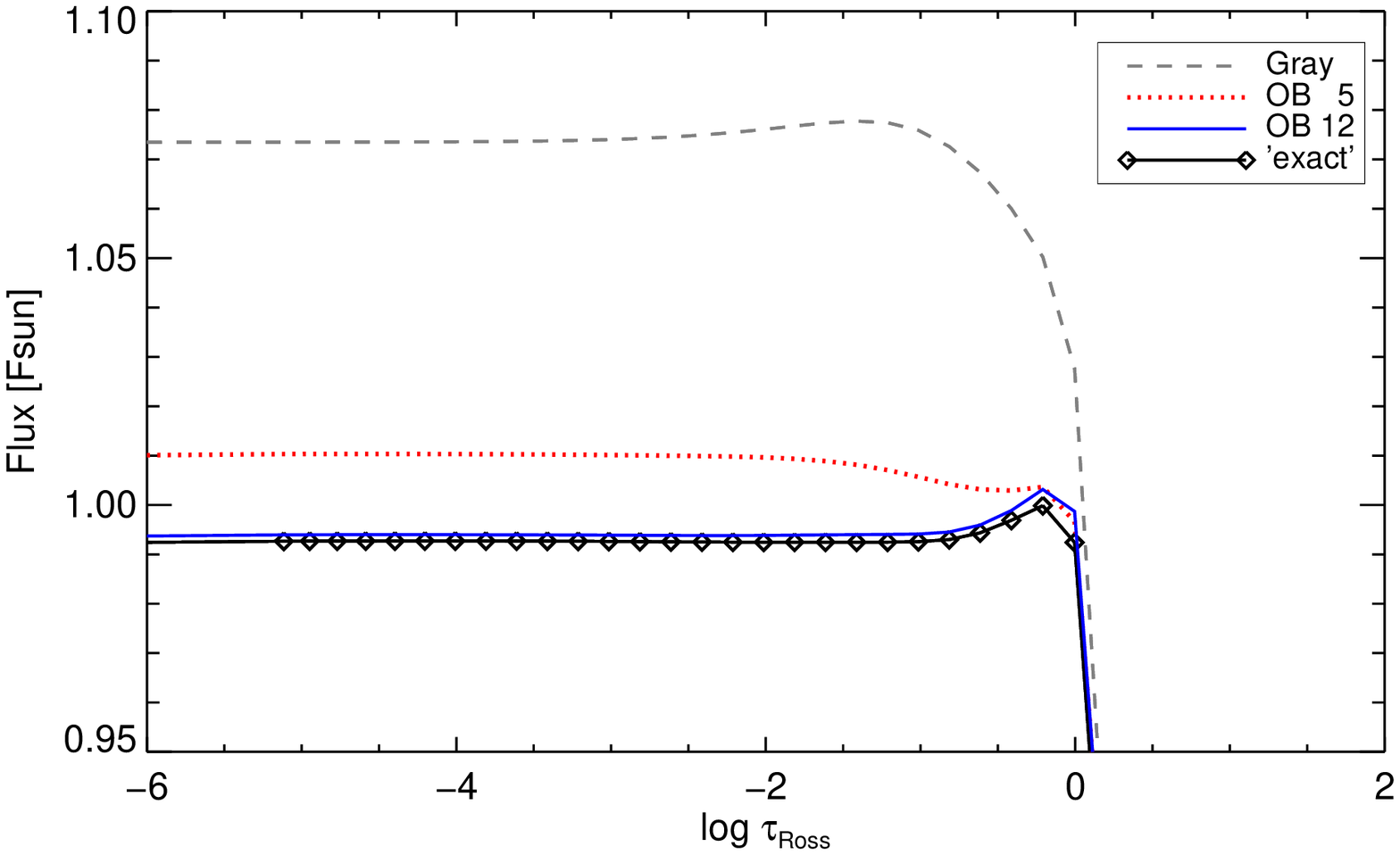}}
\caption{Performance of the opacity-binning scheme, illustrated for
a 1D solar model atmosphere. The net radiative heating 
rate per unit mass, $Q_{\rm rad}/\rho$ (top), and the bolometric radiative 
flux, $F_{\rm rad}/F_\odot$ (bottom), and are shown
as a function of Rosseland optical depth. In each panel, the results
from gray (dashed), \textcolor{tcol}{5-bin/5-group} (dotted), and \textcolor{tcol}{9-bin/12-group} (solid) 
radiative transfer are compared with the ``exact'' solution (diamonds), 
obtained with very high frequency resolution.
}
\label{f:binning}
\end{figure}

At present, typically between four and twelve 
frequency groups $\Omega_i$ are used, depending on the desired precision. An
estimate of the precision is obtained by comparing the integral radiative
heating (or cooling) rates obtained from the binned opacities with the result
obtained at high frequency resolution, both as a function of depth in the 
reference structure
used for defining the opacity bins. The estimate relies on the assumption that
the reference structure is
indeed representative of the conditions encountered in the flow simulation.
Some refinements to this basic scheme are nowadays often added. For instance,
it is sometimes advantageous to split \textcolor{tcol}{an opacity bin} as defined before into
frequency \textcolor{tcol}{sub-groups}, with the idea to separate frequency points
which systematically heat or cool particular atmospheric layers. This helps to
improve the overall energy exchange budget.

An example is given in Fig.\,\ref{f:binning}, illustrating the results
obtained for the 1D solar reference atmosphere. The basic \textcolor{tcol}{5-bin/5-group} scheme
is clearly superior to the gray approximation. The more sophisticated
\textcolor{tcol}{9-bin/12-group} scheme, in which three \textcolor{tcol}{opacity bins} are split into two 
\textcolor{tcol}{frequency sub-groups}, performs very satisfactory and almost perfectly 
reproduces the ``exact'' heating rate.

The binned opacities are obtained from a suitable average of the opacities
in a particular frequency group and stored in look-up tables as a function of
thermodynamic variables -- in \COBOLD\ as a function of gas pressure and
temperature. In addition, the Planck function (as source function), integrated
over the frequencies of a group, is stored as a function of
temperature. This approach only works if the opacities and the source function
can be calculated from the thermodynamic conditions alone, i.e., are
thermodynamic equilibrium quantities. While this is often fulfilled to good
approximation, there are exceptions. For instance, the formation of dust clouds
in cool stellar atmospheres is a non-equilibrium process (Sect.\,\ref{sec:Dust}),
and actual particle properties are only known after solving the governing kinetic equations,
taking into account the history of the evolution of a particular mass element in the
flow. In \COBOLD, we proceed by separating the equilibrium part (gas
opacities) from the non-equilibrium part (dust opacities). The gas opacities
are binned into \textcolor{tcol}{frequency groups} in the usual way, and the dust opacities are 
calculated during the simulation on-the-fly and added to the gas
opacities. Obviously, this increases the computational demands.

All in all, opacity binning has been and still is working perhaps better
than one might expect from the numerous approximations behind the
construction of the scheme. Opacity binning has proved to be an
efficient way to include the frequency dependence of the radiative transfer in
multi-dimensional simulations. However, as alluded to already before, the 
increased computing power might allow to re-consider the approach trading 
greater computational costs for higher physical fidelity.
\textcolor{tcol}{The path to largest gains} needs to be identified yet.

\subsubsection{Long-characteristics radiation transport}
\label{sec:LongCharRad}

\begin{figure}[tb]
\centering
\mbox{\includegraphics[bb=70 60 570 310, width=9cm]{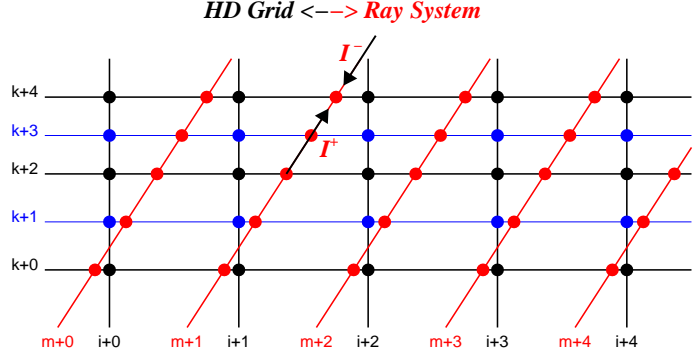}}
\caption{Schematic illustration of the different grids used with the 
long-characteristics radiative transfer method.
\textcolor{tcol}{The hydro\-dynamics equations are} solved on
a Cartesian grid (HD grid, black, dots representing cell centers), while the
radiative transfer equation is solved on a system of inclined rays 
(red, dots representing the mesh points used with the Feautrier scheme).
The HD grid can be refined in vertical direction by additional $z$-planes
(thin, blue) to provide sufficient resolution for strongly inclined rays.
\textcolor{tcol}{The cell centers of the refined HD grid have indices ($i,k$), the 
mesh points along the rays have indices ($m,k$).}
}
\label{f:raysys}
\end{figure}

The purpose of this algorithm is to compute the net radiative heating rate
per unit volume, $Q_\mathrm{rad}(x_i,y_j,z_k)$,
at the center of each cell of the hydro\-dynamical 
grid (HD grid). The basic idea is to solve the equation of 
radiative transfer on a system of straight \emph{long} rays 
\textcolor{tcol}{(long-characteristics, LC)} running from the upper to the lower
model boundary at a number of different azimuthal angles $\phi$ and 
inclinations with respect to the vertical ($0 \le \theta < \pi/2)$. As
a result, we obtain for each \textcolor{tcol}{frequency group} $\nu$ and for all bundles of 
rays with orientation $(\theta,\phi)$
the quantity $Q_{\mathrm{rad},\nu}(\theta,\phi) = \overline{\rho \kappa_\nu} \; 
\left( u_\nu(\theta,\phi) - \Source_\nu \right)$ at the mesh points along the 
rays, where the mean-intensity-like variable $u_\nu(\theta,\phi)$ is the 
average of incoming ($I_\nu^-$) and outgoing ($I_\nu^+)$ intensity, 
$u_\nu=(I_\nu^+ + I_\nu^-)/2$ (see Fig.\,\ref{f:raysys}), \textcolor{tcol}{$S_\nu$ is the group
source function, and $\overline{\rho \kappa_\nu}$ is the group opacity
averaged over the neighboring mesh points along the ray 
(see Eq.\,\ref{eq:qray}).}
$Q_\mathrm{rad}(x_i,y_j,z_k)$ is then constructed by interpolating 
$Q_{\mathrm{rad},\nu}(\theta,\phi)$ from the ray system to the cell centers 
of the hydro\-dynamics grid, and appropriate angular averaging and summation
over \textcolor{tcol}{frequency groups.} 

\textcolor{tcol}{
Note that the technique described here basically evaluates $Q_\mathrm{rad}$
according to Eq.\,(\ref{eq:J-S}). It overcomes the difficulties explained
in the context of Eq.\,(\ref{eq:J-S}) by solving the transport equation for 
the difference between mean intensity and source function, 
$p_\nu \equiv u_\nu-S_\nu$ (see Eq.\,\ref{eq:feautrier}), which gives 
accurate values of $(u_\nu-S_\nu)$ for arbitrarily large optical depth. 
At the same time, it allows $Q_{\mathrm{rad},\nu}$ to be computed such that 
energy conservation is enforced (see Eq.\,\ref{eq:divfrad}). The 
procedure is very similar to that described in 
\cite{Nordlund1982A&A...107....1N}.
}

To simplify matters, $\phi$ is restricted to $0, (1/2)\pi, \pi, (3/2)\pi$,
i.e., we consider only 2D ray systems in vertical slices along the $x$
and $y$ axis of the hydro\-dynamical grid.
The $\theta$ angles are given by Lobatto's quadrature formula
\cite{Abramowitz1972hmfw.book.....A}; typically, 2--4 non-zero inclination 
angles are sufficient, in addition to a set of vertical rays.
All rays start at the cell centers of the uppermost level of the HD grid
and follow the specified direction, assuming periodic lateral boundary 
conditions, until they reach the bottom of the computational domain.

As indicated in Fig.\,\ref{f:raysys}, the mesh points along the rays
are defined as the intersection points with the $z$-planes of the
HD grid. As this recipe would imply a rather coarse sampling along
strongly inclined rays, we introduce additional horizontal planes
such that the geometrical separation of mesh points along the \textcolor{tcol}{inclined}
rays remains comparable to the vertical resolution of the original HD grid.
The coordinates of the ray points, $(x_{mk},z_k)$, where
$m$ is the ray index and $k$ is the depth index
of the refined HD grid, are equidistant in $x$.

The main steps of the whole procedure may be summarized as follows:
first, the source function, $\Source_\nu$, and the opacity per unit volume,
$\rho\,\kappa_\nu$, are interpolated from the HD grid to the mesh points of 
the ray system. Linear interpolation of $\Source_\nu$ and 
$\log (\rho\,\kappa_\nu)$ is adopted for the vertical direction 
(additional $z$-planes), while linear interpolation of $\Source_\nu$ and 
$\rho\,\kappa_\nu$ is used in horizontal direction. 
\textcolor{tcol}{Note that only a 1D interpolation along the Cartesian grid lines is required.}
Given $\rho\,\kappa_\nu$ on the mesh points along the rays, we 
represent $\rho\,\kappa_\nu$ between two mesh points by a monotonic cubic
polynomial \cite{Steffen1990A&A...239..443S} to obtain the optical depth 
increments $\Delta\tau_\nu$ by analytical integration. Next, we solve the 
equation of radiative transfer along bundles of rays in the form of 
the second-order differential equation:
\begin{equation}
\label{eq:feautrier} 
\dededtaunu{p_\nu}=p_\nu - \dededtaunu{\Source_\nu}  \enspace , \quad   
p_\nu \equiv u_\nu - \Source_\nu \,
  \enspace ,
\end{equation}
where $\tau_\nu$ is measured along the (inclined) rays.
This modified Feautrier equation is solved by the 
forward-elimination and back-substitution formalism originally described by 
\citet{Feautrier1964CR....258.3189F} (see also 
\citet{Mihalas1978stat.book.....M}), giving $p_\nu(x_{mk}, z_k)$ at the mesh 
points of the ray system. 

At the lower boundary, where conditions are optically very thick in
general, we can choose between two basic options: if the bottom layer
is located in a radiative zone, and we want to enforce a given radiative 
flux $F_{\rm rad,\nu}$ through the lower boundary, the condition is
\begin{equation}
\label{eq:feautrier_lb1} 
\dedtaunu{p_\nu} = \frac{3}{4}\,\frac{F_{\rm rad,\nu}}{\pi}\,\cos(\theta)
 - \dedtaunu{\Source_\nu}
\enspace .
\end{equation}
If the bottom layer is located in a convective zone, where the radiative 
flux through the lower boundary is negligible compared to the energy flux
carried by the flow, a reasonable boundary condition is to require 
the net radiative energy exchange to vanish in each frequency group,
\begin{equation}
\label{eq:feautrier_lb2} 
{\nabla}\cdot F_{\rm rad,\nu} = 0 \quad \mathrm{or} \quad  p_\nu=0
  \enspace .
\end{equation}
Note that this does not imply $F_{\rm rad,\nu}=0$.

At the uppermost layer, the optical depth is computed as
\begin{equation}
\label{eq:tau0} 
\tau_{\nu,0} = H_\tau\,\rho_0\,\kappa_{\nu,0}
  \enspace ,
\end{equation}
where $H_\tau$ is the mean optical depth scale height at the top of
the model, \textcolor{tcol}{$(H_\tau)^{-1}=-(\mathrm{d} \ln (\rho\,\kappa_\nu)/\mathrm{d} z)_0
\approx -(\mathrm{d} \ln \tau_\nu/\mathrm{d} z)_0$.} 
The incident radiation is given by
\begin{equation}
\label{eq:Iminus} 
I_\nu^-(\tau_{\nu,0}) = (1-{\rm e}^{-\tau_{\nu,0}})\,\Source_{\nu,0} + I_\nu^\ast
  \enspace ,
\end{equation}
where $\Source_{\nu,0}$ is the mean source function of the upper layer,
and $I_\nu^\ast$ denotes the incident intensity due to an arbitrary
 external source (usually zero). In terms of $p_\nu$, the upper boundary
condition for radiation can be formulated as
\begin{equation}
\label{eq:feautrier_ub1} 
p_\nu-\dedtaunu{p_\nu} = (1-{\rm e}^{-\tau_{\nu,0}})\,\Source_{\nu,0} + 
I_\nu^\ast - \Source_\nu + \dedtaunu{\Source_\nu}
  \enspace .
\end{equation}
Next, the quantity $q_\nu$ is computed at all mesh points of the ray system as
\begin{eqnarray}
\label{eq:qray} 
q_\nu(x_{mk},z_k) &=& \cos \theta\,
\frac{\tau_\nu(x_{m,k+1},z_{k+1})-\tau_\nu(x_{m,k-1},z_{k-1})}{z_{k+1}-z_{k-1}}\,
p_\nu(x_{mk},z_k) \nonumber \\
&\equiv& \left\{\overline{\rho\,\kappa_\nu}\,(u_\nu-\Source_\nu)\right\}_{mk} 
\enspace .
\end{eqnarray}
Finally, the partial heating rates $q_\nu$ are interpolated back 
onto the HD grid in a conservative way, such that for all height levels $k$
\begin{equation}
\label{eq:qconserv} 
\sum_m{q_\nu({x}_{mk}, z_k)} = \sum_i{q_\nu(x_i, z_k)}
  \enspace .
\end{equation}
$Q_\mathrm{rad}(x_i,y_j,z_k)$ is then built up by adding the individual
contributions $q_\nu(x_i, y_j, z_k)$ of the different ray directions 
$(\theta,\phi)$ with their appropriate integration weights, and summation
over all frequency groups $\nu$.

By virtue of the definition of $q$ according to Eq.\,(\ref{eq:qray}), and the
requirement of a conservative back interpolation as expressed by
Eq.\,(\ref{eq:qconserv}), our long-characteristics radiative-transfer
scheme conserves energy in the sense that for each \textcolor{tcol}{frequency group}
\begin{equation}
\label{eq:divfrad} 
\int_x\int_y F_{\rm rad,\nu}^{\rm top} \,\mathrm{d}x\, \mathrm{d}y - 
\int_x\int_y F_{\rm rad,\nu}^{\rm bot} \,\mathrm{d}x\, \mathrm{d}y =
\int_x\int_y\int_z Q_{\rm rad,\nu}(x,y,z)\, 
\mathrm{d}x\, \mathrm{d}y\, \mathrm{d}z
  \enspace .
\end{equation}
Here, $F_{\rm rad,\nu}^{\rm top}$ and  $F_{\rm rad,\nu}^{\rm bot}$ are, 
respectively, the net radiative energy flux through the upper and 
lower boundaries of the model, computed directly from the ray system
intensities at the top and bottom level. Note that Eq.\,(\ref{eq:divfrad})
holds only if the volume integral includes the $Q_{\rm rad,\nu}$ obtained
at the additional horizontal sub-levels introduced for grid refinement.
The final $Q_{\rm rad}$ on the original HD grid must therefore be computed
as a suitable average over the neighboring $z$ sub-levels to ensure
energy conservation.

A \textcolor{tcol}{distinct} advantage of the long-ray approach is that it allows an efficient
solution of the transfer equation for beams of parallel rays by means 
of the Feautrier scheme, which is very fast and elegant, \textcolor{tcol}{automatically
ensures the correct asymptotic diffusion limit at large optical depth, 
and} could easily account for scattering along single rays \textcolor{tcol}{(for an early 
example of this approach see \citet{Cannon1970ApJ...161..255C}).
In principle, the LC method can also be combined with 
integral-operator techniques (e.g.\ \cite{Avrett1971JQSRT..11..559A},
\cite{Jones1973ApJ...185..167J}), which, however, are numerically less 
efficient and suffer from interpolation issues
(\cite{Jones1973ApJ...185..183J}, \cite{Mihalas1978ApJ...220.1001M}).
In contrast to what is assumed in \citet{Kunasz1988JQSRT..39...67K},
the computing time of our LC scheme scales linearly 
with the number of HD cells and the number of frequency groups, 
as for the short-characteristics scheme. It scales in a non-linear way with 
the number of $\theta$-angles, since more-inclined rays are longer and have
a larger number of mesh points. The computing time can be reduced by
computing $Q_{\rm rad}$ from the diffusion approximation in the 
lower, optically very thick layers of the model. Compared to the
ray-system solution, the computation of the diffusion approximation comes 
almost for free.}

A disadvantage of the LC method is the necessity of extensive interpolation 
from the HD grid onto the ray system and back. This procedure is prone 
to problems with ``leaking'' of heating or cooling to neighboring cells in 
the presence of localized ``hot spots'', as described in the following 
Section\,\ref{sec:ShortCharRad} (cf.\ Fig.\,\ref{f:sevol}).
To some degree, such problems may be abated, at the expense of higher
computational cost, by increasing the number of rays per unit length in 
horizontal direction.

\subsubsection{Short-characteristics radiation transport}
\label{sec:ShortCharRad}

\begin{figure}[hbt]
  \centering
  \includegraphics[width=8.8cm]{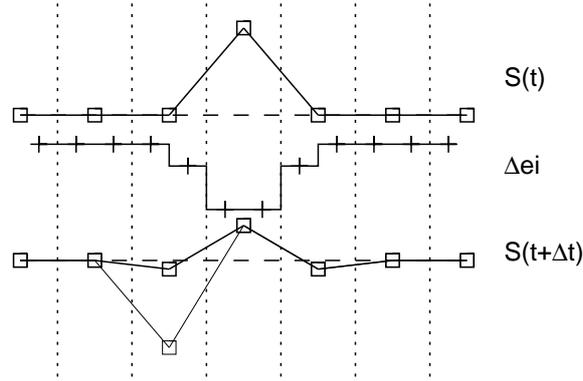}
  \caption{Initial run of the source function (top \textcolor{tcol}{curve}),
induced changed of energy per cell (center \textcolor{tcol}{curve}),
and run of the source function after one time step (bottom \textcolor{tcol}{curve}).
The three quantities are plotted as function of optical depth
for a few grid cells whose boundaries are depicted by vertical dotted lines.
The values at \textcolor{tcol}{the} cell centers are marked by squares.
The two sub-intervals in each cell can have different values of the energy change.
The lower curve shows the source function after one time step for
a constant heat capacity per grid cell (thick line) and
the case where the heat capacity in the left neighbor of the hot cell is
smaller by a factor 10.}
  \label{f:sevol}
\end{figure}

\begin{figure}[hbt]
  \centering
  \includegraphics[width=6.0cm]{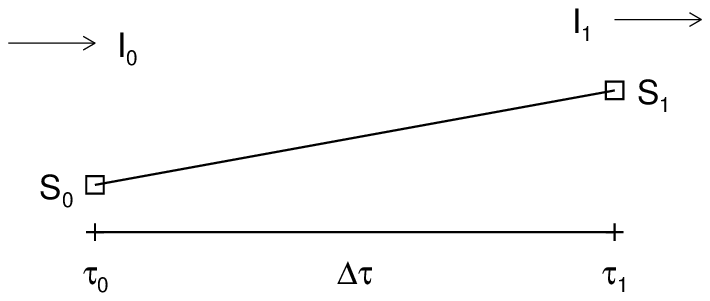}
  \includegraphics[width=6.0cm]{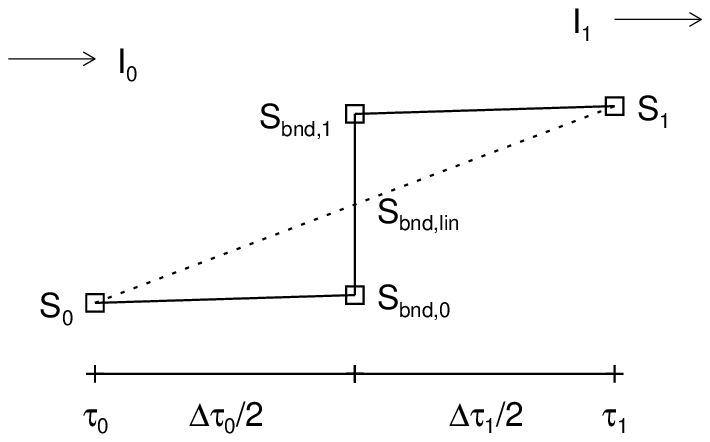}
  \caption{The sketch on the left illustrates the naming convention
used for the case of a linear dependence of source function $\Source$
on optical depth $\tau$ within a single interval
as opposed to the case of a piecewise linear source function
as in the plot on the right, where
the interval $[\tau_{\rm 0}, \tau_{\rm 1}]$ is split into
two sub-intervals with width $\dtau_{\rm 0}/2$ and $\dtau_{\rm 1}/2$,
respectively.
The source function varies linearly in each sub-interval.
However, it is allowed to have a jump at the transition.}
  \label{f:s0s1}
\end{figure}


\begin{figure}[hbt]
  \centering
  \includegraphics[width=8.8cm]{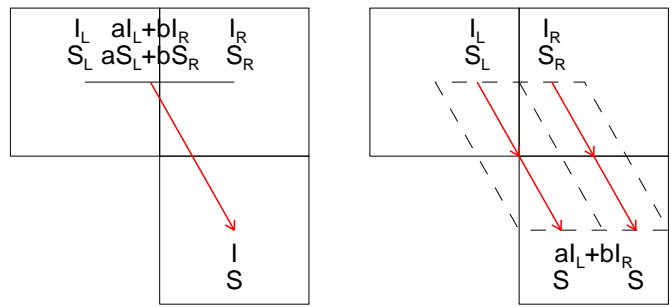}
  \caption{Short-characteristics step to get the intensity
in the target cell (bottom) from the values at the previous cell plane (top)
in two dimensions:
left: standard integration with one short ray and interpolation of intensity
      and source function at the top grid plane;
right: separate rays for each neighbor cell with a summing up of the intensity
       in the bottom grid cell,
       the splitting of each ray gives
       the intensity change within each cell.}
  \label{f:shortinter}
\end{figure}




\textcolor{tcol}{
The LC scheme described in the previous section is part of \COBOLD\ since the very beginning.
It is adapted to the conditions of plane-parallel atmospheres in local models:
e.g., it heavily makes use of periodic side boundary conditions.
The angular distribution of rays is chosen to optimize the vertical radiative flux.
The diffusion approximation used in the deeper layers can save some computational time.}

\textcolor{tcol}{
While for local models there was no reason to spend time on experimenting with
another radiation-transport scheme, this changed for global models where
the conditions are different:
the vertical direction is not preferred anymore.
Instead, all sides of the computational domain are open for radiation.
The numerical resolutions is in general worse than for local models
and the violent flows and give rise to large local temperature and opacity fluctuations.
This means that errors caused by the interpolation in LC schemes
would become more apparent.}

\textcolor{tcol}{
The short-characteristics (SC) scheme in \COBOLD\ overcomes these stability problems
at the possible expense of the accuracy of the vertical radiative energy flux.
The basic idea of not following rays through the entire volume is the
same as in \citet{Kunasz1988JQSRT..39...67K}.
But a different way of interpolating the intensity and the source function
makes it better adapted to the use within an RHD code.
}

The main emphasis during the development of \textcolor{tcol}{the SC} scheme
has been put on stability by
preventing local peaks of the source function
from ``leaking'' into neighbor cells and causing an unwanted smearing
of the cooling or heating term \textcolor{tcol}{(see Fig.\,\ref{f:sevol})}.
This requires a special reconstruction of the source function within optically
thin cells in the 1D radiation transport operator
and a carefully chosen interpolation within the SC scheme.
%

Instead of a Feautrier scheme as in Sect.\,\ref{sec:LongCharRad},
the analytic solution of the 1D version of the radiation transport equation\,(\ref{eq:deqdIdtau})
with linear source function (Fig.\,\ref{f:s0s1}, left) is used as atomic operator,
\begin{equation}
\label{eq:IntFromLinS}
  \Int_1 = \Int_0 \, e^{- \dtau}
             \, + \,
           \Source_1
             \, - \,
           \Source_0 \,
               e^{- \dtau}
             \, + \,
           \frac{\dSource}{\dtau}
             \left[
               e^{- \dtau} - 1
             \right]
  \enspace ,
\end{equation}
which guarantees the positivity of the source function everywhere.

The energy change has to be computed accurately
in the optically very thick
(e.g., in the center of a toy stellar model with $\dtau > 10^8$)
and in the optically very thin
(e.g., in some regions far away from the surface of a red supergiant model
with $\dtau < 10^{-20}$).
Both cases pose no problem for the formal solution because in the former case
the intensity is essentially given by the local source function.
And in the latter case, the changes to the radiation field due the contribution
of the extremely thin regions can be safely ignored -- or simply added to the
much larger intensity along a ray and therefore absorbed by the limited
machine precision.
However, optically very thick or thin regions still interact with the
radiation field and the local heating or cooling is significant
and has to be computed in a time-dependent code.
The \textcolor{tcol}{SC} scheme in \COBOLD\ uses different arrangements of terms
in optically thin and thick regimes
to account for round-off errors,
giving accurate values for optically very thick or thin regions --
even running only in single precision.

Separate integration steps are employed for the interval from the cell center to the boundary
and from the cell boundary to the next cell center (Fig.\,\ref{f:s0s1}, right)
to get the intensity change within each cell, from which the energy change
per cell is computed.
%
In the optically thin regime, the slope of the source function is reduced (Fig.\,\ref{f:s0s1}, right)
to suppress leaking of cooling or heating from one cell to the next (Fig.\,\ref{f:sevol}).
Each ray inclination requires an integration of the intensity in both directions.
For inclined ray directions,
there is one pair of intensity-integration steps for each pair of neighbor cells
to avoid the leaking  of cooling or heating associated with the spatial interpolation
of source function and/or intensity (Fig.\,\ref{f:shortinter}).
\textcolor{tcol}{
That means, that in contrast to the well-known SC scheme in \citet{Kunasz1988JQSRT..39...67K},
even for a single direction there might be more than one ray connecting a cell with its neighbors.
}

The numerical scheme \textcolor{tcol}{proceeds} as follows:

At the beginning of each radiation-transport sub step,
the temperature $T$ is computed from density $\rho$ and internal energy $\ei$ for every mesh point
of an equidistant \textcolor{tcol}{3D} Cartesian grid.
For every frequency \textcolor{tcol}{group} (Sect.\,\ref{sec:OpacityBinning})
opacity $\kappa$ and source function $S$ are calculated
by interpolating in precompiled tables.
%
Next, for every ray inclination the optical thickness $\dtau$ of each cell is calculated.

\textcolor{tcol}{At the beginning of each integration step,
the boundary values of the intensity have to be set.
}
For the \textcolor{tcol}{SC} scheme,
only open boundary conditions (zero infalling intensity)
are implemented
(for simulations with periodic side boundaries,
the \textcolor{tcol}{LC} scheme (Sect.\,\ref{sec:LongCharRad})
is used, instead).

\textcolor{tcol}{
The integration proceeds then layer by layer along the axis that is closest to
the inclined ray direction.
For each ray direction, the intensity at each cell does not depend on its neighbors within
a layer but only on cells in the previously computed layer.
That means, that
the innermost loop in each layer can be efficiently vectorized.
The next loop is parallelized with OpenMP directives
and the outermost loop performs the integration.
}

Each complete 3D radiation-transport step includes directions
according to the coordinates of the corners
of regular polyhedrons, which results in equal weights for all rays.
After the loop over all inclinations and the loop over all frequency groups,
the energy change per time is derived from all the accumulated intensity changes
and used to update the internal energy $\ei$ in each cell
for given time step $\Delta t$.
Here, ``conservation of intensity'' translates into ``conservation of energy''
and ensures the conservativity of the radiation-transport update step
(except for the losses through the outer boundaries).

From one sub time step to the next,
the orientation of the polyhedron can change randomly
to give some coverage of the entire sphere.
However, some simulations are restricted to rays aligned with axes or diagonals
resulting in a considerable speed-up while loosing some angular resolution.
There are several radiative time steps per hydro\-dynamics time step possible to compensate
for the short radiative time scale compared to the hydro\-dynamic one.

In cool supergiants close to the Eddington limit, radiation pressure
plays an important role in the stellar atmosphere and
the wind of asymptotic giant branch (AGB) stars is driven by radiative pressure on dust.
With the scheme presented above,
the three components of the radiative acceleration can easily by computed
from the intensity change per cell.


\subsection{MHD}
\label{sec:MHD}


\textcolor{tcol}{In \COBOLD, the numerical scheme used for the solution of the equations 
of magneto-hydrodynamics is \textcolor{tcol}{quite} different from the one employed
for the case of pure hydrodynamics described in Sect.\,\ref{sec:HD}.}
In the case of solar and stellar magneto\-convection, the scheme must be able
to deal with highly stratified flows where the plasma-$\beta$ (i.e., the ratio of
the thermal to the magnetic energy density of the plasma) varies over several 
orders of magnitude.
A special requirement of MHD calculations is the 
enforcement of the divergence-free condition 
$\mathbf{\nabla}\cdot\mathbf{B} = 0$ for the
magnetic field. Violating this condition can lead to unphysical forces, which
can degrade the solution \cite{1980JCoPh..35..426B}. Several methods have been 
developed to enforce
this condition either to roundoff error or approximately to the order of the
scheme. One method is to use the eight-wave formulation
of the MHD equations
\citep{1994JGR....9921525G, 1994arsm.rept.....P}.
The additional wave is
associated with the propagation of magnetic monopoles. In the eight-wave
formulation, additional source terms proportional to 
$\mathbf{\nabla}\cdot\mathbf{B}$ appear, i.e., the equations are no longer
conservative. Another method uses a cleanup step at the end of each time step,
removing the errors in $\mathbf{\nabla}\cdot\mathbf{B} = 0$. 
This requires the solution of a Poisson equation at each time step.
A third possibility, which is used in the MHD module of \COBOLD,
is the constrained-transport method of Evans and Hawley
\cite{1988ApJ...332..659E}. It uses a special finite-difference
discretization of the induction equation on a staggered grid such that a
discrete formulation of the
divergence-free condition for the magnetic field is maintained to machine
accuracy.
All of these methods can be treated as modifications of an underlying
base scheme. A detailed comparison of these methods can be found in
\citet{2000JCoPh.161..605T}.

Another difficulty in MHD simulations is to keep the thermal gas pressure positive
\cite{1999JCoPh.148..133B, 2000JCoPh.160..649J}.
Since the gas pressure is a dependent variable when using the conservative
form of the MHD equations, it is computed by subtracting the potential, the kinetic,
and the magnetic energy from the total energy, $\etot$. When the magnetic energy is
much larger than the internal energy, i.e., for small values of the plasma-$\beta$,
small errors in the total energy can drive the gas pressure to negative values.
This can be a problem in the solar chromosphere,
where values of $\beta \approx 10^{-4}$ are common,
whereas the gas pressure dominates in the sub-photospheric layers where $\beta$ is huge.
In the MHD module of \COBOLD, several provisions are made to avoid a negative
gas pressure. To keep the magnetic field solenoidal,
\COBOLD\ uses the constrained-transport method in combination with a
Godunov-type finite-volume scheme as the base scheme. In the following,
each component of the scheme is described in more detail.

\subsubsection{Spatial and temporal discretization}

\textcolor{tcol}{The spatial discretization of the MHD equations is
similar to the hydro\-dynamic case, i.e., the hydro\-dynamic variables
are cell centered. The magnetic fields are located at the cell interfaces.}
The cell-centered magnetic field components,
which are required by the Riemann solver of the base scheme, 
are computed from the magnetic field
components at the cell interfaces by linear interpolation.
Then all cell-centered variables are updated by the base scheme.
The extension to second order in space is 
done by linear reconstruction of the primitive variables $\rho$, $\mathbf{v}$,
$\mathbf{B}$, $P$, and $\rhoei$. Second order in time is achieved either by
a Hancock predictor step \textcolor{tcol}{\cite{vanLeer1984, Toro}}
or by a second-order TVD-Runge-Kutta time-integration scheme
\textcolor{tcol}{\cite{1998MaCom..67...73G}.}
In some situations,
where the second-order scheme would result in negative gas pressure, 
the scheme is locally reduced to first order.

\subsubsection{The approximate Riemann solver}

In the hydro\-dynamic scheme of \COBOLD, a Roe solver is used (Sect.\,\ref{sec:HD}). 
However, the Roe solver does not guarantee positivity of the density and
the pressure. This problem, which is also present in the hydro\-dynamic case
gets worse for MHD. Whereas in the hydro\-dynamic case, reducing the time step often
helps to overcome the problem, in MHD simulations, the problem remains,
even if the time step is reduced considerably.

It can be shown that the HLL solver \textcolor{tcol}{\cite{harten:35}} ensures
positivity of the gas pressure and the density if the exact solution
of the Riemann problem is positive \cite{1991JCoPh..92..273E}. 
For MHD, this is the case only
if there is no jump in the normal component of the magnetic field.
In 1D, the divergence-free condition enforces the normal component
of the magnetic field to be constant. For multi-dimensional problems
however, using cell-centered magnetic fields, jumps in the
normal component of the magnetic field occur even if the
divergence-free condition is fulfilled in a discrete sense.
It was shown by \citet{2000JCoPh.160..649J} that allowing magnetic monopoles,
which arise from these jumps,
and taking into account their contribution to the
Lorentz force, an additional source term occurs in the
induction equation only. Using a special discretization of this
source term, \citet{2000JCoPh.160..649J} demonstrated numerically
that the HLL solver for MHD always provides positive gas pressures.

We use the method of Janhunen for the MHD module of \COBOLD. However,
it should be noted that this source term is only used for the computation
of the fluxes by the HLL-solver of the base scheme. For the update of the
magnetic field by the constrained-transport method, this source term is not
used so that the magnetic field stays divergence-free.

\subsubsection{The constrained-transport step}

\COBOLD\ uses the flux-interpolated constrained-transport method of
\citet{1999JCoPh.149..270B}. First, the electric field at the
cell edges is computed from the fluxes at the centers of the cell interfaces,
provided by the base scheme. 
%
The magnetic field at the cell interfaces is then updated with
this electric field, applying Stokes theorem to every face of
a cell. The updated cell-centered magnetic field from the base
scheme is discarded. 
The new cell-centered magnetic field is computed from the updated magnetic field 
at the cell interfaces by linear interpolation.

Since the new cell-centered magnetic field is different from the magnetic field 
provided by the base scheme, the internal energy, $\ei$, must be modified after the
constrained-transport step according to
\begin{equation}
\ei = \ei^* + 
\frac{\mathbf{B}^*\cdot\mathbf{B}^* - \mathbf{B}\cdot\mathbf{B}}{2 \rho}
  \enspace ,
\end{equation}
where $\mathbf{B}^*$ and $\ei^*$ are the magnetic field and the internal energy
provided by the base scheme. If this correction would result in negative gas pressure,
it is not performed, i.e., $\ei = \ei^*$ (see also \citet{1999JCoPh.149..270B})
and the total-energy conservation is sacrificed in favor of improved robustness.

\subsubsection{Dual-energy method and Alfv{\'e}n-speed reduction}
\label{sec:DualEnergyMethod}

Even if a scheme guarantees positivity of \textcolor{tcol}{the} gas pressure,
this does not necessarily  mean that the gas pressure is computed accurately.
In fact, by using the total 
energy equation for the computation of the internal energy, the discretization errors
in the total energy, the kinetic energy, and the magnetic energy of the scheme tend
to be imposed on the internal energy. One could use the entropy equation \textcolor{tcol}{or the 
equation for the thermal energy itself, instead. Another possibility is to use
the total energy equation in combination with one of these equations.
For example \citet{1999JCoPh.148..133B}
use the entropy equation for the update of the internal energy in regions with strong
magnetic fields.
For the MHD module of \COBOLD, the so-called dual-energy method, i.e.,
a combination of the equation for the total energy
and the equation for the thermal energy is used.}
In regions with \textcolor{tcol}{a} large $\beta$, the internal energy is updated
with the equation of the total energy.
In turn, when $\beta$ is small
($\beta \lessapprox 10^{-3}$),
the equation for the internal energy is used at the expense of strict energy conservation.
\textcolor{tcol}{Since typically $\beta$ is small in very restricted regions of the 
computational box only, conservation of total
energy is still maintained in most parts of the computational domain.}

In order to avoid extremely small time steps due to the CFL condition when the 
Alfv{\'e}n speed is high, the Alfv{\'e}n speed can be limited by artificially reducing 
the strength of the Lorentz force by a factor
\begin{equation}
f = \frac{v_\mathrm{A max}^2}{v_\mathrm{A}^2 + v_\mathrm{A max}^2}
\enspace ,
\end{equation}
where $v_\mathrm{A}$ is the actual Alfv{\'e}n speed and $v_\mathrm{A max}$ is the desired 
upper limit of the Alfv{\'e}n speed. The method is similar to that used by
\citet{2009ApJ...691..640R}. Of course, caution is indicated when using this
method. Obviously, it can hardly be used for the study of magneto\-acoustic wave 
propagation. However, it may be perfectly admissible in situations, where the
low-$\beta$ regime is merely included as a buffer region to the (upper) boundary of the
physical domain.

\subsubsection{\textcolor{tcol}{Ohmic diffusivity}}
\label{sec:OhmicDiffusion}

\textcolor{tcol}{
While it is not necessary for stability, the MHD-scheme of \COBOLD\ can
also handle \textcolor{tcol}{explicit} magnetic diffusion. It is treated explicitly in
the scheme by modifying the electric field in the
constrained-transport-step. A constant magnetic diffusivity and the
artificial magnetic diffusivity according to
\citet{2001MNRAS.322..461S}
are currently implemented in \COBOLD. The constant magnetic
diffusivity can be used to specify the magnetic Reynolds number
\begin{equation}
R_m = \frac{v L}{\eta}
\enspace ,
\end{equation}
where $\eta$ is the magnetic diffusivity, $L$ is a typical length scale and
 $v$ is a typical velocity of the flow. The artificial magnetic
diffusivity is given by
\begin{equation}
\eta = C \frac{\left(\Delta x\right)^2 |\mathbf{j}|}{\sqrt{\rho}}
\enspace ,
\end{equation}
where $\Delta x$ is the grid spacing, $\mathbf{j} = \mathbf{\nabla}\times\mathbf{B}$ 
is the current density, and $C$ is a dimensionless parameter.
}

\subsubsection{Magnetic boundary conditions}

The boundary conditions for the magnetic field can be specified
independently from the hydro\-dynamic settings
for the top \textcolor{tcol}{and the bottom boundaries}, and for 
each of the horizontal directions.
Typical horizontal boundary conditions used for simulating
magneto\-convection in a local box are periodic. Another
boundary condition, mostly applied to the bottom and the top of the box, consists
in setting the magnetic field tangential to the boundary to zero, so that the
magnetic field lines stay normal to the boundary. A generalization of this
boundary condition specifies the obliquity of the magnetic field at the
boundary.
There is also a special condition for the open lower boundary,
which allows upflows to advect horizontal magnetic field into the computational box.
Another boundary condition consists in setting the electric field to zero at the
boundary.
This means that the normal component of the magnetic field at the boundary does not change.
In this case the magnetic field lines are effectively anchored at the boundary.

Conditions which keep the magnetic field vertical at the top and
bottom boundaries are typically used for the simulation of intense,
vertically directed magnetic flux tubes in the photosphere of the Sun
as they occur in magnetically active regions such as plages and
enhanced network regions. The advection of weak horizontal field
across the bottom boundary is used for the simulation of magnetically
inactive, very quiet regions on the Sun. \textcolor{tcol}{With this boundary 
condition, it} is assumed that convective
updrafts transport magnetic fields from deep layers of the convection
zone to the solar surface. Anchored fields may be useful for anchoring
an entire sunspot at the bottom boundary or for anchoring horizontal
fields at the side boundaries for the simulation of horizontally
directed penumbral filaments.

\subsection{Optional modules} 
\label{sec:OptionalModules}

\textcolor{tcol}{The numerical treatment of the source terms $S_i$ in Eq.\,(\ref{eq:ni})}
dealing with different types of dust and chemical-reaction networks
is implemented as a separate step (see below),
following the general concept of operator splitting. 
The optional modules are called after the (magneto)hydro\-dynamics step for each 
computational time step. 
These modules treat the mass or number densities of the dust particles or chemical species 
as additional quantities, which are included in the in- and output of the simulation data.
Only up to one of these extra modules can be used at a time, so far.

During the (magneto)hydro\-dynamics solver step,
the additional densities are advected with the flow field
analogously to the gas density.
Their transport velocity across each cell boundary is computed from the gas mass flux
divided by the upwind density,
in some cases modified to account for the gravitational settling of dust.

The contribution of the additional components to the total opacity can be added to
the standard equilibrium gas opacities (Sect.\,\ref{sec:OpacityBinning}).
The boundary conditions are made consistent with the hydro\-dynamics part of the code. 
%
%

%

\subsubsection{Chemical-reaction networks}
\label{sec:chemnet}

Apart from advection across the cell boundaries, the number density $n_i$ of 
a chemical species in a grid cell can be changed due to chemical reactions:  
\begin{equation}
\label{eq:dndt}
\left(S_i\right)_\mathrm{chem}
      =  - n_i      \sum_j k_{2,ij}\ n_j 
       \; + \;      \sum_j \sum_l k_{2,jl}\ n_j n_l
       \; - \; n_i  \sum_j \sum_l k_{3,ijl}\ n_j\ n_l
       \; + \;      \sum_j \sum_l \sum_m k_{3,jlm}\ n_j\ n_l\ n_m
\end{equation}
with the index~$i$ for the included species. 
So far, the implementation is restricted to two- and three-body 
reactions, which is a reasonable assumption for comparatively hot stellar 
atmospheres.
The losses (negative sign) and gains (positive sign) due to two-body reactions
are described by the first and second right-hand terms 
with the corresponding reaction rates $k_{2,ij}$ and $k_{2,jl}$, respectively. 
Analogously, the third and fourth term describe the change due to 
three-body reactions. 
Such an equation is imposed for each included chemical species, resulting 
in a system of ordinary differential equations of first order.  
In \COBOLD, the chemical reactions are handled locally for each grid 
cell separately by solving the  system of differential equations.
It starts with the calculation of the reaction rates $k$, which are functions of 
the local gas temperature and (for catalytic reactions) also of the number density 
of a representative metal.  
The influence of the radiation field has been neglected so far. 
The functions are parametrized with prescribed coefficients that are provided in the 
form of a table \citep[see][for details]{Wedemeyer2005A&A...438.1043W}.

The chemical-reaction rates,
the number densities of the involved species,
and thus their derivatives can differ by many orders of magnitude, 
which can cause the  system of equations to be stiff.
Thus, an implicit scheme is used for the numerical solution. 
We based our solver on the DVODE package \citep{brown89} 
with an implicit BDF (backward differentiation formula) method and 
an automatic internal time step. 
The solution finally provides the number densities of the involved species 
after the overall (global) computational time step.
For the numerical simulation of carbon monoxide, 7 chemical species and a 
representative metal are considered, which are connected through 27 chemical 
reactions \cite{Wedemeyer2005A&A...438.1043W}.

Carbon monoxide is a non-negligible opacity source in the solar atmosphere,
so that the opacity is in principle affected by the deviations from equilibrium 
of the CO number density. 
To account for this effect, the back-coupling \textcolor{tcol}{to} the radiative transfer was 
implemented \cite{2007A&A...462L..31W}. 
It follows the approach by \citet{steffen88}, which uses two frequency groups. 
The first comprises the gray Rosseland opacity $\kappa_\mathrm{R}$ 
without the wavelength region around the CO fundamental vibration-rotation 
band in the infrared at a wavelengths around $\sim 4.6~\mu\mathrm{m}$. 
This wavelength range is simulated with the second band, which is constructed 
from the gray Rosseland opacity $\kappa_\mathrm{R}$ and an 
additional opacity $\kappa_\mathrm{CO}$. 
The latter is directly connected to the CO number density that is derived 
from the preceding solution of the chemical-reaction network.


\subsubsection{Time-dependent hydrogen ionization} 
\label{sec:hionmodule}

A detailed treatment of the time-dependent ionization of hydrogen is 
important for atmospheric layers, where significant deviations from the 
ionization equilibrium occur, e.g., in the solar chromosphere
(see Sect.\,\ref{sec:swb_solarchrom}). 
Current applications for the Sun
use a hydrogen model atom with 5 bound energy levels and a level representing ionized hydrogen.
The number densities of the individual level populations $n_i$ enter as 
additional quantities in \COBOLD. 
The levels are connected by collisional transitions and 10~radiative transitions
(5 bound-bound and 5 bound-free).  
The rate $P_{\ij}$ of a transition between a level~$i$ and a level~$j$ is given 
by $P_{\ij} = C_{\ij} + R_{\ij}$, where $C_{\ij}$ and $R_{\ij}$ are the rates due 
to collisional and radiative transitions, respectively. 
First, these rates are calculated from the local gas density and gas 
temperature, the imposed radiation field, and the level populations
and electron densities, that are available from the previous time step. 
Apart from advection, the change of the population number densities and thus the 
ionization degree of hydrogen in a grid cell is then described by a set of time-dependent 
rate equations of the form 
\begin{equation}
\left(S_i\right)_\mathrm{hion} = 
      \sum_{j \ne i}^{n_l} n_j P_{ji} - n_i \sum_{j \ne i}^{n_l} P_{ij}
  \enspace ,
 \label{eq:RateEq} 
\end{equation}
where the terms on the right-hand side are the rates into and out of level~$i$. 
In \COBOLD, the set of rate equations for all considered energy 
levels is solved with the DVODE package like that for the chemical-reaction 
networks (see Sect.\,\ref{sec:chemnet}).

An important simplification concerns the usage of fixed radiative rates. 
In principle, the radiation field and the radiative transitions for hydrogen 
(both bound-bound and bound-free) are connected in such a way that a 
detailed solution has to be found by iteration, which makes it   
computationally expensive. 
For the implementation in a multi-dimensional radiation MHD code, the rates 
are fixed and calibrated so that they reproduce the full solution as it is 
implemented in time-dependent 1D simulations \citep{sollum1999, 2002ApJ...572..626C}. 
There is no back-coupling to the equation of state and the opacities in the 
current implementation in \COBOLD, a point, which will be worked on 
in the future. 
More details are given in \citet{2006A&A...460..301L}.

\subsubsection{Dust}
\label{sec:Dust}

If the atmospheric temperatures are low enough,
not only molecules but larger particles -- dust -- can form.
In the Earth's atmosphere, possible types \textcolor{tcol}{of such particles} 
are for instance aerosols of different compositions,
rain, snow, or hail, mostly made of water,
or particles from volcanic ashes.
Many cool stars or substellar objects may contain dust in their atmospheres, too
(see Fig.\,\ref{f:Teffloggcontrast}).
In the hotter objects, the dust will mostly be made of minerals (e.g., forsterite).
At lower temperatures (e.g., on the Earth) water plays an important role.
For the dust chemistry \textcolor{tcol}{of the warmer objects},
it is crucial whether oxygen or carbon is more abundant,
because these two elements first form carbon monoxide gas (CO) and only the
remainder participates in dust formation.
The formation, interaction, and destruction of grains of
different chemical composition, size, and shape
is difficult to model \cite{Helling2001A&A...376..194H, Woitke2003A&A...399..297W}.
Compared to the possible complexity of the processes
occurring in real objects, the various dust modules in \COBOLD\ are simple.
They are permanently under development.
Two examples will be outlined in the following:

Stars at the tip of the asymptotic giant branch lose part of their mass
in form of a stellar wind, likely driven by radiation pressure on dust.
The formation of carbon-rich dust around such stars
was investigated by \citet{Freytag2008A&A...483..571F}
with \COBOLD\ and with the 1D-RHD code of \citet{Hoefner2003A&A...399..589H}.
%
The \COBOLD\ dust model includes a time-dependent description of
dust grain growth and evaporation using a method developed by
\citet{Gail1988A&A...206..153G} and \citet{Gauger1990A&A...235..345G}.
In this approach, the dust component is described in terms of four moments $K_j$
of the grain-size distribution function, weighted with a power $j$ of the grain radius.
The moment $K_0$ represents the total number density of grains
(integral of the size distribution function over all grain sizes),
while the ratio $K_3 / K_0$ is proportional to the average
volume of the grains.
The equations, which determine the evolution of the dust components, are solved
considering spherical grains consisting of amorphous carbon.
The nucleation, growth, and evaporation of grains is assumed to
proceed by reactions involving C, C$_2$, C$_2$H, and C$_2$H$_2$.
The four moments $K_j$ are number densities and are advected with the gas
as described in Sect.\,\ref{sec:OptionalModules}.
The gas and dust opacities in this case are gray.
Some results are shown in Sect.\,\ref{sec:SupergiantsAGBstars}
and Fig.\,\ref{f:SurfaceImagesAlongMS}.

In contrast to the cool giants,
the conditions for the formation of (oxygen-rich) dust in 
M \textcolor{tcol}{dwarfs} and brown dwarfs 
are fulfilled even in standard 1D atmosphere models.
However, the comparatively heavy dust grains should sink under the
influence of gravity and vanish from the visible photosphere,
leaving no direct trace in emergent spectra,
which is \textcolor{tcol}{at variance with} observations.
The scheme used in \cite{Freytag2010A&A...513A..19F} to investigate the question
why the dust does not settle or how the material comes back up
is based on a simplified version of the dust model used in
\cite{Hoefner2003A&A...399..589H}, adapted to forsterite (Mg$_2$SiO$_4$, 3.3\,g/cm$^3$).
In this method, there are only two density fields, one is used to
specify the mass density of dust particles and the other to describe
the monomers (the dust constituents), instead of four density fields
for the dust and none for the monomers as in
\cite{Hoefner2003A&A...399..589H} and
\cite{Freytag2008A&A...483..571F}.
Therefore, the ratio of the sum of dust and
monomer densities to the gas density is allowed to change,
in contrast to the dust description in \cite{Hoefner2003A&A...399..589H}
used for the AGB simulations mentioned above.
Instead of modeling the nucleation and the detailed evolution of the
number of grains, a constant ratio of the number of seeds (dust nuclei)
to the total number of monomers (in grains or free) per cell \textcolor{tcol}{is assumed}.
If all the material in a grid cell were to be condensed into dust, the
grains would have the maximum radius $r_\mathrm{d,max}$, which \textcolor{tcol}{is}
set to a typical value of 1\,$\mu$m. This is close to the
typical particle sizes found for the optically thick part of the
cloud deck in solar-metallicity brown dwarfs according to the
\texttt{DRIFT-PHOENIX} models of \citet{Witte2009A&A...506.1367W}.


Condensation and
evaporation are modeled as in \cite{Hoefner2003A&A...399..589H},
\textcolor{tcol}{with} parameters and saturation vapor curve adapted to forsterite.
In the hydro\-dynamics module, monomers and dust densities are advected
with the gas density,
with the terminal velocities given by the low-Reynolds-number case of Eq.\,(19) in
\citet{Rossow1978Icar...36....1R} as settling speed
added to the vertical advection velocity of dust grains.
In contrast to the sophisticated treatment of the gas opacities,
a simple formula for the dust opacities is used,
which assumes that the large-particle limit is valid for all grain sizes
and treats scattering as true absorption.
The dust opacity in each cell of the simulated atmosphere is added to the gas opacity
(Sect.\,\ref{sec:OpacityBinning}).
Experiments have been made with another dust model,
that uses in addition to one density field for the monomers
a number of further fields, one for each possible grain size.

\section{Results}

\subsection{Code comparison: the solar benchmark}
\label{sec:codecomp}

\begin{table}[tb]
\caption{Setup and emergent radiation of solar models computed with 
different codes. \label{T1}}
\begin{center}
  \scriptsize
 \begin{tabular}{l|rrrr}
  \hline\noalign{\smallskip}
            & STAGGER & MURaM & \COBOLD  & \COBOLD   \\
            &         &       & standard & high resolution \\
  \noalign{\smallskip}\hline\noalign{\smallskip}
Box size [Mm$^3$] ($ x\times y\times z$)
                         & $6.0$$\times$$6.0$$\times$$3.6$ & 
                           $9.0$$\times$$9.0$$\times$$3.0$ & 
                           $5.6$$\times$$5.6$$\times$$2.3$ &
                           $5.6$$\times$$5.6$$\times$$2.3$ \\
Grid dimension           & $240$$\times$$240$$\times$$230$ & 
                           $512$$\times$$512$$\times$$300$ & 
                           $140$$\times$$140$$\times$$150$ &
                           $400$$\times$$400$$\times$$300$ \\
Cell size [km$^3$] ($\Delta x\times\Delta y\times\Delta z$)
                         & $25.1$$\times$$25.1$$\times$$\Delta z^\ast$ & 
                           $17.6$$\times$$17.6$$\times$$10.0$ & 
                           $40.0$$\times$$40.0$$\times$$15.1$ &
                           $14.0$$\times$$14.0$$\times$$ 7.5$ \\
Height range [Mm] ($z=0$ at $\langle\tau\rangle=1$)
                         & $-2.72 \ldots +0.88$ & 
                           $-2.00 \ldots +1.00$ &
                           $-1.38 \ldots +0.88$ &
                           $-1.38 \ldots +0.88$ \\
upper boundary condition & transmitting   & 
                           closed         &
                           transmitting   &
                           transmitting   \\
lower boundary condition & open   & 
                           open   &
                           open   &
                           open   \\
\# snapshots used        & $ 19$ & 
                           $ 19$ &
                           $ 19$ &
                           $ 60$ \\
\# frequency groups used & $ 12$ & 
                           $  4$ &
                           $ 12$ &
                           $ 12$ \\
  \noalign{\smallskip}\hline\noalign{\smallskip}
Effective temperature [K]
                         & $5762$ & 
                           $5768$ &
                           $5781$ &
                           $5763$ \\
Bol.\ intensity contrast ($\mu$=1) [\%] 
                         & $14.9$ & 
                           $15.4$ &
                           $14.4$ &
                           $14.1$ \\
  \noalign{\smallskip}\hline
 \end{tabular}\\[2mm]
\end{center}
\begin{flushleft}
$^\ast$ The STAGGER code uses a non-equidistant grid in the vertical 
direction, with spacings ranging from $\Delta z=7$~km near the optical 
surface to $\Delta z=32$~km in the deepest layers.
\end{flushleft}
\end{table}

The natural benchmark for the comparison of different codes is of
course the solar atmosphere. On the one hand, its mean thermal stratification 
is well known empirically, and its velocity field and associated temperature 
fluctuations have been studied in great detail based on a large body of 
observations. On the other hand, many numerical simulations have
been carried out to study solar surface convection with a variety of different
computer codes. Here, we compare some basic quantities obtained from numerical
simulations of the solar surface layers with three different codes: STAGGER, 
MURaM, and \COBOLD. All three codes solve the time-dependent equations of
compressible (magneto)hydro\-dynamics for a gravitationally stratified,
radiating fluid in a Cartesian box in 3 spatial dimensions, taking into 
account partial ionization and non-gray radiative energy exchange, the latter
treated with the opacity-binning scheme (see Sect.\,\ref{sec:OpacityBinning}).

The codes have been developed independently and use different
numerical methods. \textcolor{tcol}{STAGGER and MURaM are similar in that both use a method 
of lines for the hydrodynamics part as well as artificial (hyper)diffusivities
to stabilize the numerical solution.}
The STAGGER code \cite{Nordlund+Galsgaard1995}, 
\cite{Stein2006ApJ...642.1246S} (see also 
Sect.\,\ref{sec:introduction}) uses a sixth-order finite-difference
method to determine the spatial derivatives on a staggered mesh,
while the equations are stepped forward in time using an explicit 
third-order predictor-corrector procedure, conserving mass, momentum, 
energy, and magnetic-field divergence. Radiative energy exchange is 
found by the formal solution of the Feautrier equations on long rays.
\textcolor{tcol}{Similarly,} MURaM \cite{Voegler2003}, \cite{Voegler2005A&A...429..335V}
uses a \textcolor{tcol}{fourth-order} central-difference scheme in space, and a
fourth-order Runge-Kutta time stepping; radiation transport is computed
with a short-characteristics method. \textcolor{tcol}{On the other hand,} \COBOLD\ is based 
on a finite-volume approach and employs an approximate Riemann solver of
Roe type \textcolor{tcol}{to advance the hydrodynamics in time, relying on second-order
monotonic reconstruction schemes to achieve numerical stability without 
the need to invoke artificial viscosities. Directional splitting reduces 
the 3D problem to one dimensional sub-steps. Similar to STAGGER,} radiative 
transfer is treated with a Feautrier method on long characteristics \textcolor{tcol}{(see
Sect.\,\ref{sec:LongCharRad}).}

\begin{figure}[tb]
\centering
\mbox{\includegraphics[width=\textwidth]{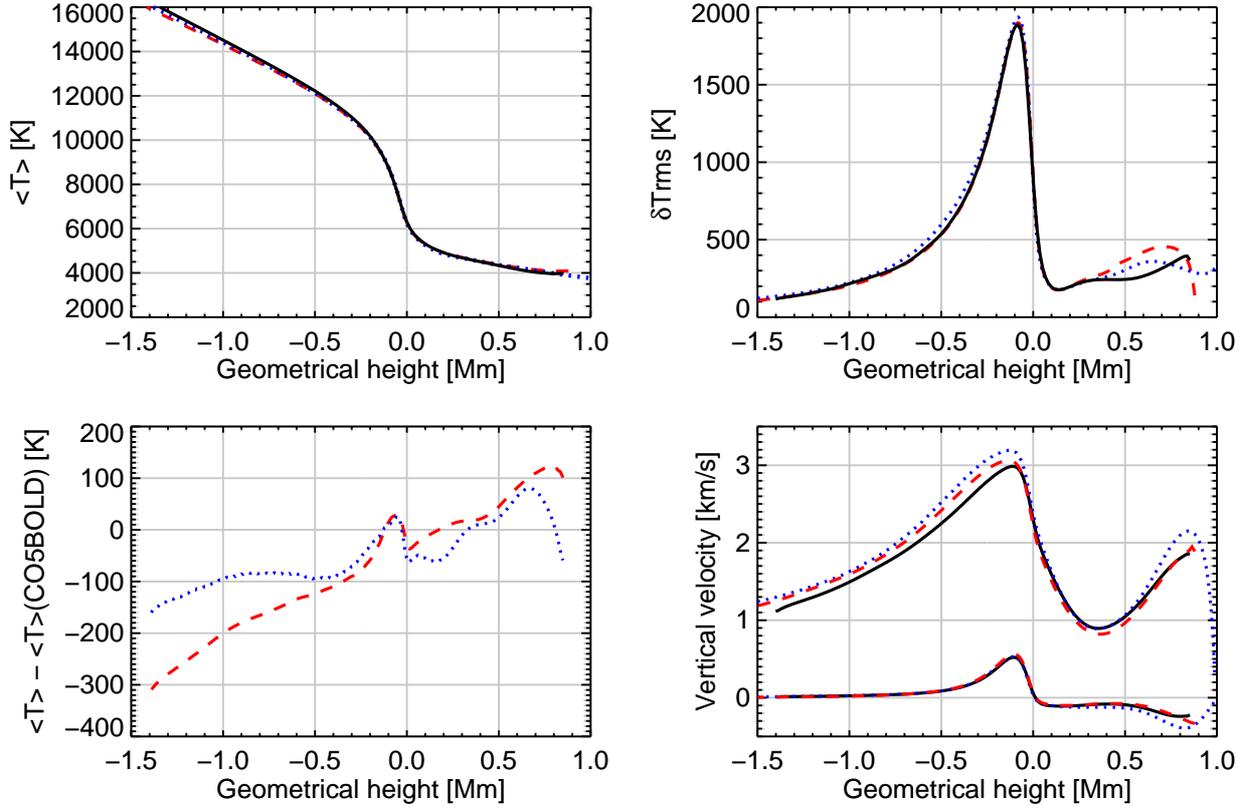}}
\caption{Comparison of the average vertical temperature structures, 
$\langle T\rangle(z)$ (left panels), the rms horizontal temperature 
fluctuations 
$\delta T_{\rm rms}=\sqrt{\langle T^2\rangle(z)-\langle T\rangle^2(z})$ 
(top right), and the mean and rms vertical velocity  $\langle V_z\rangle(z)$, 
and $\sqrt{\langle V_z^2\rangle(z)}$, respectively (lower and upper set of 
curves in bottom right panel, respectively), as obtained with different codes 
for the solar simulations described in Table\,\ref{T1}: STAGGER (dashed), 
MURaM (dotted), \COBOLD\ standard (solid). Here $\langle .\rangle$ denotes 
averaging over horizontal planes of the numerical grid (constant geometric 
height $z$) and over selected snapshots in time.
}
\label{f:cmp_SMC}
\end{figure}

The basic setup of the different solar simulation runs is summarized in 
Table\,\ref{T1}. The two \COBOLD\ models differ only in their spatial 
resolution.
Since the models of the different groups have not been constructed
for the purpose of this comparison, they differ in many aspects, such as
horizontal box size, vertical extent, spatial resolution, boundary conditions, 
opacity tables, number of \textcolor{tcol}{frequency groups,} and equation of state (EOS), apart 
from the different numerical methods used to solve the equations of 
hydro\-dynamics and radiative transfer. Despite these substantial differences, 
the mean vertical structure, obtained from the various simulations by
horizontal and temporal averaging, turns out to be remarkably similar, 
as demonstrated in Fig,\,\ref{f:cmp_SMC}.
Obviously, the mean thermal structure is the most robust quantity.
Ignoring the layers influenced by the top boundary, the temperature
differences are everywhere below 2\%; deviations seen in the deeper
layers are probably related to differences in the EOS. Except for the
photospheric layers above $\approx 300$~km, where the details of the
opacity-binning recipe play a major role, the predicted amplitude of
the horizontal temperature fluctuations is also amazingly similar.  As
a consequence, the predicted continuum-intensity contrast (see
Sect.\,\ref{sec:Contrast}) is found to be in very good agreement (last
row of Table\,\ref{T1}).

The depth-dependence of the mean vertical velocity obtained from the three 
different simulations agrees closely (lower set of curves in lower right 
panel of Fig,\,\ref{f:cmp_SMC}). As theoretically expected, $\langle
V_z\rangle$ is positive in the convectively unstable layers below the
surface, and negative in the overshoot region. Somewhat larger
deviations among the different models are found in the velocity
dispersion, $\sqrt{\langle V_z^2\rangle(z)}$ (upper curves). 
Is seems that both the location of the lower boundary and the spatial
resolution have some influence on the resulting velocity amplitude.
Nevertheless, the overall agreement is very satisfactory.

We have to keep in mind that the different codes are largely based on the
same physical assumptions and approximations. It may therefore not be too 
surprising that the resulting atmospheric structures are similar.
And it does not prove that all details of the models are physically
correct.


\begin{figure}[tb]
\centering
\mbox{\includegraphics[width=\textwidth]{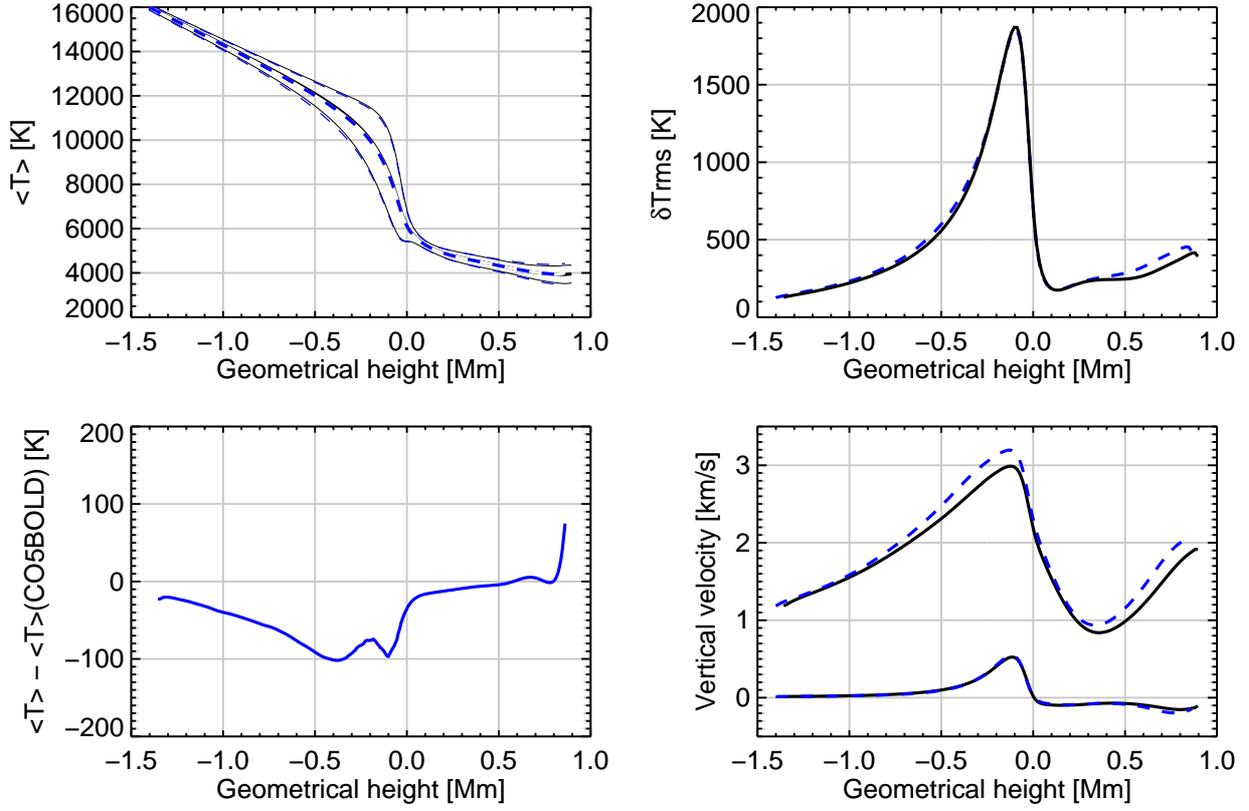}}
\caption{Comparison of the same quantities as in Fig.\,\ref{f:cmp_SMC},
but for two \COBOLD\ solar simulations which differ only in spatial 
resolution:
\COBOLD\ \emph{standard} (solid) and \COBOLD\ \emph{high resolution} (dashed).
The thin curves in the upper left panel refer to $\langle T\rangle(z)\,\pm\,
\delta T_{\rm rms}(z)$.
}
\label{f:cmp_x16}
\end{figure}

The role of the spatial resolution is illustrated in
Fig.\,\ref{f:cmp_x16}, where we compare two \COBOLD\ models that differ
only in spatial resolution: in the high-resolution model 
\textcolor{tcol}{(cf.\ Fig.\,\ref{f:granulation}),} the
horizontal cell size is reduced by a factor $2\,\sqrt{2}$ with respect
to the standard \COBOLD\ model, while the vertical cell size is reduced
by a factor of $2$. The mean thermal structure is practically
unchanged, as is the amplitude of the T-fluctuations up to the mid
photosphere. As a consequence, the intensity contrast is not significantly
affected by the increased grid resolution (see Table\,\ref{T1}). 
However, in the upper photosphere 
above $z \approx 300$~km, the amplitude of both the temperature and velocity
fluctuations increases somewhat with increasing spatial resolution. 
The question whether the moderate spatial resolution of the standard
hydro\-dynamical models is fully sufficient to account for the ``turbulent''
character of the solar photosphere, and hence for correctly capturing the 
non-thermal Doppler broadening of spectral lines, is currently under 
investigation.  Obviously, this is an important issue in the context of 
accurate chemical abundance determinations based on 3D model atmospheres 
(cf.\ Sect.\,\ref{sec:Abundances}).

\subsection{Granular intensity contrast}
\label{sec:Contrast}

\begin{figure}[htbp]
\centering
\includegraphics[width=10cm]{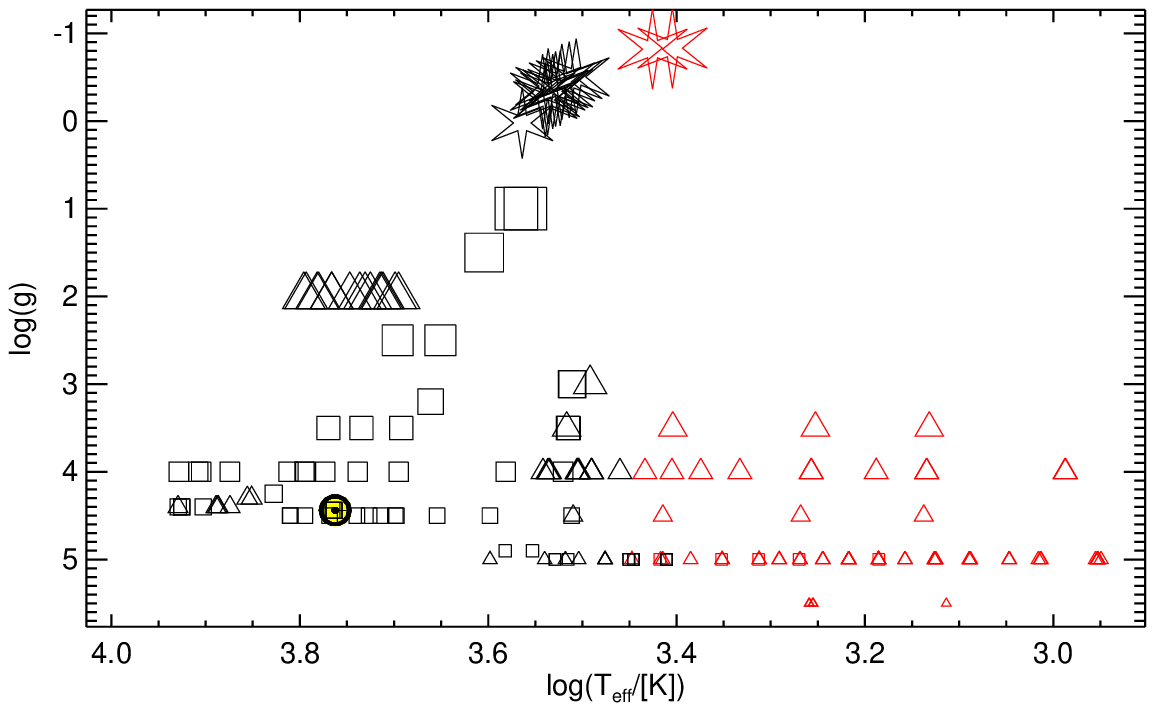}
\includegraphics[width=10cm]{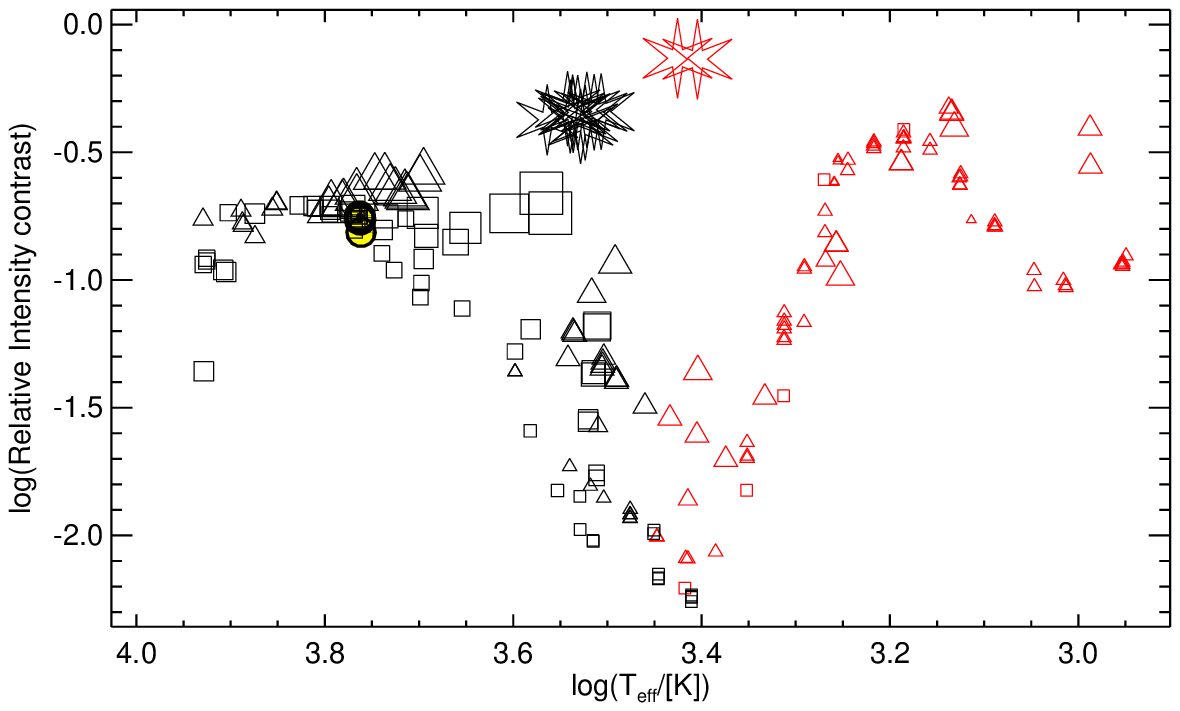}
\caption{Top: grid of \COBOLD\ models in the $\log T_\mathrm{eff}$-$\log g$ diagram.
Bottom: logarithm of rms bolometric intensity contrast versus $\log T_\mathrm{eff}$.
The squares mark 3D local models, the triangles 2D local models.
The stars indicate global models (with very low surface gravity).
Larger symbols indicate lower gravity.
The Sun has its own \textcolor{tcol}{standard symbol $\odot$
with yellow background.}
A red symbol (at lower effective temperatures) shows that some treatment of dust
is included in the simulation.
}
\label{f:Teffloggcontrast}
\end{figure}

%
The granulation pattern visible at the solar surface is a manifestation of
convection in the sub-photospheric layers: bright granules correspond to
hot rising gas, while the dark intergranular lanes consist of cooler 
downward-sinking material.
The relative continuum-intensity contrast,
\begin{equation}\label{e:Idrms}
\delta I_{\rm rms} = {\left\langle\frac{\sqrt{\langle I(x,y,t)^2\rangle_{x,y} - 
                           \langle I(x,y,t)\rangle^2_{x,y}}}
                     {\langle I(x,y,t)\rangle_{x,y}}\,\right\rangle_t}{}
  \enspace ,
\end{equation}
of this granulation pattern and its
variation from the centre of the solar disk to its limb are important
tests for the degree of realism of numerical models.
For many years, the values derived from observations were
significantly lower than those calculated on the basis of numerical
simulations \cite[e.g.,][]{2008PhST..133a4016K}.
Recently, it was shown that synthetic continuum-intensity maps based
on \COBOLD\ \cite{2007IAUS..239...36S} and also the code by
Nordlund and Stein \cite{Stein2006ApJ...642.1246S} can indeed reproduce the
empirical values quite well, if the instrumental image degradation is
taken into account properly \cite{2009A&A...503..225W}.
The necessary image reconstruction is a very demanding task,
which on the other hand turns out to be crucial,
as was shown for observations obtained
with the Solar Optical Telescope (SOT) onboard the Hinode satellite
\cite{2008SoPh..249..167T}.
\citet{Hirzberger2010ApJ...723L.154H}
find very good agreement between the rms contrast of solar
granulation obtained from measurements with a balloon-borne 1-m solar
telescope and simulations at wavelengths of 388\,nm and 312\,nm. At
shorter wavelengths, discrepancies between observations and
simulations seem to persist.

By analogy, the surface of cool stars must be covered by a similar
pattern, the \emph{stellar granulation}. Its intensity contrast
cannot be measured directly. Numerical simulations
are necessary to infer how the intensity contrast depends on the
stellar parameters: the effective temperature $T_\mathrm{eff}$, the 
surface gravity $\log g$, \textcolor{tcol}{and the chemical composition.}

The top panel in Fig.\,\ref{f:Teffloggcontrast} shows the \COBOLD\ models on the main sequence
and above it
(i.e., with gravities around $\log g$ 4 to 5 and lower: white dwarf models are not plotted)
in a $\log T_\mathrm{eff}-\log g$ diagram.
The displayed models comprise the solar-metallicity part of
the CIFIST grid of solar-like 3D models \cite{Ludwig2009MmSAI..80..711L},
3D M-dwarf models \cite{Wende2009A&A...508.1429W},
local 2D ``dusty'' brown dwarf models \cite{Freytag2010A&A...513A..19F},
global 3D red supergiant \cite{Freytag2002AN....323..213F,Chiavassa2009A&A...506.1351C,Chiavassa2010A&A...515A..12C}
and AGB star models \cite{Freytag2008A&A...483..571F},
as well as more experimental models of e.g., A-type stars (in 2D or 3D) and cepheids (in 2D).
Larger symbols mean lower gravity (and usually a larger stellar radius).
Squares depict 3D models, triangles 2D models.
Solar models have the $\odot$ symbol.
Global 3D models of red supergiants and AGB stars are
marked as \textcolor{tcol}{star symbols} at the top.
Red symbols indicate that the simulations have accounted for dust in some form.
Models with non-solar metallicities are not shown.

The bottom panel shows the (bolometric) relative intensity contrast
according to Eq.~(\ref{e:Idrms}) versus 
$T_\mathrm{eff}$ for the same models and with the same symbols as in the top panel.
On the main sequence (smallest symbols),
the contrast decreases for stars cooler than the Sun
since the stellar energy flux decreases and convection can transport it with
smaller temperature fluctuations.
The contrast does not increase \textcolor{tcol}{further} but has a plateau
for stars a bit hotter than the Sun
because convection does not carry the entire stellar flux anymore.
Below a minimum at around 2600\,K, the contrast increases again because fluctuating
dust clouds start dominating the surface contrast (see Sect.\,\ref{sec:MSstars}).
The contrast decreases at the very cool end due to the decreasing overall flux.

In general, lowering the gravity has a similar effect as increasing the effective temperature,
but result\textcolor{tcol}{s} in slightly more vigorous convective flows.
The largest surface contrast is seen in the global AGB-star models,
followed by the global red-supergiant models.
2D models have a larger contrast than 3D models.
Other types of dust (and/or dust schemes) as well as global fluctuations might change
the picture for the cooler models.

\subsection{Solar and stellar abundances}
\label{sec:Abundances}

One important application of \textcolor{tcol}{multi-dimensional (multi-D)}
radiation-(mag\-neto)hydro\-dynamics models
is the determination of chemical abundances in late-type stars.  Under most
circumstances, information about the thermal and kinematic structure of a
stellar atmosphere is necessary to interpret measured strengths of spectral
lines in terms of chemical abundances. To this end, simulated time series of
the evolution of the stellar photospheric flow field are serving as input for
detailed spectral-synthesis calculations. The result of these calculations are
time series of spatially resolved synthetic spectra which, after suitable
averaging in space and time, can be compared to the observations. For
\COBOLD, we developed the spectral-synthesis code
Linfor3D~\citep{Steffen2010Linfor3D-Manual} which is on the one side
adapted to the particular data formats and structures of \COBOLD, and on the
other side designed to facilitate the abundance analysis.

Historically, the application of multi-D models for deriving abundances
started out on the Sun already early on. However, in the beginning mainly
structural properties of surface convection and associated magnetic fields
were in the modeling focus so that abundance analyses with multi-D models
remained sparse.  A turning point came with the work of Asplund and
collaborators
\citep{Asplund1999A&A...346L..17A,Asplund2000A&A...359..743A}
suggesting that multi-D effects are important in the Sun and metal-poor stars
if one wishes to obtain high-fidelity abundances.  Since then efforts are
directed towards improving multi-D modeling aspects specific to abundance
analysis work, and extending the model basis covering successively larger
regions of the Hertzsprung-Russell diagram \citep{Ludwig2009MmSAI..80..711L}.

Hitherto, \COBOLD\ models were applied to derive abundances of twelve elements
in the Sun (see \citet{Caffau2010SoPh..tmp...66C} and references therein)
including the CNO elements, which are important for the overall solar
metallicity; work on further elements is in progress. In the field of
metal-poor stars, \COBOLD\ models were used to derive abundances from atomic
(e.g., \citep{Caffau2010AN....331..725C,Bonifacio2010A&A...524A..96B})
as well as molecular lines
(e.g., \citep{Hernandez2010A&A...519A..46G,Behara2010A&A...513A..72B}). An
element of particular interest in metal-poor stars is lithium due to its
connection to nucleosynthetic processes in the \textcolor{tcol}{big bang and early
universe.} The lithium abundance is commonly derived from the Li\,I resonance
line at 6707\,\AA. Since lithium is mostly ionized in the stars of interest,
the formation of the line is highly temperature sensitive, which makes the
resulting abundances strongly model-dependent.
Hoping for lithium abundances of higher fidelity,
multi-D models were rather extensively applied.
\COBOLD\ models were used to obtain lithium abundances in the most metal-poor dwarf
stars known \citep{Hernandez2008A&A...480..233G}, to investigate the structure
of the so-called ``Spite plateau'' at lowest
metallicities \citep{Sbordone2010A&A...522A..26S}, and to study the evolution
of lithium in the globular cluster
NGC\,6397 \citep{Hernandez2009A&A...505L..13G}. A related aspect is the
abundance ratio between the lithium isotopes \mbox{$^6$Li/$^7$Li} in
metal-poor stars. \COBOLD\ models were applied to argue against claims of a
non-zero isotopic ratio \citep{Cayrel2007A&A...473L..37C,Steffen2010IAUS..268..215S}.
The aforementioned investigations focused on dwarf or subgiant stars.
There are also ongoing efforts to extend the application of \COBOLD\ simulations to giant
stars including studies of their abundances
\citep{Kucinskas2010IAUS..265..209K,Dobrovolskas2010nuco.confE.288D,
Ivanauskas2010nuco.confE.290I}.

Besides conducting actual abundance analyses, \COBOLD\ models were instrumental
in a number of studies more indirectly linked to the derivation of stellar
abundances from spectroscopy: the long-lasting issue of how small-scale velocity
fields in stellar atmospheres give rise to the spectroscopically derived
microturbulence \citep{Steffen2009MmSAI..80..731S,Wende2009A&A...508.1429W},
and the influence of thermal inhomogeneities on effective temperatures derived
from Balmer-line profiles
\citep{Behara2009A&A...502L...1L}.   

The application of multi-D models to stellar-abundance studies is in its early
stages and modeling challenges still exist: the need for a precise thermal
structure of the optically thin, line-forming regions demands for a
detailed representation of the radiation field.
Sufficient wavelength resolution (Sect.~\ref{sec:OpacityBinning}),
and inclusion of scattering processes are current challenges in the
simulations proper. In the post-processing, line-formation calculations
including departures from LTE are necessary to fully exploit the model
potentialities but are demanding in terms of computational resources and
amount of necessary atomic input data.

\subsection{The magnetic Sun}
\label{sec:MagneticSun}

\subsubsection{Current status} 

\begin{figure}[tb]
\centering
\includegraphics[width=\textwidth]{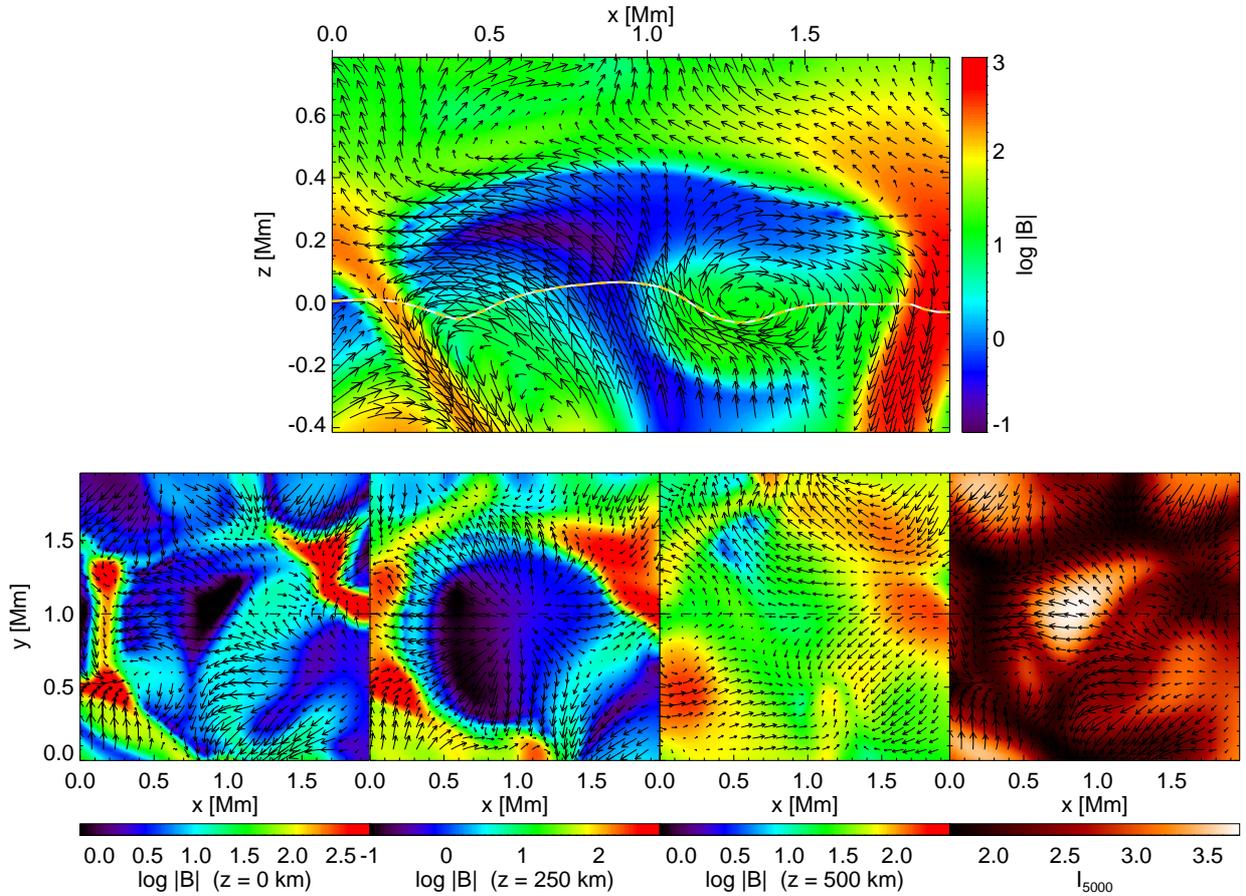}
\caption{Flux expulsion in a close-up from a simulation of solar magneto\-convection: 
  Logarithmic magnetic field strength in a vertical cross section (top) and in three 
  horizontal cross sections (bottom) at heights of 0\,km, 250\,km, and 500\,km. The 
  emergent intensity is displayed in the rightmost panel. The arrows represent the 
  velocity field projected in the respective coordinate planes. 
  The \textcolor{tcol}{yellow/white} curve in 
  the top panel marks the height of visible optical depth unity, i.e., the ``solar
  surface''. From \citet{2009SSRv..144..317W}.  
}
\label{f:mhd_fluxexpulsion}
\end{figure}

Figure\,\ref{f:mhd_fluxexpulsion} 
exemplifies the type of MHD applications that are currently 
performed with \COBOLD. It illustrates the magnetic field structure at the 
interface between the convection zone and the overlying atmosphere of the Sun. The 
top panel shows \textcolor{tcol}{a close-up of} a vertical cross section through 
the three-dimensional computational 
domain, where colors represent the magnetic field strength and arrows the velocity 
field. The dashed yellow/white curve indicates optical depth unity, i.e., the 
``solar surface'' as seen in the visible part of the spectrum. Below this surface, 
the atmosphere is convectively unstable and energy is transported mainly by convection. 
Above this surface adjoins the stably stratified photosphere, where energy is mainly 
transported by radiation and where convective overshoot motions are rapidly damped. The panels 
in the bottom row show horizontal cross sections \textcolor{tcol}{of corresponding size}
at three selected height levels and 
the emerging bolometric intensity in the rightmost panel (intensity map). The location 
of the vertical cross section is indicated by the dashed horizontal line in the bottom 
panels.

We can see a strong central updraft in the vertical cross section of 
Fig.\,\ref{f:mhd_fluxexpulsion}, which corresponds to the central granule visible 
in the intensity map. This granule is a typical representative for real solar
granulation with respect to intensity contrast and size. Since the diffusion 
length scale of the magnetic field is small compared to the size of a granule, 
it is useful to think of the magnetic field to be ``frozen into the plasma'' 
so that the flow field transports the magnetic field from the granule center 
to its boundaries where it gets concentrated. This process is called the
flux expulsion process, as magnetic flux is expelled from the granule interior to 
its boundaries. Correspondingly, the magnetic field in the central part of the 
granule is weak (dark blue) and it gets concentrated in the intergranular 
lanes (red), where plasma flows back into the convection zone again.

As the updraft runs into the stable stratification of the photospheric layer,
it loses the driving buoyancy force and buoyancy starts to act in the opposite 
direction.
Consequently,
\textcolor{tcol}{and also because of the strong density stratification,} the flow
must deviate in \textcolor{tcol}{the} horizontal direction and it drags the magnetic 
field with it as a consequence of the frozen-in condition.
Hence, the magnetic field assumes a predominantly horizontal direction in the upper part 
of the photosphere, above the mushroom shaped void in Fig.\,\ref{f:mhd_fluxexpulsion}
(top panel).
While MHD simulation have since long predicted the existence of this 
prevalently horizontal field \citep{grossmann-doert+al1998}, it is only very recently 
that it got observationally detected with polarimetric measurements from the Hinode
space observatory \citep{lites+al2008}. More details about MHD simulations with
regard to this horizontal field can be found in \citet{schuessler+voegler2008},
\citet{steiner+al2008}, and \citet{steiner+al2009}.

\begin{figure}[tb]
\centering
\includegraphics[width=\textwidth]{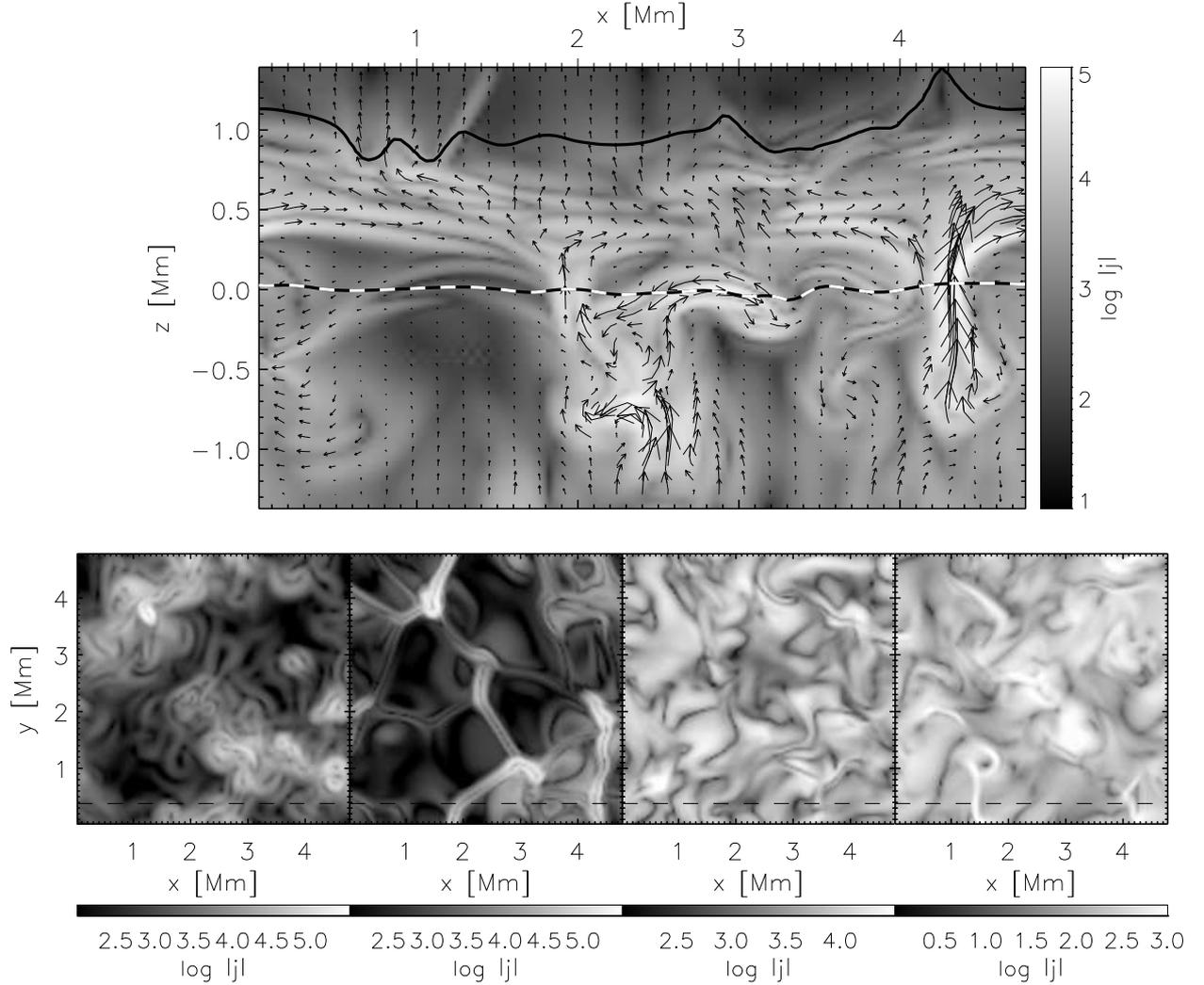}
\caption{Logarithmic current density, $\log |j|$, in a vertical cross section 
  (top panel) and in four horizontal cross sections (bottom panels) in a depth 
  of 1180\,km below, and at heights of 90\,km, 610\,km, and 1310\,km above the 
  mean surface of optical depth unity from left to right, respectively.
  The arrows in the top panel indicate the magnetic field strength and
  direction. The dashed line in the bottom row indicates the position of the 
  vertical section.
  $j$ is given in units of $3\,\times\,10^5$ A/m$^{2}$.
  From \citet{Schaffenberger2006ASPC..354..345S}.
}
\label{f:mhd_currentsheet}
\end{figure}

Figure\,\ref{f:mhd_currentsheet} shows the electric current density, 
\textcolor{tcol}{$\mathbf{j} = \mathbf{\nabla}\times\mathbf{B}$}, that forms as a 
consequence of the process discussed above, in a similar but larger domain than that
of Fig.\,\ref{f:mhd_fluxexpulsion}. Again, the top panel is a vertical cross section
through the simulation domain, and the bottom row shows four horizontal cross sections 
at various heights. The magnetic flux concentrations in the intergranular lanes, which
form as a consequence of the flux expulsion process, give rise to the conspicuous
pairs of current sheets, visible in the horizontal cross section at $z = 90$~km.
Another system of current sheets forms in the region of predominantly horizontal
magnetic fields above granules, in a height range from approximately 400--900~km,
as can be seen from the vertical cross section. Higher up in the atmosphere, shock
waves form where the supersonic plasma flow sweeps magnetic field into the 
compression zone downstream of and along the shock front. There again, current
sheets form, e.g., the one extending from $x=1.1$ to $x=1.5$~Mm in the top part of the
vertical cross section, which is also visible as a thin filament at $x=1.4$~Mm in 
the rightmost panel which corresponds to $z=1310$~km. More details about
MHD simulations with \COBOLD\ with regard to chromospheric shock waves can
be found in  \citet{2005ESASP.596E..65S} and \citet{Schaffenberger2006ASPC..354..345S}

Interestingly, the mean vertical Poynting flux, $\langle S_3\rangle$, where
$\mathbf{S} = \mathbf{B} \times (\mathbf{v}\times\mathbf{B}) =
 \mathbf{B}^2 \mathbf{v} - (\mathbf{v}\cdot\mathbf{B})\mathbf{B}$
is all of the magnetic contribution to the total energy flux (see Eq.(\ref{eq:MHD-equations})),
changes sign near optical depth unity.
Below this depth, in the convection zone, 
the intense downdrafts in the intergranular lanes pump magnetic fields
in the downward direction---cool and dense plumes compress and drag the magnetic 
field with them. This leads to a net Poynting flux in the downward direction.
Above $\tau_c = 1$, the Poynting flux that is connected to the magnetic field 
carried by the convective overshoot prevails, leading to a net Poynting flux in 
the upward direction.
We expect that at least a part of this flux is turned into heat via
ohmic dissipation by the current sheets that form on top of the
overshoot in the chromospheric layers (see
Fig.\,\ref{f:mhd_currentsheet}).

One purpose of performing \emph{realistic} simulations is the
\textcolor{tcol}{synthesis} of observable quantities
from the simulation data, which then become directly comparable 
to actual observations of the Sun. This typically involves the integration of the 
radiation transfer equation along lines of sight across the computational domain in 
order to obtain (two-dimensional) synthetic intensity maps. 
\textcolor{tcol}{This analysis step is performed post factum, after
completion of the simulation.} In case of 
magneto\-hydro\-dynamics, it requires the integration of 
the Unno-Rachkovsky equation for polarized radiation for obtaining intensity maps of the 
Stokes parameters. For direct comparison with observations from space-based or ground-based 
observatories, application of the corresponding instrumental point spread-function is 
necessary for degrading the synthetic intensity maps to the spatial resolution limits
of the actual observation. Degradation in frequency space and addition of noise may be 
required for taking into account the frequency resolution of the spectrometer and the 
photon noise of the recording device, respectively. 
Quantitative comparisons of synthesized spectropolarimetric maps from \COBOLD\ simulation 
data with corresponding intensity maps from the Hinode space observatory were performed
by \citet{steiner+al2008} who focus on the above discussed horizontal magnetic fields 
and by \citet{rezaei+al2007} with respect to intergranular magnetic flux concentrations. 

MHD simulations with \COBOLD\ are also performed for studying the excitation 
and propagation of magneto\-acoustic waves in the magnetically structured solar 
atmosphere. Effects of mode conversion, refraction, and transmission are studied
for application in solar atmospheric seismology
\cite{Steiner2007AN....328..323S,nutto+al2010,nutto+al2011,kato+al2010}.

\subsubsection{Next steps} 

Most if not all simulations of stellar magneto\-convection rely on the MHD
approximation. But the plasma in the photosphere of the Sun is weakly 
ionized so that the frozen-in condition may not really apply after all, 
despite the large scales of the magneto\-convective processes under 
consideration \citep[see, e.g.,][]{petrovic+al2007,singh+krishan2010}.
In the tenuous plasma of the chromosphere and corona, effects become
important which are not included in the single-fluid model of the standard 
MHD equations, and a multi-fluid model or even a kinetic description of the 
plasma may become necessary.

Two consequences of a multi-fluid description are the Hall effect and the 
ambipolar diffusion. Since the electrons and ions are moving on curved 
trajectories between collisions, the current density vector is no longer 
colinear with the electric field vector, i.e., the electric conductivity 
is a tensor. Ambipolar diffusion occurs in partially ionized plasmas, where 
collisions between charged and neutral particles produce new diffusion 
mechanisms. Since only the charged particles are coupled to the magnetic 
field, the \textcolor{tcol}{forces} acting on the different particles are different. This leads 
to a drift between charged and neutral particles modifying the transport of 
magnetic flux in the plasma.

In the limit of a weakly ionized plasma consisting of electrons, ions, and
neutrals, the single fluid description can still be applied along with an
appropriate modification of the induction equation. This modification
includes a term representing the Hall effect and a term representing
the ambipolar diffusion. We plan to implement an optional inclusion of 
these terms in the computation of the numerical fluxes for the induction
equation of \COBOLD.

Ohmic dissipation of electric currents is probably the most important
process in the heating of the outer solar atmosphere. In the present
version of the MHD module of \COBOLD, there is only an explicit implementation
of a turbulent subgrid-scale magnetic diffusion 
\textcolor{tcol}{(see Sect.~\ref{sec:OhmicDiffusion})}. 
However, for taking a 
significant resistivity into account, an implicit treatment must be implemented. 
Also the occurrence of anomalous resistivity should be accounted for.

Currently, a much discussed topic in solar physics is the origin of
the ubiquitous weak magnetic field outside sunspots. It is thought to
be generated through induction due to the turbulent motion of the
plasma near the solar surface, viz., by the turbulent dynamo
\cite{Cattaneo1999ApJ...515L..39C}.
On the  other hand, turbulent pumping transports this magnetic field in the downward 
direction, away from the surface into the deep, less turbulent layers of the 
convection zone
\cite{Tobias1998ApJ...502L.177T}.
It is unclear, which effect prevails.
\textcolor{tcol}{
First MHD simulations have been designed and carried out with the aim to improve our
understanding of this riddle (\citet{voegler+schuessler2007,moll+al2011}).
For a review see
\citet{stein2011} who argues that the solar dynamo has no preferred scale but 
rather acts throughout the convection zone over a wide range of scales.
}

Regarding global models of solar magneto\-hydro\-dynamics,
the long term goal should be the global, three-dimensional, numerical 
simulation of the entire solar convection zone including at least the 
overshoot layers and a proper radiative transfer at its boundaries,
but ideally also photosphere, chromosphere (see below), and corona. 
Thus, we seek a \emph{``Sun simulator''} as a laboratory for the holistic 
simulation of the Sun on scales from the stellar radius to the size of 
granules of about 1000~km. The outcome should be a virtually
self-consistent simulation of the differentially rotating convection 
zone including the self-exciting dynamo, which is thought to be at 
the \textcolor{tcol}{origin} of solar magnetic activity. Simulations should shed light 
on the functioning of the dynamo, the solar rotation law, the 
torsional oscillation, the luminosity variability, the sunspot cycle, 
and the global solar oscillation. Despite the apparent sphericity of 
stars, these processes are truly three-dimensional and therefore, 
they require a three-dimensional treatment unlike the traditional
one-dimensional approach in stellar-evolution modeling.

The development of such a simulation tool is a formidable task
(see Sect\,\ref{sec:BasisConsiderations}).
The main challenge at the beginning consists in the recognition of the relevant 
physics to be included and in finding the proper physical approximations, 
numerical scheme, and adaptive meshing for achieving sufficient spatial 
resolution. So far, global MHD simulations have not been carried out 
with \COBOLD. However, with \COBOLD\ it should be possible to
collect first experiences on the way to a true ``Sun simulator''.

\subsection{Solar chromosphere}
\label{sec:swb_solarchrom}

The chromosphere is the thin atmospheric layer between the photosphere and the 
transition region and corona above. 
Although these layers are coupled by magnetic fields and waves,
the properties of the atmospheric gas differ significantly. 
Compared to the photosphere below, the chromospheric gas is 
relatively thin, which has a number of important implications for the modeling
(see Sect.\,\ref{sec:swb_solarchromchall}) and also for observations.
There are only a few diagnostics suitable for probing the chromosphere, which 
makes it hard to derive constraints and viable reality checks for numerical models. 
Even worse, the interpretation of most of these diagnostics is
complicated by the fact that non-equilibrium effects must be taken
into account.
Advances in instrumentation both for ground-based and space-borne observations
during the recent years made it nevertheless possible to access the dynamic and 
intermittent fine structure at small spatial scales like it is seen in current 
chromosphere simulations (see, e.g., the review by \citet{2009SSRv..144..317W}).
The coexistence of magnetic fields and propagating waves and interaction 
of these makes
the modeling of the chromosphere a challenging task and a true hardness 
test for the stability of the code.

\subsubsection{Challenges in chromospheric modeling}
\label{sec:swb_solarchromchall}
%
%
\textit{Radiative transfer:} 
The gas becomes optically thin (i.e., essentially transparent) in a substantial 
wavelength range, leading to a strongly non-local coupling of regions within this 
layer and also with the layers below and above. 
However, the chromosphere is neither completely optically thin nor completely 
optically thick. 
The often used simplifying assumption of local thermodynamic equilibrium (LTE) 
breaks down in the chromosphere and effects like scattering become important. 
All this makes the numerical treatment of radiation challenging.
Detailed (non-LTE) calculations are today usually done for (a number of) single 
simulation snapshots but are still computationally too expensive to be included 
in 3D radiation magneto\-hydro\-dynamic simulations. 
Simplifications are unavoidable, so far.

\textit{Non-equilibrium effects:} 
Many more processes depart from equilibrium conditions in the thin chromosphere. 
For instance, the ionization degree of hydrogen can no longer be modeled under the 
assumption of an instantaneous equilibrium (see Sect.~\ref{sec:hionmodule}). 
Some processes become so slow that the detailed time evolution must be followed. 
A detailed treatment, however, is computationally expensive. 
Current approaches \citep{2002ApJ...572..626C,2006A&A...460..301L,
2007A&A...473..625L} use a hydrogen model atom with 6 energy levels and 
10~radiative transitions.  
The corresponding rate matrix can be stiff and is solved implicitly for each grid 
cell in the model chromosphere. 
Another example for non-equilibrium modeling is the application of 
chemical-reaction networks for carbon monoxide \cite{Wedemeyer2005A&A...438.1043W,
2007A&A...462L..31W,2006ASPC..354..301W}. 
%
See Sect.\,\ref{sec:OptionalModules} for a short description of the implementation 
of chemical-reaction networks and hydrogen ionization in \COBOLD. 
 
\textit{Computational time step:} 
The thermal pressure is so low that the magnetic pressure can become larger 
already at relatively low magnetic field strengths. 
The plasma-$\beta$, which is defined as the ratio of thermal to magnetic 
pressure, is consequently less than one above heights of $\sim 1000$\,km above 
the bottom of the photosphere. 
Under these conditions, the magnetic field is no longer advected passively and 
magnetic wave modes become important. 
A full magneto\-hydro\-dynamic approach is therefore necessary. 
The computational time step is then determined by the Alfv{\'e}n speed, which 
easily can result in steps of the order of milliseconds
(in simulations that cover a few hours),
depending on the magnetic 
configuration in the chromosphere.
This is a reduction by one to two orders of magnitude compared to purely 
hydro\-dynamic simulations. 
Under certain circumstances, an artificial reduction of the Alfv\'en speed can be
used, as is discussed in Sect.\,\ref{sec:DualEnergyMethod}.

\textit{Numerical stability:} 
Acoustic waves, which are continuously excited by the non-stationary surface convection
below, grow in amplitude while propagating into the thinner 
chromosphere. 
There, they develop (MHD) shocks with high peak temperatures of the order of 
7000\,K or more,
and the dynamical pressure exceeds the gas pressure.
The physical conditions in a grid cell can change drastically during the 
passage of a shock wave, which requires a high degree of stability of the numerical scheme. 
In the MHD case, the \textcolor{tcol}{occurrence} of strong gradients in thermal and magnetic 
properties can lead to exceptional situations in small parts of the computational 
domain.
In this respect, the HLL solver (see Sect.\,\ref{sec:MHD}) has been proven a 
good choice.


\subsubsection{Chromospheric modeling in the recent years}

The many complications in modeling the chromosphere made it inevitable to begin 
with simplified models and increase the degree of realism step by step.  
A very prominent example is the pioneering study by \citet{1994chdy.conf...47C, 1995ApJ...440L..29C}. 
They restricted the simulations to one spatial dimension but 
implemented a detailed radiative transfer treatment
\citep[cf.][]{1977A&A....54...61U}. 
\citet{2000ApJ...541..468S} succeeded to produce a 3D hydro\-dynamic model 
with a relatively coarse spatial resolution by using a simplified description of 
the radiative transfer, which nevertheless included scattering.
The 3D hydro\-dynamic simulations by \citet{Wedemeyer2004A&A...414.1121W}, which were carried out 
with \COBOLD, did not include scattering but used a higher spatial 
resolution. 
This type of local 3D models is restricted to a relatively small part of the 
atmospheric layers extending from the chromosphere into the upper convection 
zone.  
The latter is important as it provides an intrinsic driver for the atmospheric 
dynamics and thus avoids the need for an artificial driver like it is necessary in 
1D simulations. 

The step to multi-dimensional magneto\-hydro\-dynamic simulations of the chromosphere
has been performed only a few years ago.
%
The (non-local) radiative transfer is still a limiting factor.  
Consequently, simplifications of the radiative transfer are still necessary for 
3D MHD simulations of the solar chromosphere. 
\citet{2005ESASP.596E..65S} therefore used a frequency-independent (``gray'') radiative 
transport and a weak initial magnetic field for the first 3D MHD simulations 
with \COBOLD\ (see also \citet{Schaffenberger2006ASPC..354..345S}).
Further 2D numerical experiments \cite{2005ESASP.596E..16W}
combined higher magnetic field strengths ($B_0 = 100$\,G) with the treatment 
of chemical-reaction networks including carbon monoxide and the methylidyne 
radical CH in view of their diagnostic potential. 

The models mentioned above focus on the small-scale structure and dynamics 
of the solar chromosphere, while another class of models also includes 
the corona above
\cite{2002ApJ...572L.113G,2005ApJ...618.1031G, 2007ApJ...665.1469A, 
Martinez-Sykora2008ApJ...679..871M, Hansteen2010AAS...21630505H, Gudiksen2011A&A...531A.154G}. 
These models have a larger spatial extent so that the large-scale magnetic field 
structure can be fitted into the computational box. 
In order to keep the simulations feasible, compromises such as a lower 
spatial resolution were unavoidable for the earlier models. 
However, the progress in computational performance and efficient
numerical methods allows for higher spatial resolution and at the same time 
a larger number of implemented physical processes, producing models 
with a increasing degree of realism.  

%
\begin{figure} 
\centering
\includegraphics[]{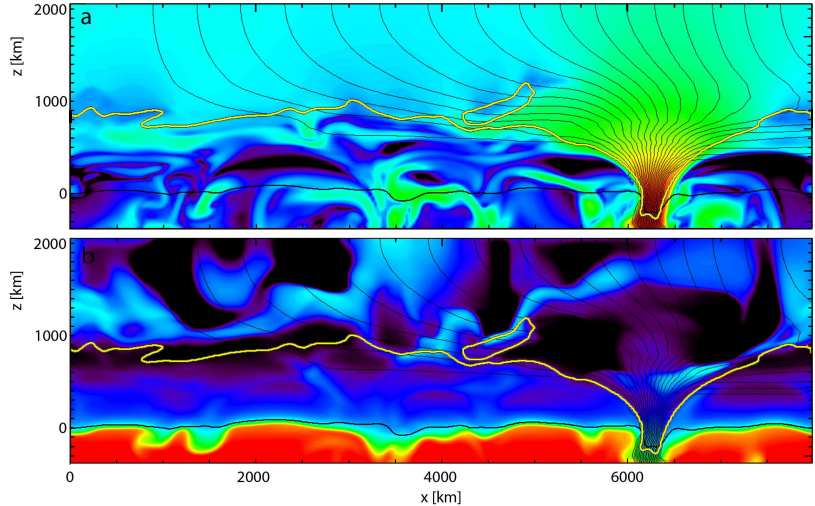}
\caption{Vertical cross sections through the upper layers of a 3D MHD model by 
	\citet{wedemeyerinprep}. 
	\textbf{a)} logarithmic magnetic field strength (color range 0.2 - 3.5 [G]), 
	\textbf{b)} gas temperature (color range 2500 - 12000 K). 
	The solid \textcolor{tcol}{curves} in both panels represent (projected) magnetic 
	field lines (thin black \textcolor{tcol}{curves}), 
	\textcolor{tcol}{the} contour for plasma-$\beta = 1$ 
	\textcolor{tcol}{(thick yellow contour)}, and 
	the height where the optical depth is unity \textcolor{tcol}{(horizontally
	running black curve around $z=0$)}. 
	The latter defines the bottom of the photosphere. 
  }
\label{fig:chromospherexz}
\end{figure}

The chromospheric layer of the hydro\-dynamic \COBOLD\ models by 
\citet{Wedemeyer2004A&A...414.1121W} exhibits a very dynamic and intermittent pattern made of 
propagating hot shock fronts and cool post-shock regions (cf. 
Fig.\,\ref{fig:chromospherexz}b).
The resulting fluctuations of the gas properties are substantial like it was found 
already from 1D simulations. 
As the shock fronts are very narrow, the peak temperatures of 7000--8000\,K 
in \COBOLD\ simulations depend to some degree on the resolution of the 
computational grid. 
Adiabatic expansion of the post-shock regions produces gas temperatures 
down to $\sim 2000$\,K. 
A similar shock pattern can already be perceived in the gas-temperature maps by 
\citet{2000ApJ...541..468S} but much less clearly due to the lower spatial resolution. 
\citet{Martinez-Sykora2008ApJ...679..871M}, who employed the Oslo Staggered Code for 3D 
simulations of magnetic-flux emergence, find a shock-induced pattern in their 
model chromosphere, too. 
The range in chromospheric gas temperature is similar to the
\COBOLD\ results, whereas there are differences in the temperature amplitudes.
%
This is presumably caused by the different numerical treatment of the radiative 
transfer in the upper layers.  
The existence of a chromospheric small-scale pattern in quiet regions of the Sun 
is now supported by recent observations 
\citep[e.g.,][]{2008A&A...480..515C, 2006A&A...459L...9W}.

Also  the MHD simulations carried out with
\COBOLD\ exhibit strong shock fronts in the chromosphere 
(see Fig.\,\ref{fig:chromospherexz}b). 
Compared to the photosphere below, the magnetic field in the model chromospheres is 
less concentrated and reaches a higher filling factor; it has a lower average field 
strength and evolves faster \cite{2005ESASP.596E..65S,Schaffenberger2006ASPC..354..345S}.
See Sect.\,\ref{sec:MagneticSun} for a description of the photosphere in this 
type of models. 
The topology of the chromospheric magnetic field is yet complex and features 
shock-induced compression and amplification into magnetic field filaments.
Fig.\,\ref{fig:chromospherexz}a, shows the upper layers of a current 3D simulation.
The vertical cross section is intersecting a magnetic flux concentration.
The magnetic field lines (thin solid lines, projected into the view plane) show that 
the field is highly concentrated in the photosphere and expands in the chromosphere 
above. 
The \textcolor{tcol}{thick yellow} curve represents the surface where plasma-$\beta = 1$.  
Although the height of this surface varies strongly, it is typically found around 
$z \sim 1000$\,km outside strong magnetic flux concentrations. 
For plasma-$\beta = 1$, sound speed and Alfv{\'e}n speed are 
\textcolor{tcol}{similar}, which has 
important implications for the occurrence, conversion, and propagation of 
different wave modes \citep[e.g.,][]{2002ApJ...564..508R,2003ApJ...599..626B,%
Cally2007AN....328..286C,Steiner2007AN....328..323S,nutto+al2011}.
The simulations indeed show a different behavior in the domains with 
$\beta < 1$ and $\beta > 1$, resulting in a slowly evolving lower part 
and a highly dynamic upper part. 
Current sheets exist below and above the $\beta = 1$ surface but 
differ in their orientation (see Fig.\,\ref{f:mhd_currentsheet}).  
They are stacked with predominantly horizontal orientation in the lower atmosphere 
(see Sect.\,\ref{sec:MagneticSun}) but are aligned with shock fronts in the low-$\beta$ regime in the 
chromosphere, resulting in oblique or even vertical orientation.

\subsubsection{Next steps} 
The next steps towards realistic models of the solar chromosphere concern an 
improvement of the radiative transfer under chromospheric conditions and the 
detailed treatment of non-equilibrium processes that have a significant impact on 
the equation of state and the opacities. 
However, the detailed modeling of non-equilibrium effects
might increase the computational costs to a degree that it can become impractical. 
For instance, important opacity sources that deviate from their equilibrium state
can prevent the usage of numerically efficient opacity look-up tables
(Sect.~\ref{sec:OpacityBinning}) and require a costly detailed line-by-line treatment
of the radiation transfer.
The inclusion of large-scale magnetic fields is another important point as most of the 
models discussed above resemble rather quiet Sun internetwork conditions with 
comparatively weak magnetic fields and thus cannot be applied to more
active regions on the Sun.
\textcolor{tcol}{Chromospheric simulations with \COBOLD\ have been restricted to heights below 
the transition region, where thermal conduction can still be neglected. 
Simulations that include the transition region and low corona have not been performed 
so far because it would require the implementation of the computationally expensive 
treatment of thermal conduction. }

\subsection{Local models of other stars}
\label{sec:MSstars}

\begin{figure}[htbp]
\centering
\includegraphics[width=4.5cm]{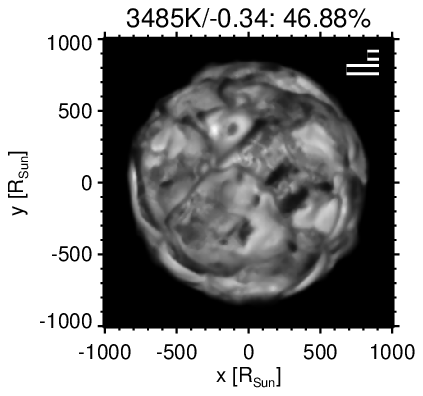}\includegraphics[width=4.5cm]{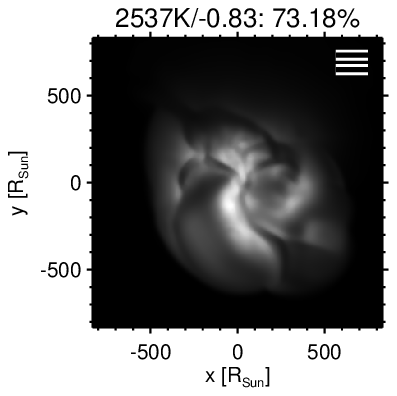}
\includegraphics[]{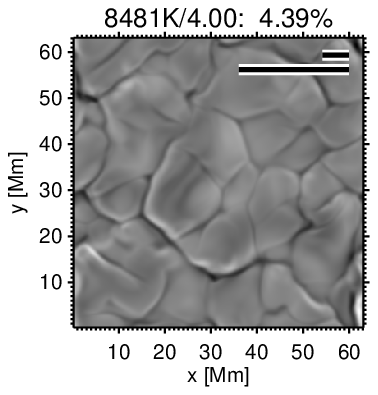}\includegraphics[]{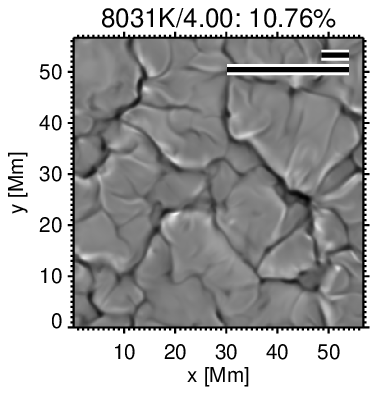}\includegraphics[]{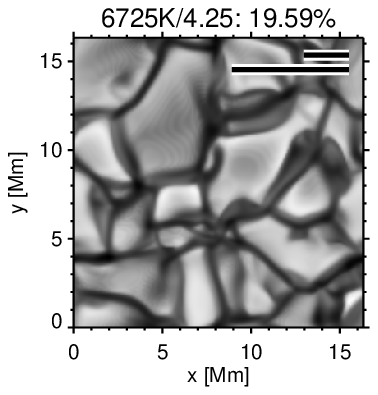}
\includegraphics[]{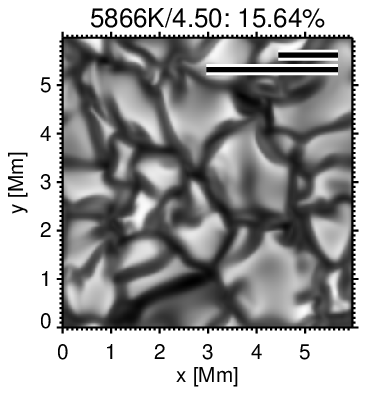}\includegraphics[]{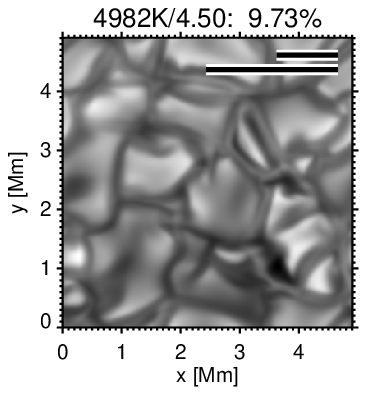}\includegraphics[]{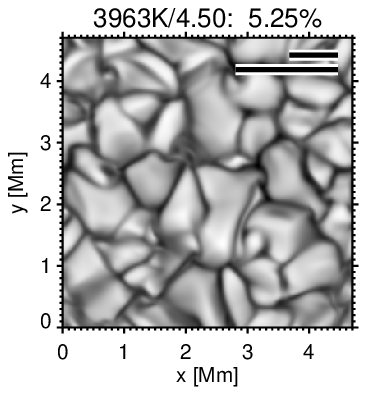}
\includegraphics[]{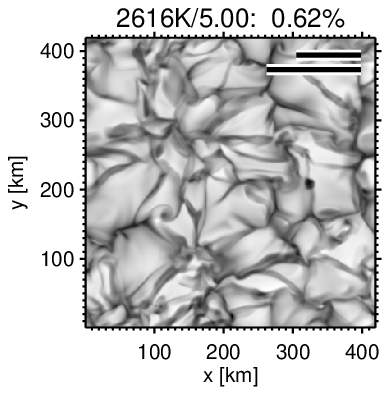}\includegraphics[]{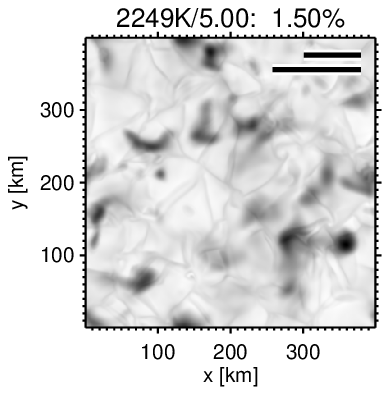}\includegraphics[]{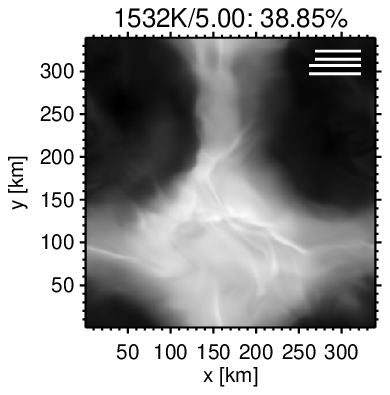}
\caption{Frequency-integrated-intensity snapshots of global models of a red supergiant and
an AGB star with low surface gravity (top row)
and local models of stars near the main sequence
with larger gravities.
The title lines show the effective temperature,
\textcolor{tcol}{the decadic} logarithm of  \textcolor{tcol}{the} surface 
gravity \textcolor{tcol}{in cm\,s$^2$},
and  \textcolor{tcol}{the} relative gray intensity contrast --
averaged over a representative time span.
The length of the upper bar in the top right of each frame
is 10~times the surface pressure scale height.
The bar below is 10~times the pressure scale height
but measured 3 pressure scale heights below the other level.
}
\label{f:SurfaceImagesAlongMS}
\end{figure}

\begin{figure}[htbp]
\centering
\includegraphics[width=9cm]{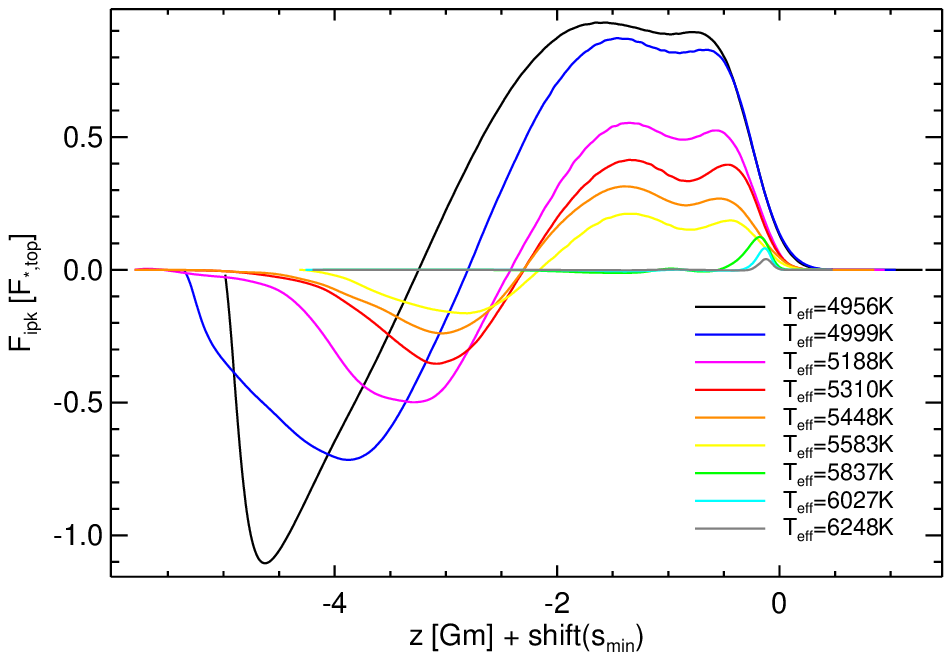}
\caption{
Convective fluxes (normalized by \textcolor{tcol}{the} flux emitted at 
\textcolor{tcol}{the} top of \textcolor{tcol}{the} model) versus
height for a temperature sequence of cepheid models at $\log g$=2.
The curves are shifted in $z$, so that the entropy minimum at the top
of the convection zone lies at $z$=0.
}
\label{f:seqavg_gt50g20n01_feipk0_xc3}
\end{figure}

Surface-intensity snapshots
from the hot A-type star regime (about 8500\,K)
over solar-type stars ($T_\mathrm{eff, Sun}$=5775\,K)
to brown dwarfs (about 1500\,K)
are shown in Fig.\,\ref{f:SurfaceImagesAlongMS}.
\emph{A-type stars}
\cite{Freytag2004IAUS..224..139F,Steffen2005ESASP.560..985S,Kochukhov2007IAUS..239...68K}
have only thin surface convection zones, where convection
carries only a small fraction of the total energy flux.
Of interest in these stars is the amount of overshoot \emph{below} the convection zone,
that causes a mixing of elements counteracting the separating effect of gravitational settling
and radiative acceleration \cite{Freytag1996A&A...313..497F}.
The shapes of spectral lines differ from solar counterparts \cite{Kochukhov2007IAUS..239...68K},
suggesting different atmospheric flow patterns
or deviating correlations between temperature fluctuations and velocity fields.
Due to the shorter radiative time scales,
caused by efficient radiative energy exchange,
the steeper and stronger subphotospheric temperature jump,
and the larger convective cells,
the simulations are numerically more challenging than those of solar-type stars,
requiring more numerical grid cells and many more time steps.
Therefore, the required CPU time per model goes up by a factor of 100 or more --
depending on the stellar parameters -- and an implicit treatment (at least of the
radiation transport) seems appropriate \cite{Viallet2011A&A...531A..86V}.
%
The transition from a thin, inefficient convection zone
to a deeper zone, where the convection carries in some layers almost
all the energy flux,
occurs in a similar way in the
temperature sequence of 2D cepheid models at $\log g$=2
in Fig.\,\ref{f:seqavg_gt50g20n01_feipk0_xc3}.
Remarkable is the extended overshoot region with significant \emph{negative} convective flux.

\emph{F, G, and K dwarfs} form a temperature sequence,
in which the convection zone gets deeper,
the Mach number of the convective velocity declines,
the relative efficiency of convection compared to radiation increases,
and the granular contrast decreases (Fig.\,\ref{f:Teffloggcontrast}).
While the amplitude of pressure waves drops rapidly,
the amplitude of gravity waves decreases more slowly: in solar models they are
more difficult to detect than pressure waves \cite{Straus2008ApJ...681L.125S},
but they dominate in brown dwarfs \cite{Freytag2010A&A...513A..19F}
and influence the shape of the dust clouds (coolest models in Fig.\,\ref{f:SurfaceImagesAlongMS}).
The change in the amount of photospheric overshoot
and the optical depth \textcolor{tcol}{at} the top of the convection zone
affect the appearance of granules in Fig.\,\ref{f:SurfaceImagesAlongMS}.
The scale of the granules seems related to the surface -- or rather the sub-surface --
pressure scale height \cite{Freytag1997svlt.work..316F}.
Both are indicated by the horizontal bars in Fig.\,\ref{f:SurfaceImagesAlongMS}, which
have lengths of 10\,$H_p$,
measured at two different heights.

The first multi-D radiation-hydro\-dynamical model atmospheres for \emph{M-type stars}
were calculated by Ludwig and collaborators
\cite{Ludwig2002A&A...395...99L,Ludwig2006A&A...459..599L}
using the simulation code of Nordlund and
Stein \cite{Stein1998ApJ...499..914S}.
An important issue,
that needed to be settled in the model construction,
was the handling of molecular opacities in
the opacity-binning scheme (see Sect.~\ref{sec:OpacityBinning}). It turned out
that no particular treatment is necessary, as long as the molecular opacities
dominate the atmospheric opacities.
Subsequently, \citet{Wende2009A&A...508.1429W} used \COBOLD\ with
the previously developed opacity set-up
to calculate a sequence of models
covering the main-sequence in the temperature range
$2600\,\mathrm{K}\le T_\mathrm{eff}\le 4000\,\mathrm{K}$, and probing surface gravities
$3.0\le\log(g)\le 5.0$ at fixed effective temperature of $\approx
3300\,\mathrm{K}$. Motivated by observational demands, the authors
investigated the impact of the velocity field and thermal structure on
properties of FeH lines.

While the 2616\,K model of an M-dwarf in Fig.\,\ref{f:SurfaceImagesAlongMS}
does not show visible amounts of dust,
in \emph{brown dwarfs} at even lower temperatures,
dust clouds begin to form in the atmospheres \cite{Freytag2010A&A...513A..19F},
that show up as small dark patches (shaped by gravity waves)
above the low-contrast granules in the 2249\,K model in Fig.\,\ref{f:SurfaceImagesAlongMS}.
The coolest model in Fig.\,\ref{f:SurfaceImagesAlongMS} has completely opaque dust clouds,
that hide the underlying gas convection zone.
A long-wavelength and high-amplitude gravity wave
creates the large-scale pattern in the plot,
while convection within the dust clouds causes the thin bright filaments.
The numerical challenges in these simulations
come from the -- hardly known -- non-equilibrium dust chemistry,
the long relaxation times of dust settling and mixing,
and the need to account for the likely interaction of small ``cloud'' scales,
covered by the local models in \citet{Freytag2010A&A...513A..19F},
and global ``weather front'' scales, far beyond the size of the computational domain
of the local models.

Cool hydrogen-rich \emph{white dwarfs} are old stars, that have used up their nuclear fuel
and cool down slowly.
The large surface gravity causes high atmospheric pressures,
small convective scales,
and a settling of elements heavier than hydrogen from the atmosphere into subsurface layers.
Recently, first models for those objects were computed
with \COBOLD\ \citep{Tremblay2011A&A...531L..19T}.
The work follows up on earlier investigations by some of the
authors (HGL, BF, MS) and
collaborators \citep{Ludwig1994A&A...284..105L,Steffen1995A&A...300..473S,Gautschy1996A&A...311..493G}
using a code which was a precursor of \COBOLD. Similar to the case of M-type
models, the intention is to study the influence of multi-D effects on the
formation of Balmer lines.


\subsection{Global models of supergiants and AGB stars}
\label{sec:SupergiantsAGBstars}

Early explanations of the irregular light curves of \emph{red supergiants}
as due to giant convection cells by
\citet{Stothers1971A&A....10..290S}
and
\citet{Schwarzschild1975ApJ...195..137S}
got more recently support by interferometric detections of 
spatial inhomogeneities on the surface of Betelgeuse
(e.g., \cite{Buscher1990MNRAS.245P...7B}).
When convective scales are not very small compared to the stellar diameter,
global star-in-a-box simulations are possible (Sect.\,\ref{sec:GlobalModels})
making these stars the easiest targets for global models,
that include a major part of the convective
envelope as well as the near environment of the star.
The top row of Fig.\,\ref{f:SurfaceImagesAlongMS} shows snapshots
of the emergent intensity of such models:
on the left from a red-supergiant \cite{Freytag2002AN....323..213F}
and on the right from an AGB-star simulation \cite{Freytag2008A&A...483..571F}.
The supergiant model confirms that convective scales are indeed huge,
with a few very large, deep, long-lived envelope cells
and many small, short-lived surface cells.
The surface contrast is enormous (see Fig.\,\ref{f:Teffloggcontrast}),
due to violent convective flows and in addition waves, that have already in the lower
photosphere a large amplitude, in contrast to the solar case.
Large scales and contrast values render the features observable with current interferometers:
the models compare favorably with VLTI observations
\cite{Chiavassa2009A&A...506.1351C,Chiavassa2010A&A...511A..51C,Chiavassa2010A&A...515A..12C}
indicating that these global models start to become ``realistic'', too.

RHD simulations of an \emph{AGB star}
(Fig.\,\ref{f:SurfaceImagesAlongMS}, top right \cite{Freytag2008A&A...483..571F})
demonstrate that
convection can excite pressure waves with amplitudes sufficient
to turn them into shocks,
which then push out dense material into layers cool enough that dust can form
(included in the 3D models, see Sect.\,\ref{sec:Dust}).
This allows radiation pressure on dust to accelerate the material outward
causing a stellar wind (not included yet in the 3D models of \citet{Freytag2008A&A...483..571F}
but in the 1D simulations of
\citet{Hoefner1995A&A...297..815H, Hoefner2003A&A...399..589H}).
Major challenges for the simulations are posed
by the molecular opacities varying strongly with frequency,
that cause -- together with large dust opacities --
very small radiative time steps during the time-explicit treatment
of the radiative energy exchange.
So far, only 1D RHD models (e.g., \cite{Hoefner2003A&A...399..589H})
include important ingredients like
scattering,
radiation pressure,
and a sufficiently large computational volume to account for the
extended wind acceleration region.
While the properties of the simulated surface granulation seem already quite realistic,
there are discrepancies further out:
The models have a too steep density drop and show no ``molsphere'', chromosphere,
or wind.
Future generations of models will help to investigate the role of
radiation pressure on molecular lines and dust, magnetic fields, and rotation
for these outer layers.

\section{Conclusions}

For \textcolor{tcol}{stellar} parameters close to the solar values,
the transition from
1D static stellar-atmosphere models to
3D dynamic local stellar-atmosphere simulations
is in full swing.
RHD simulations of surface convection of such stars are
routinely performed by codes like \COBOLD, as presented in this paper.
There is a good consistency between the results of similar codes
\textcolor{tcol}{and with solar observations}.
\textcolor{tcol}{The simulations} provide insight into processes related 
to stellar surface convection,
and deliver high-accuracy model atmospheres for spectrum-synthesis
and abundance-determination work
for a variety of stellar parameters.

However, many physical effects are not properly incorporated by the current models:
small-scale convective structures, covered by local-box simulations,
interact with their environment
via, for instance, large-scale convective flows,
magnetic fields,
waves,
or global dust flows (the latter in cool substellar objects).
%
The inclusion of \textcolor{tcol}{the} chromosphere, \textcolor{tcol}{the} corona, 
\textcolor{tcol}{the} wind-formation zone, etc., requires
to cover an even wider range of densities and temperatures
than the previous models.
At lower densities, the detailed treatment of non-equilibrium processes
(molecule formation, radiation transport not in local thermal equilibrium)
are a major challenge, requiring algorithms with a complexity far beyond the current treatment.
The modeling of magnetic phenomena needs appropriate MHD or plasma-physics simulations.
Objects significantly cooler than the Sun require a detailed non-equilibrium treatment of dust
and ``weather phenomena''.
Varying efficiency ratios between radiation and convection
\textcolor{tcol}{as a function of}  stellar parameters
have to be considered:
the extremely small radiative relaxation time scale in hotter stars
causes small numerical time steps and
slows down simulations significantly.
Cool objects on the other hand need extended simulation runs
because of their long thermal relaxation time scales.
Low-gravity objects have extended atmospheres and can produce winds.
Magnetic-field phenomena exist on very large and very small scales
and couple the stellar interior \textcolor{tcol}{to the} photosphere 
and \textcolor{tcol}{the} envelope.
In stars more active than the Sun, the fields are stronger and can form
very different configurations.

In the future, we will see a refinement of local simulations and
more and more extended model grids,
providing reliable stellar model atmospheres.
However, the main challenges arise from the need to
extend the simulations in terms of
stellar parameters (from A-type stars to planetary objects and 
from supergiants to white dwarfs),
\textcolor{tcol}{from} physical effects, and 
\textcolor{tcol}{from the extension of}
spatial and temporal scales
towards 3D large-scale or even global dynamic models.
These \textcolor{tcol}{models} should not only include the photosphere
but the stellar interior and the outer atmospheric layers \textcolor{tcol}{as well},
covering short and long time scales
(many rotation periods, dynamo cycles, stellar oscillation periods, climate cycles).

Realistic global 3D MHD simulations for cool stars
will remain \textcolor{tcol}{a dream} for the foreseeable future.
Nevertheless, numerical simulations will continue to be 
\textcolor{tcol}{indispensable tools} for \textcolor{tcol}{the} 
understanding \textcolor{tcol}{of} the various complex dynamical 
processes in stars.







\bibliographystyle{model1-num-names}
\bibliography{aa_redsg,MHD-papers,jcp_swb,jcp2010steiner}







\end{document}